\documentclass[a4paper,11pt]{article}
\usepackage{jheppub} 

\usepackage{amsfonts}
\usepackage{amsmath}
\allowdisplaybreaks[4]        
\usepackage{amssymb}
\usepackage{euscript}     
\usepackage{color}         
\usepackage{tensor}        
\usepackage{amsthm}
\usepackage{graphicx}
\usepackage{subfigure}
\usepackage{bbm}
\usepackage[header,title,page,titletoc]{appendix}  

\usepackage[numbers]{natbib}  
\usepackage{float}

\newcommand{\be}{\begin{equation}}
\newcommand{\bea}{\begin{eqnarray}}
\newcommand{\eea}{\end{eqnarray}}
\newcommand{\ba}{\begin{array}}
\newcommand{\ea}{\end{array}}
\newcommand{\ee}{\end{equation}}
\newcommand{\bes}{\begin{equation*}}
\newcommand{\beas}{\begin{eqnarray*}}
\newcommand{\eeas}{\end{eqnarray*}}
\newcommand{\bas}{\begin{array*}}
\newcommand{\eas}{\end{array*}}
\newcommand{\ees}{\end{equation*}}

\title{\boldmath Aspects of Hyperscaling Violating Geometries at Finite Cutoff}

\author[a]{Salomeh Khoeini-Moghaddam,}
\author[b,1]{Farzad Omidi ,\note{Corresponding author.}}
\author[c]{Chandrima Paul}

\affiliation[a]{Department of Astronomy and High Energy Physics, Faculty of Physics, 
	\\Kharazmi University, Tehran 15719-14911, Iran }
\affiliation[b]{School of Physics, Institute for Research in Fundamental Sciences (IPM),\\
	P.O. Box 19395-5531, Tehran, Iran}
\affiliation[c]{ Department of Theoretical Physics, Institute of Mathematical Science, \\
	4th Cross Street,C.I.T  campus, Tharmani, Chennai,Tamilnadu, 600113, India}

\emailAdd{skhoeini@khu.ac.ir}
\emailAdd{farzad@ipm.ir}
\emailAdd{chandrimap@imsc.res.in}
		
\abstract{
Recently, it was proposed that a $T\overline{T}$ deformed CFT is dual to a gravity theory in an asymptotically AdS spacetime at finite radial cutoff. Motivated by this proposal, we explore some aspects of Hyperscaling Violating geometries at finite cutoff and zero temperature. We study holographic entanglement entropy, mutual information (HMI) and entanglement wedge cross section (EWCS) for entangling regions in the shape of strips. It is observed that the HMI shows interesting features in comparison to the very small cutoff case: It is a decreasing function of the cutoff. It is finite when the distance between the two entangling regions goes to zero. The location of its phase transition also depends on the cutoff, and decreases by increasing the cutoff. On the other hand, the EWCS is a decreasing function of the cutoff. It does not show a discontinuous phase transition when the HMI undergoes a first-order phase transition. However, its concavity changes. Moreover, it is finite when the distance between the two strips goes to zero. Furthermore, it satisfies the bound $ E_W \geq \frac{I}{2}$ for all values of the cutoff.}


\arxivnumber{2011.00305}

\begin{document} 
	
	\begin{flushright}
	 IPM/P-2020/057   \\	
	\end{flushright}
	
	\maketitle
	\flushbottom

\section{Introduction}
\label{Sec: Introduction}

In recent years some kinds of very rich irrelevant deformations of two-dimensional conformal field theories (CFT), dubbed {\it $T \overline{T}$ deformations}, have been introduced \cite{Zamolodchikov:2004ce,Smirnov:2016lqw,Cavaglia:2016oda}, under which the action is changed as follows
\bea
\frac{\partial S(\lambda)}{\partial \lambda} = \int d^2x \sqrt{g} \;T \overline{T} (x),
\eea 
where $\lambda$ is the deformation parameter and $T \overline{T}(x)$ is an irrelevant operator  of dimension $(2,2)$ which is defined by \cite{Zamolodchikov:2004ce,Smirnov:2016lqw,Cavaglia:2016oda,Taylor:2018xcy}
\bea
T \overline{T}(x) = \lim_{y \rightarrow x} \left( T^{\alpha \beta}(x) T_{\alpha \beta}(y) - T^{\alpha}_{\alpha}(x) T^{\beta}_{\beta}(y) \right)  - \sum_i A_i (x-y) \nabla_y O_i (y).
\eea 
Here $T_{\alpha \beta}$ is the stress tensor of the CFT, $A_i(x,y)$ is a function which might be divergent in the limit $y \rightarrow x$ and  $O_i$ are local operators. The remarkable feature of these deformations is that they are solvable, in the sense that some quantities of the deformed CFT such as energy spectrum
can be calculated exactly \cite{Zamolodchikov:2004ce,Smirnov:2016lqw,Cavaglia:2016oda}. Moreover, analogues of these deformations were recently explored in higher dimensions, i.e. $d>2$, \cite{Cardy:2018sdv,Bonelli:2018kik,Taylor:2018xcy,Hartman:2018tkw} and in one dimension \cite{Gross:2019ach,Gross:2019uxi}.
\footnote{In ref. \cite {Bonelli:2018kik} a method was introduced which enables one to find the exact action of the $T \overline{T}$ deformed theory. This method was applied for several QFTs including non-linear sigma, WZW and massive Thirring models.}
\\In the context of AdS/CFT \cite{Maldacena:1997re}, it has been proposed in ref. \cite{McGough:2016lol} (see also \cite{Kraus:2018xrn,Taylor:2018xcy,Hartman:2018tkw}) that the holographic dual of a $T\overline{T}$ deformed $CFT_{d+1}$ is a gravity theory in an asymptotically $AdS_{d+2}$ spacetime with a radial cutoff on which one imposes Dirichlet boundary conditions on the fields. Moreover, the deformation parameter $\lambda$ in the deformed CFT is related to the cutoff $r_c$ on the radial coordinate in the bulk spacetime (See Section \ref{Sec: TTbar-deformed-CFTd+1} for more details). Furthermore, different aspects of $T \overline{T}$ deformed CFTs both on the QFT and holography sides were studied extensively including correlation functions \cite{Kraus:2018xrn,Aharony:2018vux,Cardy:2019qao,He:2019vzf,He:2019ahx,He:2020udl,He:2020cxp}, entanglement entropy \cite{Donnelly:2018bef,Chen:2018eqk,Park:2018snf,Banerjee:2019ewu,Jeong:2019ylz,Grieninger:2019zts,Murdia:2019fax,Donnelly:2019pie}, mutual information \cite{Asrat:2020uib} and entanglement wedge cross section \cite{Asrat:2020uib}.
\\ Motivated by the proposal of ref. \cite{McGough:2016lol}, we consider Hyperscaling Violating (HV) geometries given in eq. \eqref{metric-vacuum} at finite radial cutoff and zero temperature. 
\footnote{This problem has also been studied in ref. \cite{Paul:2020gou}.}
Recently, in ref. \cite{Alishahiha:2019lng}, it was proposed that these geometries might be dual to $T \overline{T}$ like deformation of QFTs in which Lorentz and scaling symmetries are broken. In the following, we call these QFTs, HV QFTs. Moreover, the deformation operator, energy spectrum and Action-Complexity of these geometries at finite temperature were also studied in ref. \cite{Alishahiha:2019lng}. Furthermore, it should be emphasized that $T \overline{T}$ deformations of two dimensional QFTs which do not have Lorentz symmetry was also explored in ref. \cite{Cardy:2018jho}. The aim of this paper is to explore the effects of a finite radial cutoff in the bulk spacetime on some of the quantum entanglement measures such as Holographic Entanglement Entropy (HEE), Mutual Information (HMI), and Entanglement Wedge Cross Section (EWCS) in these geometries when the entangling regions are in the shape of strips. In particular, we explore the dependence of these quantities on the cutoff, and compare their behaviors to those when the cutoff is very small. In the following, we briefly review these quantities.
\\One of the remarkable and rigorous quantities which is able to measure quantum entanglement in a quantum system is entanglement entropy (EE). 
Suppose a subsystem $A$ on a constant time slice of the manifold on which a quantum field theory (QFT) lives. By knowing the density matrix $\rho$ of the QFT, one can assign a reduced density matrix $\rho_A= {\rm Tr}_A \rho$ to $A$, where the trace is taken over the Hilbert space $\mathcal{H}_A$ corresponding to $A$. Then the EE between the degrees of freedom inside and outside $A$ is defined as the Von-Neumann entropy $S= - {\rm Tr ( \rho_A \log \rho_A)}$, which can be calculated by the replica trick \cite{Ryu:2006ef,Rangamani:2016dms}. In the context of the AdS/CFT correspondence \cite{Maldacena:1997re}, there is a very brilliant prescription known as the Ryu-Takayanagi (RT) proposal for the calculation of EE \cite{Ryu:2006bv,Ryu:2006ef}, which is proved in ref. \cite{Lewkowycz:2013nqa}. According to the RT proposal, for holographic CFTs the holographic dual of EE is given by
\footnote{Here we restrict ourselves to time independent states which are dual to static bulk spacetimes. Otherwise, one should apply the covariant prescription of the RT formula, known as HRT \cite{Hubeny:2007xt}.}
\bea
S_A = \frac{{\rm Area}(\Gamma_A)}{4 G_N},
\label{RT}
\eea 
where $G_N$ is the Newton constant in the bulk spacetime. Moreover, $\Gamma_A$ is a codimension two, spcaelike minimal surface in the bulk spacetime which is homologous to the subsystem $A$, i.e. $\partial \Gamma_A = \partial A$. The above formula is valid to the leading order in the number $N$ of the degrees of freedom of the CFT, and there is a sub-leading quantum correction to eq. \eqref{RT} (see \cite{Faulkner:2013ana,Rangamani:2016dms}) where we omit it here. 
\\On the other hand, EE is a divergent quantity in QFT and depends on the UV cutoff of the theory. However, one can define other measures which are independent of the UV cutoff. One of these quantities is quantum mutual information (MI) which is a measure of both classical and quantum correlations in a bipartite system. For two subsystems $A$ and $B$, MI is defined as follows
s\bea
I(A,B) = S_A + S_B - S_{AB},
\label{MI}
\eea 
where $S_{AB} = S_{A \cup B}$. It has several interesting  properties including \cite{Casini:2004bw,Casini:2005rm,Casini:2006ws,Casini:2008wt,Headrick:2010zt,Swingle:2010jz,Tonni:2010pv,Fischler:2012uv}: it is always non-negative as a result of subadditivity. It is finite and independent of the UV cutoff.
\footnote{It should be pointed out that for singular entangling regions whose geometrical singularities are coincident with each other, the HMI depends on the UV cutoff \cite{Mozaffar:2015xue}.}
It shows a first-order phase transition when the two subsystems becomes far enough from each other. Moreover, when the distance between the entangling regions goes to zero, it diverges.
\\On the other hand, when the QFT is in a mixed state, EE is not an appropriate measure of quantum entanglement, and one should apply other measures. One of these quantities is Entanglement of Purification (EoP) \cite{EoP-1}, which measures both classical and quantum correlations. Suppose that the density matrix $\rho_{AB}$ of two subsystems $A$ and $B$ is mixed. By enlarging the Hilbert space to $\mathcal{H}_{A A^\prime} \otimes \mathcal{H}_{B B^\prime}$ in which $A^\prime$ and $B^\prime$ are two arbitrary subsystems, one can find a pure state $| \psi \rangle \in \mathcal{H}_{A A^\prime} \otimes \mathcal{H}_{B B^\prime}$, such that $\rho_{AB} = {\rm Tr}_{A^\prime B^\prime} | \psi \rangle \langle \psi |$. In this case, $| \psi \rangle $ is called a purification of $\rho_{AB}$. Then one can define the EoP as follows \cite{EoP-1}
\bea
E_P(\rho_{AB}) = \min_{\rho_{AB} = {\rm Tr}_{A^\prime B^\prime} | \psi \rangle \langle \psi |} S(\rho_{A A^\prime}),
\label{EoP}
\eea 
where $\rho_{AA^\prime} = {\rm Tr}_{B B^\prime}  | \psi \rangle \langle \psi |$ and the minimization is done over all possible purifications of $\rho_{AB}$.
\\On the other hand, for a bipartite system consisting of $A$ and $B$, one can define another quantity called entanglement wedge cross section (EWCS) as follows \cite{Takayanagi:2017knl,Nguyen:2017yqw}
\bea
E_{W} = \frac{Area(\Sigma^{\rm min}_{AB})}{4 G_N},
\label{EW-1}
\eea 
where $\Sigma^{\rm min}_{AB}$ is a minimal, codimension two, spacelike surface anchored on the RT surface $\Gamma_{AB}$ corresponding to the region $A \cup B$ (see figure \ref{fig:HMI-EWCS}). Moreover, $\Sigma^{\rm min}_{AB}$ divides the entanglement wedge $M_{AB}$ corresponding to $A \cup B$ into two parts (see section \ref{Sec: Entanglement Wedge Cross Section} for more details). In other words, $E_W$ measures the minimal cross section of the entanglement wedge $M_{AB}$. EWCS has a variety of interesting properties including \cite{Takayanagi:2017knl,Nguyen:2017yqw}: it is non-negative, finite and independent of the UV cutoff. When $\rho_{AB}$ is pure, one has $E_W = S_A= S_B$. When the two entangling regions are far enough from each other, $M_{AB}$ becomes disconnected (see the right panel of figure \ref{fig:HMI-EWCS}). Consequently, EWCS undergoes a discontinuous phase transition when the HMI shows a first-order phase transition (see also \cite{Liu:2019qje,BabaeiVelni:2019pkw}). Furthermore, it satisfies a variety of inequalities such as
\bea
E_W(\rho_{AB}) \geq \frac{I(A,B)}{2},
\label{EW geq I}
\eea 
where the inequality is saturated, whenever $\rho_{AB}$ is pure.
It was observed in refs. \cite{Takayanagi:2017knl,Nguyen:2017yqw} that the properties of EoP \cite{EoP-1,EoP-2} are exactly the same as those of EWCS. Consequently, it was proposed that EWCS is the holographic dual of EoP \cite{Takayanagi:2017knl,Nguyen:2017yqw} 
\footnote{It should be emphasized that  minimization over all possible purifications in eq. \eqref{EoP} is a difficult task. Therefore, other candidates for the CFT counterpart of EWCS were introduced in the literature, such as logarithmic negativity \cite{Kudler-Flam:2018qjo}, odd entanglement entropy \cite{Tamaoka:2018ned} and reflected entropy \cite{Dutta:2019gen,Kusuki:2019rbk}.}
\bea
E_P (\rho_{AB}) = E_W(\rho_{AB}).
\label{EOP=EW}
\eea 
This conjecture has been explored extensively in refs. \cite{Bao:2017nhh,Hirai:2018jwy,Bao:2018gck,Agon:2018lwq,Bao:2018fso,Umemoto:2018jpc,Caputa:2018xuf,Yang:2018gfq,Liu:2019qje,Ghodrati:2019hnn,BabaeiVelni:2019pkw}. 
\\The organization of the paper is as follows: in Section \ref{Sec: Hyperscaling Violating Geometries}, we first review some features of $T \overline{T}$-deformed CFTs in d+1 dimensions. Next, we review HV geometries at zero and finite radial cutoff as well as their dual QFTs. In Section \ref{Sec: Holographic Entanglement Entropy for a Strip}, we numerically calculate the HEE for entangling regions in the shape of strips. Moreover, we find analytic expressions for very large and small entangling regions as well as very small cutoff. In Section \ref{Sec: Holographic Mutual Information}, we first briefly review the holographic prescription for the calculation of MI. Next, we compute the HMI for two parallel disjoint strips and investigate its dependence on the cutoff $r_c$. It is observed that both the HMI and the location of its phase transition depends on the cutoff. In Section \ref{Sec: Entanglement Wedge Cross Section}, we first review EWCS which is believed to be the holographic dual of EoP. Next, we calculate it for two parallel disjoint strips, and study its behavior. It is observed that the EWCS is continuous at the point where the HMI undergoes a phase transition. However, the concavity of the EWCS changes at this point. Moreover, it is verified that eq. \eqref{EW geq I} is valid for all values of the cutoff. In Section \ref{Sec: Discussion}, we summarize our results.
\section{Deformed CFTs and HV QFTs}
\label{Sec: Hyperscaling Violating Geometries}
In this section, we first review some properties of $T \overline{T}$-deformed CFTs in d+1 dimensions. Next, we discuss HV geometries at zero and finite cutoff  and their possible dual QFTs. In our conventions the bulk radial coordinate is indicated by $r$ and the boundary is located at $r=0$. In the following, the case in which the radial cutoff $r_c$ is very small, i.e. $r_c= \epsilon \rightarrow 0$,  is dubbed the {\it zero cutoff case}. 
On the other hand, when the the radial cutoff is finite is called the {\it finite cutoff} case.
\subsection{$T \overline{T}$ deformed $CFT_{d+1}$}
\label{Sec: TTbar-deformed-CFTd+1}
Consider a $T \overline{T}$-deformed $CFT_{d+1}$ on a flat manifold whose action satisfies the following equation
\bea
\frac{\partial S(\lambda)}{\partial \lambda} = \int d^{d+1}x \sqrt{g} \; X  (x).
\label{TT-CFT-d+1}
\eea 
To find the $T \overline{T}$ deformation operator $X$, one can calculate the renormalized Brown-York stress tensor in the dual gravity. After adding the appropriate counterterms, one can show that when the boundary manifold is flat and there are no matter fields inside the bulk spacetime, the radial-radial component of Einstein's equations leads to the following constraint on the components of the stress tensor $T_{ij}$ in the $T \overline{T}$ deformed $CFT_{d+1}$ \cite{McGough:2016lol,Kraus:2018xrn,Taylor:2018xcy,Hartman:2018tkw}
\footnote{Note that our radial coordinate is the inverse of the radial coordinate in ref. \cite{Hartman:2018tkw}, .i.e. $r_{\rm here} = \frac{1}{r_{\rm there}}$.}
\bea
T^i_i = - 4 \pi G_N r_c^{d+1} \left( T^{ij} T_{ij} - \frac{1}{d} \left( T^i_i \right)^2 \right).
\label{trace-flow-CFT-1}
\eea 
On the other hand, if there is only a dimensionful scale $\lambda$ in the deformed $CFT_{d+1}$, one can find the following equation \cite{McGough:2016lol,Kraus:2018xrn,Hartman:2018tkw}
\bea
T^i_i = - (d+1) \lambda X,
\label{trace-flow-CFT-2}
\eea 
which is called the {\it trace flow} equation. Next, by comparison of eq. \eqref{trace-flow-CFT-1} and \eqref{trace-flow-CFT-2}, one might conclude that \cite{McGough:2016lol,Kraus:2018xrn,Taylor:2018xcy,Hartman:2018tkw} the deformation parameter is given by
\bea
\lambda = \frac{4 \pi G_N}{(d+1)} r_c^{d+1},
\label{lambda-CFT}
\eea 
where
$r_c$ is the radial cutoff in the bulk spacetime. Moreover, one obtains the $T \overline{T}$ deformation operator as follows
\bea
X =  T^{ij} T_{ij} - \frac{1}{d} \left( T^i_i \right)^2.
\label{TT-operator-CFT}
\eea 
It should be pointed out that when the boundary manifold is curved or there are matter fields in the bulk spacetime, one should add more terms on both sides of the above equation \cite{McGough:2016lol,Kraus:2018xrn,Taylor:2018xcy,Hartman:2018tkw}.
\\Another amazing property of $T \overline{T}$-deformed CFTs is the factorization property \cite{Zamolodchikov:2004ce,Taylor:2018xcy,Hartman:2018tkw} of the deformation operator
\bea
\langle X \rangle =  \langle T^{ij} \rangle \langle T_{ij} \rangle - \frac{1}{d} \langle T^i_i \rangle^2,
\label{factorization-CFT-d+1}
\eea 
which has a prominent role in obtaining exact results in the deformed CFT, such as in the calculation of energy levels. This property was proved in two dimensions \cite{Zamolodchikov:2004ce} and expected to be valid in higher dimensions in the large N limit \cite{McGough:2016lol,Hartman:2018tkw,Grieninger:2019zts} (see also \cite{Taylor:2018xcy}). 

\subsection{HV Geometries: Zero Cutoff Case}
\label{Sec: Zero Cutoff Case}
It was shown in ref. \cite{Charmousis:2010zz,Gouteraux:2011ce,Alishahiha:2012qu} that by turning on an appropriate combination of scalar and gauge fields in the bulk spacetime, one can make geometries called HV geometries whose metrics are given by
\bea
ds^2 = \frac{R^2}{r_F^{2 \theta_e}} r^{ 2 (\theta_e -1)} \left( - r^{-2 (z-1)} dt^2  + dr^2 + \sum_{i=1}^{d} dx_i^2 \right),
\label{metric-vacuum}
\eea
where $R$ is the AdS radius and $r_F$ is a dynamical scale.
\footnote{When the dual QFT has a Fermi surface, $r_F$ is fixed by the inverse of the radius of the Fermi surface \cite{Dong:2012se}.}
Moreover, we defined an effective hyperscaling violation exponent $\theta_e = \frac{\theta}{d}$. Here $z$ and $\theta$ are the dynamical critical and hyperscaling violation exponents, respectively, whose values are restricted by the null energy conditions \cite{Dong:2012se}
\bea
(d- \theta) (d(z-1) - \theta) \geq 0, \;\;\;\;\;\;\;\;\;\;\;\;\;\;\;\;\;\;\; (z-1) (d- \theta+z ) \geq 0.
\eea 
It was observed in \cite{Dong:2012se} that, when $ \theta >d$, the HEE scales faster than the volume of the entangling region which is not consistent with the behavior of EE in a QFT. Moreover, in this case the gravity theory is not stable and it does not have a well defined decoupling limit in its string theory realizations \cite{Dong:2012se}. In the following, we restrict ourselves to the case $ d - \theta  \geq 1$,  then from the null energy conditions one has $z \geq 1$. 
\\On the other hand, it should be emphasized that this geometry is dual to a HV QFT in which the Lorentz and scaling symmetries are broken. As long as we know, for arbitrary values of $z$ and $\theta$ HV QFTs were not explored very much in the literature. However, for $\theta=0$, the metric \eqref{metric-vacuum} reduces to a Lifshitz geometry which might be considered as dual to a non-relativistic QFT, known as Lifshitz QFT. The symmetry group of Lifshitz QFT consists of time translations, spatial translations, spatial rotations, and anisotropic scaling symmetry which acts on the coordinates as follows \cite{Taylor:2008tg,Taylor:2015glc}
\bea
t \rightarrow \lambda^z t, \;\;\;\;\;\;\;\;\;\;\;\;\;\;\;\;\;\;\;\;\; x^i \rightarrow \lambda x^i.
\label{Lifshitz-scaling-sym}
\eea 
Moreover, this scaling symmetry is an isometry of the metric in the dual gravity \cite{Taylor:2015glc}. In other words, the metric is invariant under the following transformation
\bea
r \rightarrow \lambda r,  \;\;\;\;\;\;\;\;\;\;\;\;\;\;\;\;\;\;\  t \rightarrow \lambda^z t, \;\;\;\;\;\;\;\;\;\;\;\;\;\;\;\;\;\;\ x^i \rightarrow \lambda x^i.
\label{Lifshitz-scaling-sym-bulk}
\eea 
Although the symmetry group is smaller than the conformal group, there are enough symmetries that one can write a Ward identity for the stress tensor of the Lifshitz QFT as follows \cite{Taylor:2015glc}
\bea
z T^t_t + T^i_i =0.
\label{trace-Lifshitz}
\eea 
Furthermore, a simple model for Lifshitz QFTs is a scalar field with the following action \cite{Taylor:2008tg,Kachru:2008yh}
\bea
S= \int d^d x dt \sqrt{-g}  \left( (\partial_t \phi)^2 - \kappa \left( \partial_i^z \phi \right)^2 \right),
\eea 
where $\kappa$ is a constant. The above action has all of the aforementioned symmetries. 
\\On the other hand for $\theta \neq 0$, the scaling symmetry in eq. \eqref{Lifshitz-scaling-sym} is broken. In this case, the bulk metric in eq. \eqref{metric-vacuum} is not invariant under the coordinate transformations in eq. \eqref{Lifshitz-scaling-sym-bulk}, and transforms as $ds \rightarrow \lambda^{\theta_e} ds$.
It is evident that due to the lack of the scaling and Lorenz symmetries, the symmetry group of a HV QFT is smaller than the conformal group. However, for a HV QFT on $\mathbb{R}^{d+1}$ there are enough symmetries such that one can put the following constraint on the components of the stress tensor \cite{Alishahiha:2019lng}
\bea
z T^t_t + \frac{d_e}{d} T^i_i =0.
\label{trace-HV-zero-cutoff}
\eea 
It should be emphasized that the above constraint is an analogue of the trace condition, i.e. $T^{\mu}_{\mu}=0$, for a CFT which lives on $\mathbb{R}^{d+1}$. Furthermore, for $z=1$ and $\theta=0$, the Lorentz and scaling symmetries are restored and the HV QFT reduces to a $CFT_{d+1}$. In this case, the metric \eqref{metric-vacuum} becomes an $AdS_{d+2}$ spacetime in Poincar\'{e} coordinates. Another important feature of $T_{\mu \nu}$ is that it is not symmetric, i.e. $T_{t i} \neq T_{i t}$, due to the lack of  Lorenz symmetry
\cite{Kiritsis:2015doa,Cardy:2018jho}.
\\Another important point is that in theories which are hyperscaling, one can find the temperature dependence of free energy and thermal entropy by simple dimensional analysis \cite{Huijse:2011ef}. For instance, in a d+1 dimensional Lifshitz QFT, the temperature dependence of the thermal entropy is given by $S \sim T^\frac{d}{z}$. In contrast, in HV QFTs this scaling is violated and one has $S \sim T^\frac{d-\theta}{z}$ \cite{Huijse:2011ef}. That is why $\theta$ is called the hyperscaling violation exponent. An example of systems which shows this behavior is a random field Ising model in its spin-glass phases \cite{Fisher:1986zzb} (See also \cite{Dong:2012se}). However, as long as we know there is not a known action for a HV QFT and it would be interesting to study these QFTs further.
\subsection{HV Geometries: Finite Cutoff Case}
\label{Sec: HV-Finite Cutoff Case}
In this paper, we are interested in HV geometries at finite radial cutoff. One might ask what are the QFTs dual to these geometries?
\footnote{We would like to thank the referee very much for raising the question.}
Despite of the fact that our knowledge about the HV QFTs is very limited, one might speculate about the deformation operator by holographic methods applied in refs. \cite{Taylor:2018xcy,Hartman:2018tkw}. In ref. \cite{Alishahiha:2019lng}, it was proposed that a HV geometry at finite cutoff, might be considered as the gravity dual of a $T \overline{T}$ like deformation of a HV QFT.
Moreover, by considering the appropriate counterterms, the renormalized Brown-York stress tensor for a HV black brane was calculated in ref. \cite{Alishahiha:2019lng}. Next, by applying the radial-radial component of the Einstein's equation it was shown that the stress tensor of the HV QFT satisfies the following constraint \cite{Alishahiha:2019lng}
\footnote{Here we restricted ourselves to the zero temperature case where the emblankening factor is given by $f(r)=1$. Therefore, the time component $\tilde{J}^t$ of the vector current associated to the gauge field that produces the anisotropy is zero. Moreover, notice that the manifold on which the HV QFT lives is assumed to be flat.}
\footnote{Note that our radial coordinate is the inverse of the radial coordinate in ref. \cite{Alishahiha:2019lng}, .i.e. $r_{\rm here} = \frac{1}{r_{\rm there}}$.}
\bea
z T^{i}_{i} + \frac{d_e}{d} T^i_i = - \frac{8 \pi d_e G_N}{z (2 d_e+z-1)} r_c^{d_e +z} \Bigg[
z  \left( T^t_t \right)^2 + \frac{d_e}{d} T^i_j T^j_i - \frac{1}{d_e} \left( z T^t_t + \frac{d_e}{d} T^i_i \right)^2
\Bigg].
\label{trace-boundary-HV-finite}
\eea 
Notice that for $\theta=0$ and $z=1$, eq. \eqref{trace-boundary-HV-finite} reduces to eq. \eqref{trace-flow-CFT-1} for a $CFT_{d+1}$ on a flat space. In other words, eq. \eqref{trace-boundary-HV-finite} can be considered as an analogue of eq.  \eqref{trace-flow-CFT-1}. Next, by comparison of eqs. \eqref{trace-boundary-HV-finite} and \eqref{trace-flow-CFT-1}, it was proposed in ref. \cite{Alishahiha:2019lng} that 
the deformation operator $X$ might be given by 
\bea
X  =
z  \left( T^t_t \right)^2 + \frac{d_e}{d} T^i_j T^j_i - \frac{1}{d_e} \left( z T^t_t + \frac{d_e}{d} T^i_i \right)^2.
\label{TT-operator-HV}
\eea 
Notice that for $\theta=0$ and $z=1$ 
eq. \eqref{TT-operator-HV} reduces to eq. \eqref{TT-operator-CFT}.
\\On the other hand, the factorization property was also proved for non-Lorentz invariant QFTs, such as Lifshitz QFTs, in two dimensions \cite{Cardy:2018jho}. The proof is just based on translational symmetry and the scaling symmetry is not applied. Therefore, one expects that it can be generalized to HV QFTs in two dimensions. Furthermore, in higher dimensions, we use the fact that we are working with large-N QFTs. It is well known that in the large N limit, the correlation functions factorize \cite{McGreevy:2009xe}. Therefore, similar to ref. \cite{McGough:2016lol,Hartman:2018tkw,Grieninger:2019zts}, it is expected that the deformation operator to have the factorization property
\bea
\langle X \rangle =
z  \langle T^t_t \rangle^2 + \frac{d_e}{d} \langle T^i_j \rangle \langle T^j_i \rangle - \frac{1}{d_e} \langle z T^t_t + \frac{d_e}{d} T^i_i \rangle^2.
\label{TT-operator-HV-factorization}
\eea 
\section{Holographic Entanglement Entropy for a Strip}
\label{Sec: Holographic Entanglement Entropy for a Strip}
In this section, we first review the calculation of HEE for an entangling region in the shape of a strip in HV geometries at zero temperature and zero cutoff.
Next, we calculate the HEE at finite radial cutoff. 
\subsection{Zero Cutoff}
\label{Sec: Zero cutoff}
Now we review the calculation of HEE for a strip in HV geometries, which has been studied extensively  in refs. \cite{Huijse:2011ef,Dong:2012se,Alishahiha:2012cm,Alishahiha:2012qu}. One can parametrize a strip with width $\ell$ and length $L$ as follows
\bea
-l/2 \le x_1 \le l/2,   \quad\quad   0 \le x_{2,3, \cdots , d} \le L,    \quad\quad     t= const .
\label{x1p-strip}
\eea
Due to the translation symmetry along the length of the strip, the profile of the RT surface is given by $x_1= x_1 (r)$, and hence the area functional is as follows
\bea
Area (\Gamma_A) = 2 R^d L^{d-1} \int_{r_c}^{r_t} r^{ - d_e} \sqrt{1 + x'_1(r)^2 } dr,
\label{Area-1-strip}
\eea 
where $r_t$ is the radial coordinate of the turning point at which $x'_1(r_t) \rightarrow \infty$. In the above expression, we defined an effective dimension $d_{e} = d- \theta$, for convenience. Next, minimization of the area functional leads to
\bea
x'_1(r) = \pm \frac{r^{d_e}}{\sqrt{r_t^{2 d_e} - r^{2 d_e}}},
\label{eom-strip}
\eea 
from which one can find the relation between $r_t$ and $\ell$ as follows
\bea
\frac{\ell}{2} =  \int_{r_c}^{r_t} \frac{dr \; r^{d_e}}{\sqrt{r_t^{2 d_e} - r^{2 d_e}}}.
\label{l-rt-strip}
\eea 
Now one can plug eq. \eqref{eom-strip} into eqs. \eqref{Area-1-strip} and \eqref{RT} to obtain the HEE. One has
\bea
S = \frac{R^{d} L^{d-1}}{2 G_N} \int_{r_c}^{r_t} \frac{dr}{r^{d_e } \sqrt{1 - \left( \frac{r}{r_t}\right)^{2 d_e}}}.
\label{EE-strip}
\eea 
To find the HEE, one should first find $r_t$ in terms of $\ell$ from eq. \eqref{l-rt-strip} and substitute it in eq. \eqref{EE-strip}. When the cutoff $r_c$ is very small, i.e. $r_c= \epsilon \rightarrow 0$, the calculation was done in ref. \cite{Dong:2012se}.  For $d_e =1$, the RT surface is a semicircle and $r_t$ can be obtained from eq. \eqref{l-rt-strip} as follows
\bea
r_t = \frac{\ell}{2}.
\label{rt-l-de=1-zero cutoff}
\eea 
Moreover, the HEE is given by \cite{Dong:2012se}
\bea
S_0= \frac{R^d}{2 G_N} \left( \frac{L}{r_F}\right)^{d-1} \log \left( \frac{\ell}{\epsilon}\right),
\label{EE-strip-de=1-zero cutoff}
\eea 
which has a logarithmic UV divergent term. Consequently,  the well known area law behavior of the HEE is violated in this case \cite{Dong:2012se}. On the other hand, for $d_e \neq 1$ from eq. \eqref{l-rt-strip}, $r_t$ is simply given by \cite{Dong:2012se}
\bea
r_t = \frac{\ell}{2 \Upsilon},
\label{rt-zero cutoff}
\eea 
where $\Upsilon$ is defined as follows
\bea
\Upsilon = \frac{\sqrt{\pi} \Gamma \left(\frac{d_e +1}{2 d_e} \right)}{\Gamma \left(\frac{1}{2 d_e}\right)}.
\label{Upsilon}
\eea 
In this case, the HEE is given by \cite{Dong:2012se}
\bea
S_0= \frac{R^d L^{d-1}}{2 G_N (d_e-1) r_F^\theta} \left[ \frac{1}{\epsilon^{d_e-1} } - \Upsilon^{d_e} \left(\frac{2}{\ell}\ \right) ^{d_e-1} \right].
\label{EE-strip-de neq 1-zero cutoff}
\eea 
In the following, we study the HEE for the finite cutoff case. To do so, we consider the two cases $d_e=1$ and $d_e \neq 1$, separately. For $d_e=1$, one can find the HEE exactly. However, for $d_e \neq 1$, finding $r_t$ in terms of $\ell$ is a difficult task and one has to do it either numerically or by making some approximations.
\subsection{Finite Cutoff and $d_e =1$}
\begin{figure}
	\begin{center}
		\includegraphics[scale=0.34]{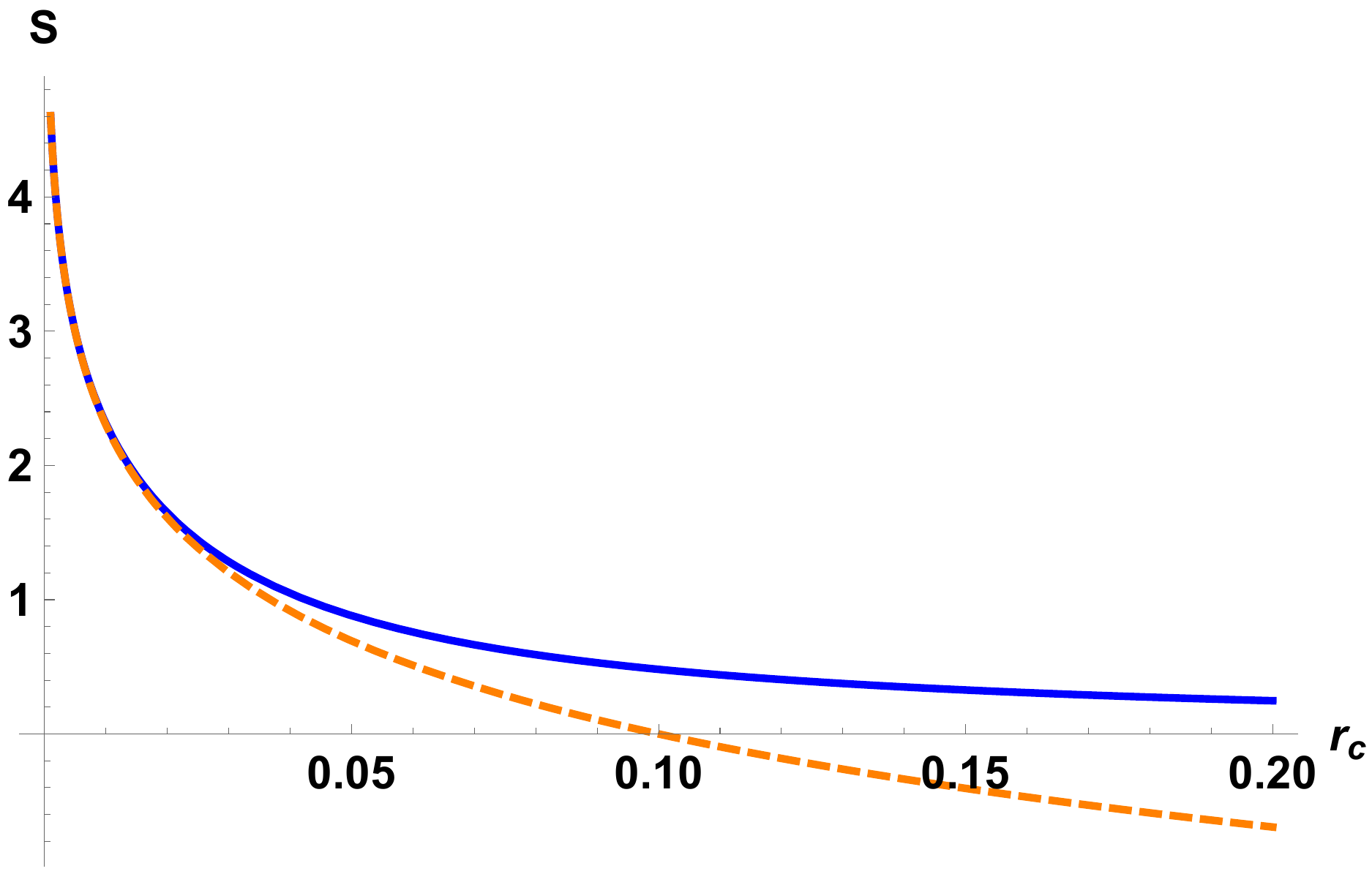}
		\hspace{1cm}
		\includegraphics[scale=0.34]{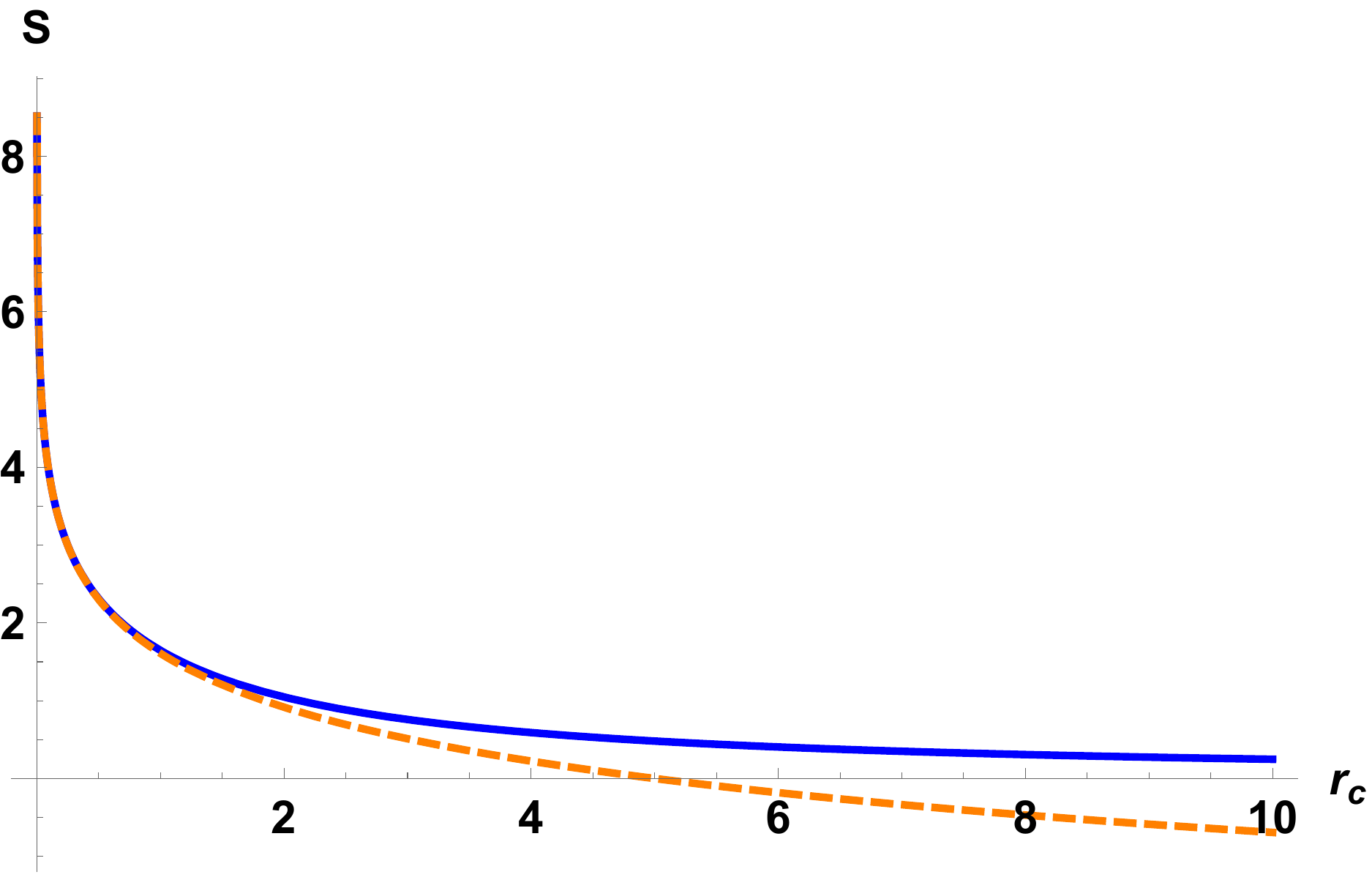}
		\vspace{-5mm}
	\end{center}
	\caption{HEE as a function of $r_c$ for $d_e=1$ and different values of $\ell$: {\it Left}) $\ell=0.1$ {\it Right}) $\ell= 5$. The solid blue curves show the HEE for the finite cutoff case given in eq. \eqref{EE-strip-de=1} and the dashed orange curves indicate the HEE for the zero cutoff case given in eq. \eqref{EE-strip-de=1-zero cutoff} which are valid for very small cutoffs, i.e. $r_c \ll \ell$. Here we set $R=1$ and renormalized $S$ as $\tilde{S}= \frac{ S}{a}$, where $a= \frac{R^d L^{d-1}}{2 G_N r_F^\theta}$.
	}
	\label{fig:S-rc-de-1}
\end{figure}

For $d_e =1$, form eq. \eqref{l-rt-strip} one can simply find that the RT surface is again a semicircle such that
\bea
r_t = \sqrt{\left( \frac{\ell}{2} \right)^2 + r_c^2}.
\label{rt-strip-de=1}
\eea 
Now in contrast to eq. \eqref{rt-l-de=1-zero cutoff}, $r_t$ depends on the cutoff $r_c$. On the other hand, from eq. \eqref{EE-strip}, one obtains
\bea
S = \frac{R^d L^{d-1}}{2 G_N r_F^\theta} \log \left( \frac{r_t + \sqrt{r_t^2 - r_c^2}}{r_c} \right).
\label{EE-de=1-finite cutoff}
\eea 
Next, by plugging eq. \eqref{rt-strip-de=1} into eq. \eqref{EE-de=1-finite cutoff}, one has
\bea
S = \frac{R^{d}}{2 G_N} \left(\frac{L}{r_F}\right)^{d-1} \log \left( \frac{\ell + \sqrt{\ell^2 + 4 r_c^2}}{2 r_c}\right).
\label{EE-strip-de=1}
\eea 
Notice that the HEE is independent of the exponent $z$. In figure \ref{fig:S-rc-de-1}, the HEE is drawn as a function of $r_c$ for different values of $\ell$. It is evident that the HEE is a decreasing function of $r_c$. Furthermore, when the cutoff is very small, i.e. $r_c \ll \ell$, one might expand eq. \eqref{EE-strip-de=1} in powers of $r_c$ to obtain
\bea
S = \frac{R^{d}}{2 G_N} \left(\frac{L}{r_F}\right)^{d-1} \left[ \log \left( \frac{\ell}{r_c}\right) + \left(\frac{r_c}{\ell}\right)^2\right] + \mathcal{O} \left(r_c^4\right) + \cdots.
\label{EE-strip-de=1-very small cutoff}
\eea 
For the zero cutoff case, i.e. $r_c = \epsilon \rightarrow 0$, one might neglect the second term, and hence the HEE shows the expected logarithmic behavior which is given in eq. \eqref{EE-strip-de=1-zero cutoff}. Therefore, imposing a finite radial cutoff, introduces sub-leading corrections to the HEE. As mentioned before, for $z=1$ and $\theta =0$, the Lorentz and scaling symmetries in the dual QFT is restored and it becomes a $CFT_{d+1}$. In particular, for $\theta=0$ and $d=1$, from eq. \eqref{EE-strip-de=1-very small cutoff}, one has
\bea
S = \frac{c}{3} \left[ \log \left( \frac{\ell}{r_c}\right) + \left(\frac{r_c}{\ell}\right)^2\right] + \mathcal{O} \left(r_c^4\right) + \cdots,
\label{EE-strip-de=1-very small cutoff-CFT}
\eea 
where $c= \frac{3 R}{2 G_N}$ is the central charge of the dual CFT \cite{Brown:1986nw}. The above expression agrees with the HEE of a $CFT_2$ at zero temperature and finite cutoff calculated in ref. \cite{Park:2018snf}.
\subsection{Finite Cutoff and $d_e \neq 1$}
\begin{figure}
	\begin{center}
		\includegraphics[scale=0.31]{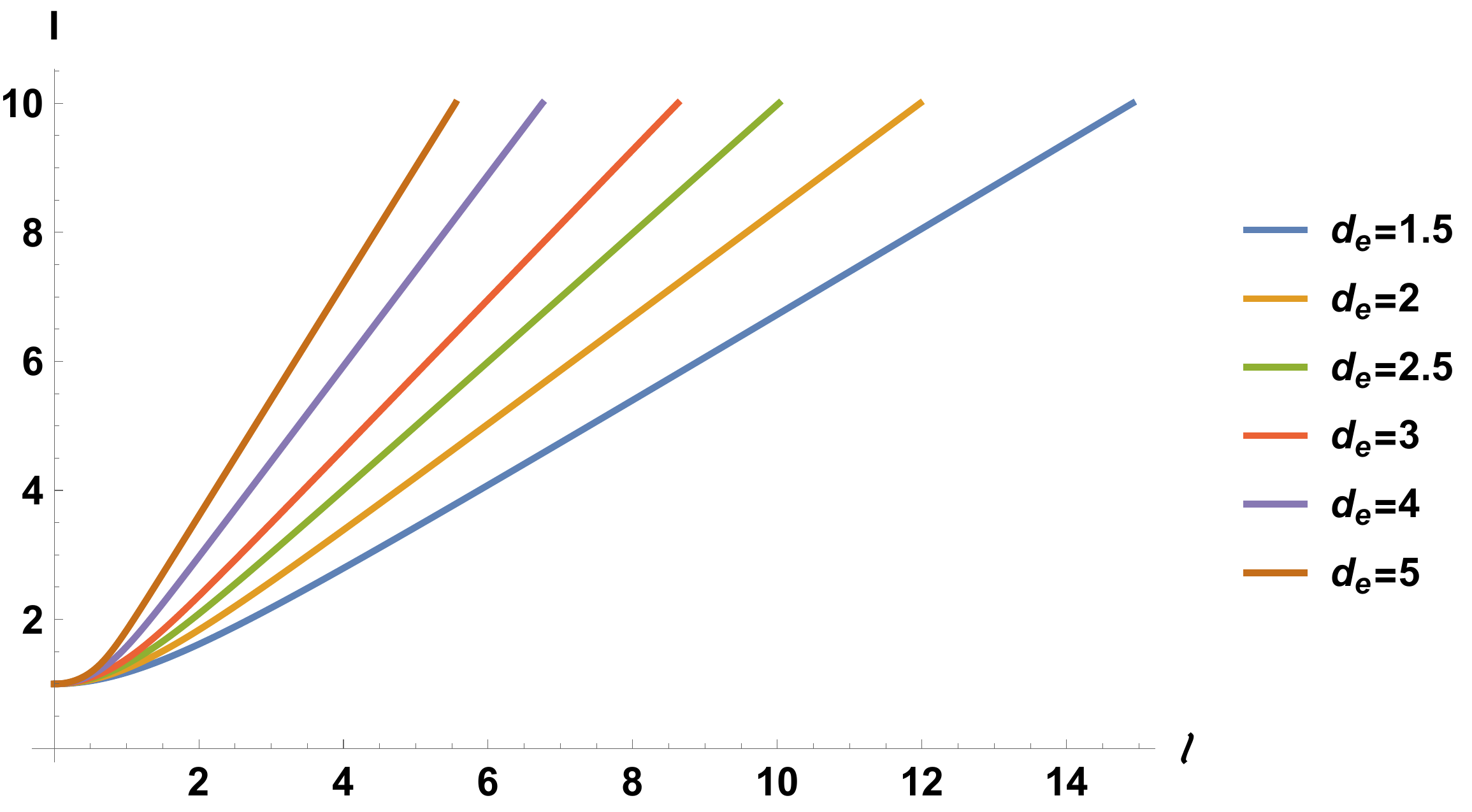}
    	\hspace{0.2cm}
		\includegraphics[scale=0.31]{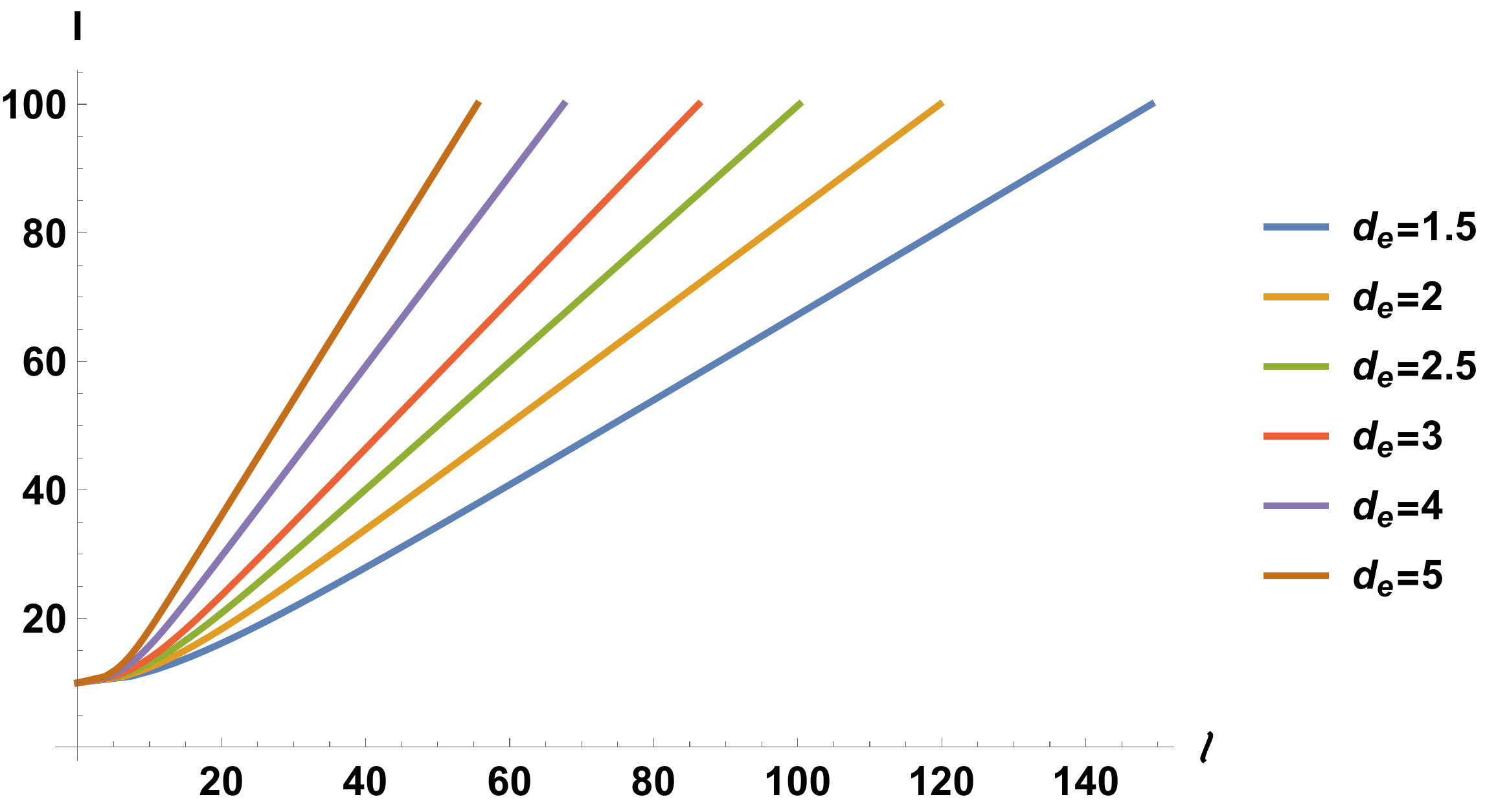}
		\vspace{-5mm}
	\end{center}
	\caption{$r_t$ as a function of $\ell$ for different values of the cutoff: {\it Left}) $r_c=1$ {\it Right}) $r_c= 10$.
	}
	\label{fig:l-rt}
\end{figure}

\begin{figure}
	\begin{center}
		\includegraphics[scale=0.31]{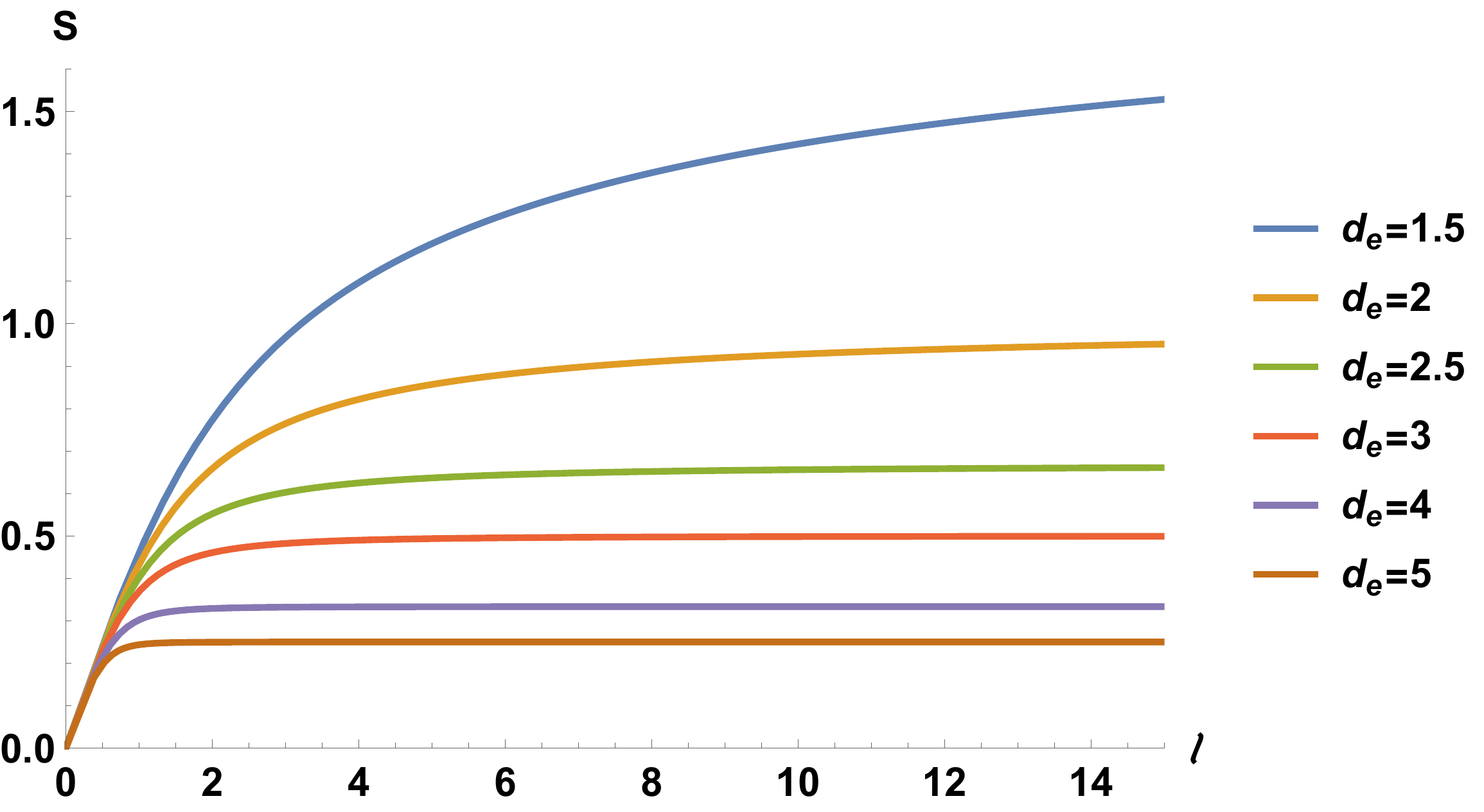}
    	\hspace{0.2cm}
		\includegraphics[scale=0.31]{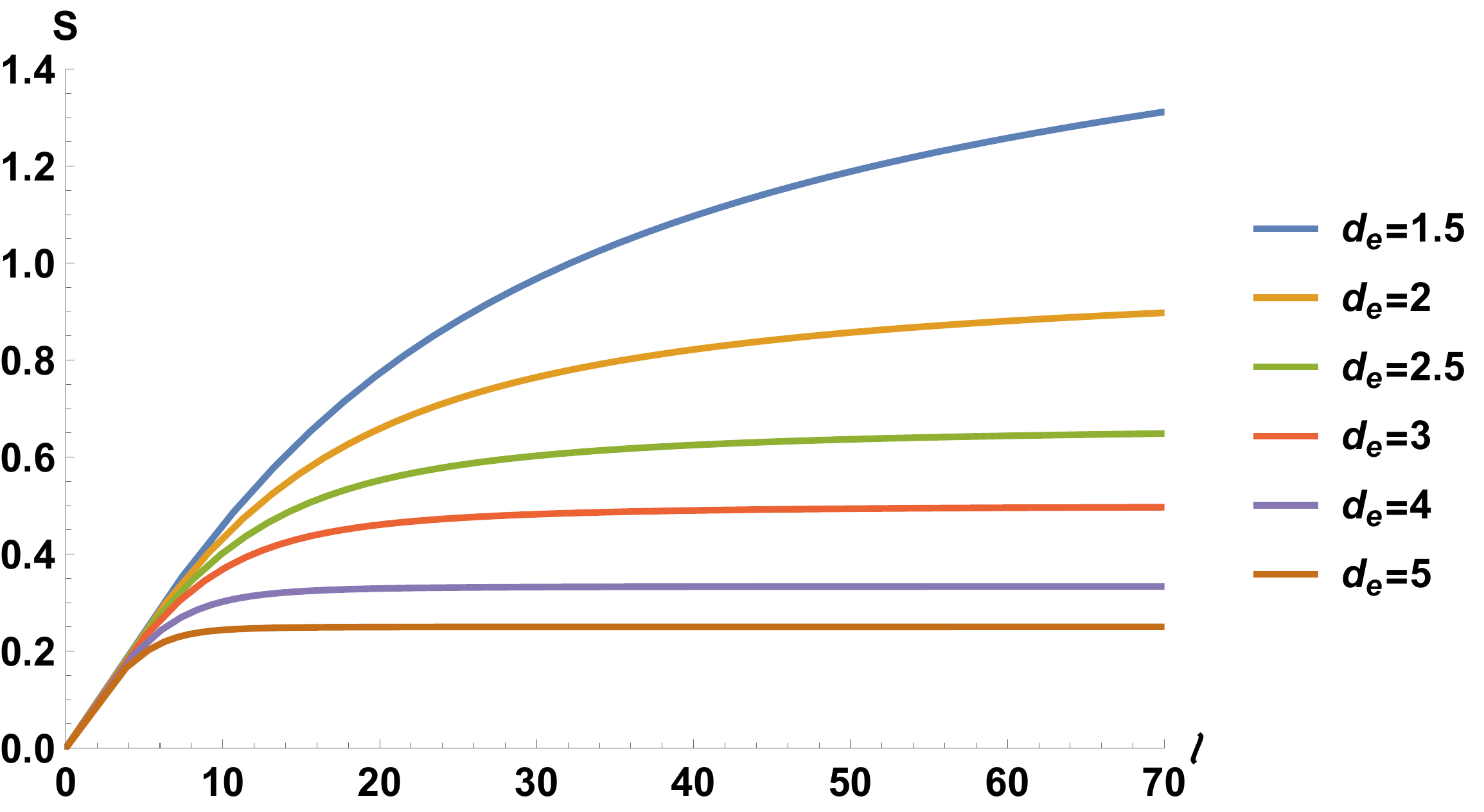}
		\vspace{-5mm}
	\end{center}
	\caption{HEE as a function of $\ell$ for different values of the cutoff: {\it Left}) $r_c=1$ {\it Right}) $r_c= 10$. Here we set $R=1$ and renormalized $S$ as $\tilde{S}= \frac{r_c^{d_e -1} S}{a}$, where $a= \frac{R^d L^{d-1}}{2 G_N r_F^\theta}$.
	}
	\label{fig:S}
\end{figure}

\begin{figure}
	\begin{center}
		\includegraphics[scale=0.34]{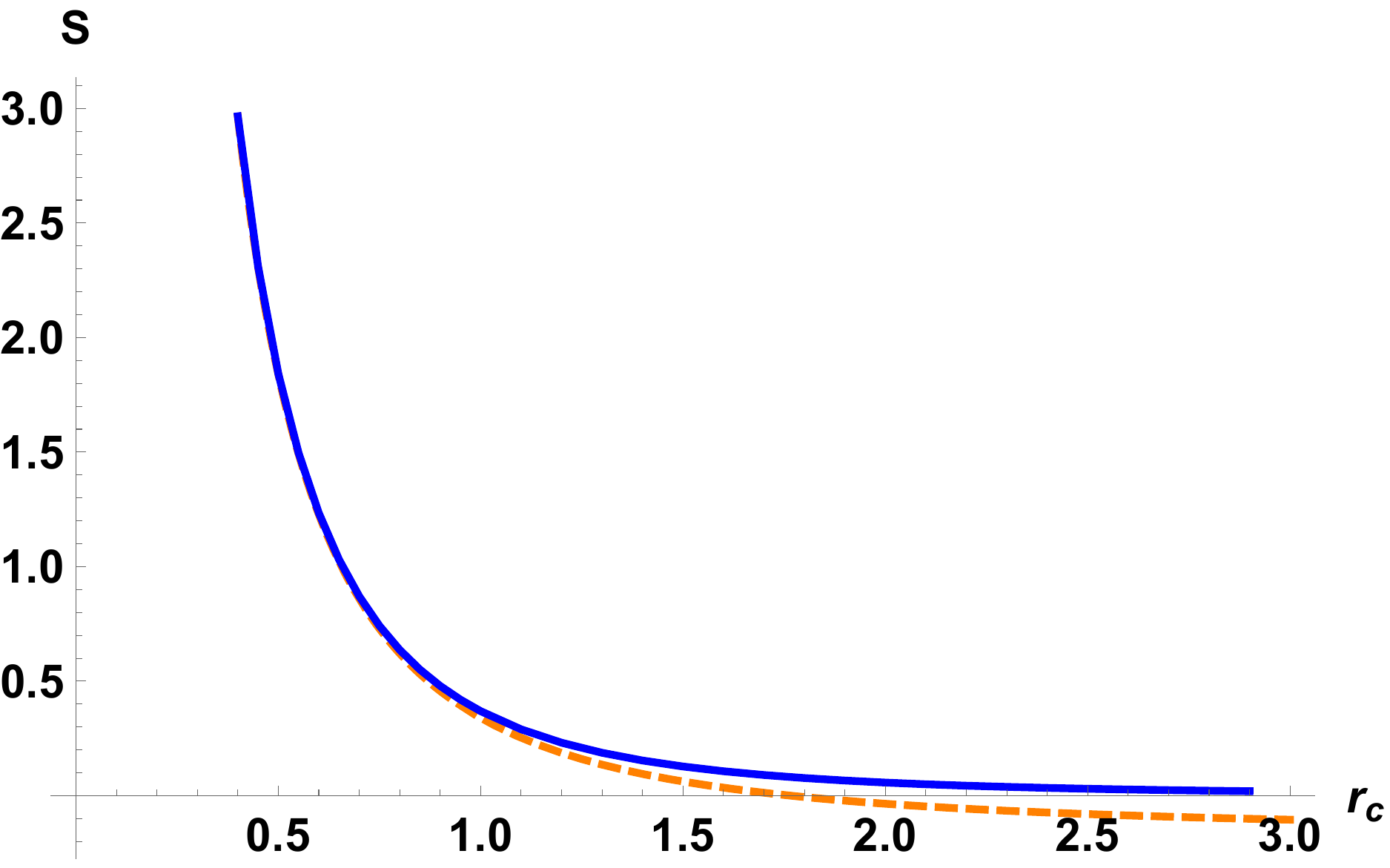}
		\hspace{1cm}
		\includegraphics[scale=0.34]{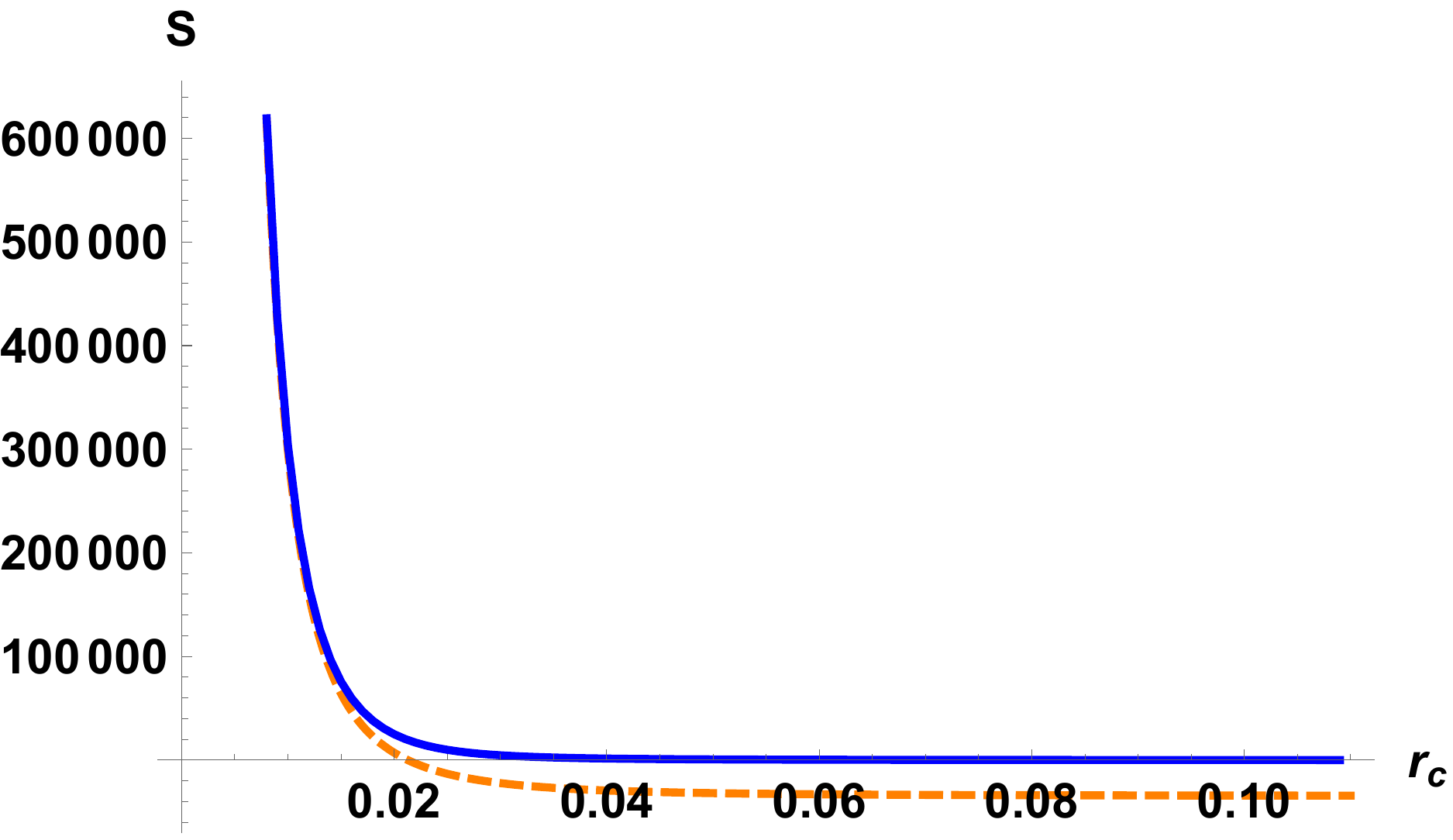}
		\vspace{-5mm}
	\end{center}
	\caption{HEE as a function of $r_c$ for: {\it Left}) $d_e = 3$ and $\ell=1$ {\it Right}) $d_e= 4$ and $\ell=0.01$. The solid blue curves show the numerical results based on eq. \eqref{EE-strip-de neq 1} and the dashed orange curves indicate the HEE for the zero cutoff case given in eq. \eqref{EE-strip-de neq 1-zero cutoff} which are valid for very small cutoffs, i.e. $r_c \ll \ell$. Here we set $R=1$ and renormalized $S$ as $\tilde{S}= \frac{r_c^{d_e -1} S}{a}$, where $a= \frac{R^d L^{d-1}}{2 G_N r_F^\theta}$.
	}
	\label{fig:S-rc}
\end{figure}

For $d_e \neq 1$, from eq. \eqref{l-rt-strip} one obtains
\bea
\frac{\ell}{2} = \Upsilon r_t - \frac{r_c^{d_e+1}}{(d_e + 1) r_t^{d_e} } \;\; {}_2F_1 \left[ \frac{1}{2} , \frac{d_e +1}{2 d_e} , \frac{3 d_e +1}{2 d_e} , \left( \frac{r_c}{r_t} \right) ^{2 d_e} \right],
\label{rt-l-strip-de neq 1}
\label{l-rt-neq 1}
\eea 
where $\Upsilon$ is given by eq. \eqref{Upsilon}. Notice that in contrast to the zero cutoff case, $r_t$ depends on the cutoff. In figure \ref{fig:l-rt}, $r_t$ is drawn as a function of $\ell$ for different values of the cutoff $r_c$. From figure \ref{fig:l-rt}, it is straightforward to see that for very small and large entangling regions, one has $r_t \approx r_c \gg \ell$ and $\ell \approx r_t \gg r_c$, respectively. On the other hand, from eq. \eqref{Area-1-strip} one has
\bea
S = \frac{R^{d} L^{d-1}}{2 G_N r_F^\theta (d_e -1)} \left[- \frac{\Upsilon}{r_t^{d_e -1}}  + \frac{1}{r_c^{d_e -1}} \;\; {}_2F_1 \left[\frac{1}{2} , \frac{1-d_e}{2 d_e} , \frac{d_e +1}{2 d_e} , \left( \frac{r_c}{r_t}\right)^{2 d_e}\right] \right].
\label{EE-strip-de neq 1}
\eea 
Next, one has to find $r_t$ from eq. \eqref{rt-l-strip-de neq 1} and plug it into eq. \eqref{EE-strip-de neq 1} which is a very tough task. However, one can easily find the HEE numerically. In figure \ref{fig:S}, the HEE is drawn as a function of $\ell$ for different values of $r_c$. Moreover, in figure \ref{fig:S-rc}, the HEE is drawn as a function of $r_c$ for different values of $\ell$ and $d_e$. It is observed that the HEE is a decreasing function of $r_c$. Therefore, one might conclude that the quantum correlations among the degrees of freedom decrease by increasing the cutoff. On the other hand, since finding an analytic expression for $r_t$ in terms of $\ell$ and $r_c$ is impossible for an arbitrary value of $r_c$, in the following we consider very large and small entangling regions and find perturbative expressions for the HEE.
\subsubsection{Very Large Entangling Regions}
\label{Sec: Very Large Entangling Regions}
\begin{figure}
	\begin{center}
		\includegraphics[scale=0.34]{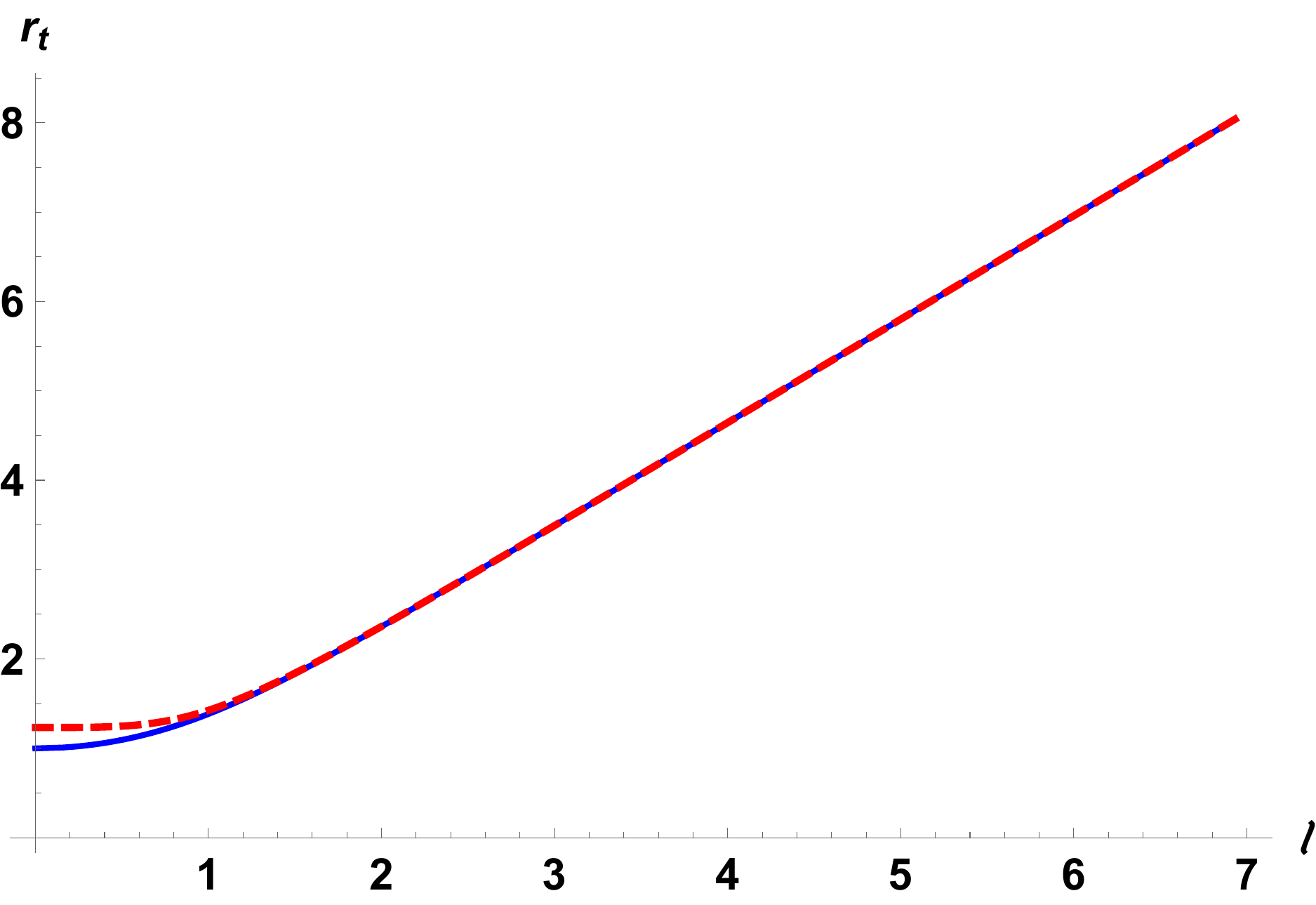}
		\hspace{1cm}
		\includegraphics[scale=0.34]{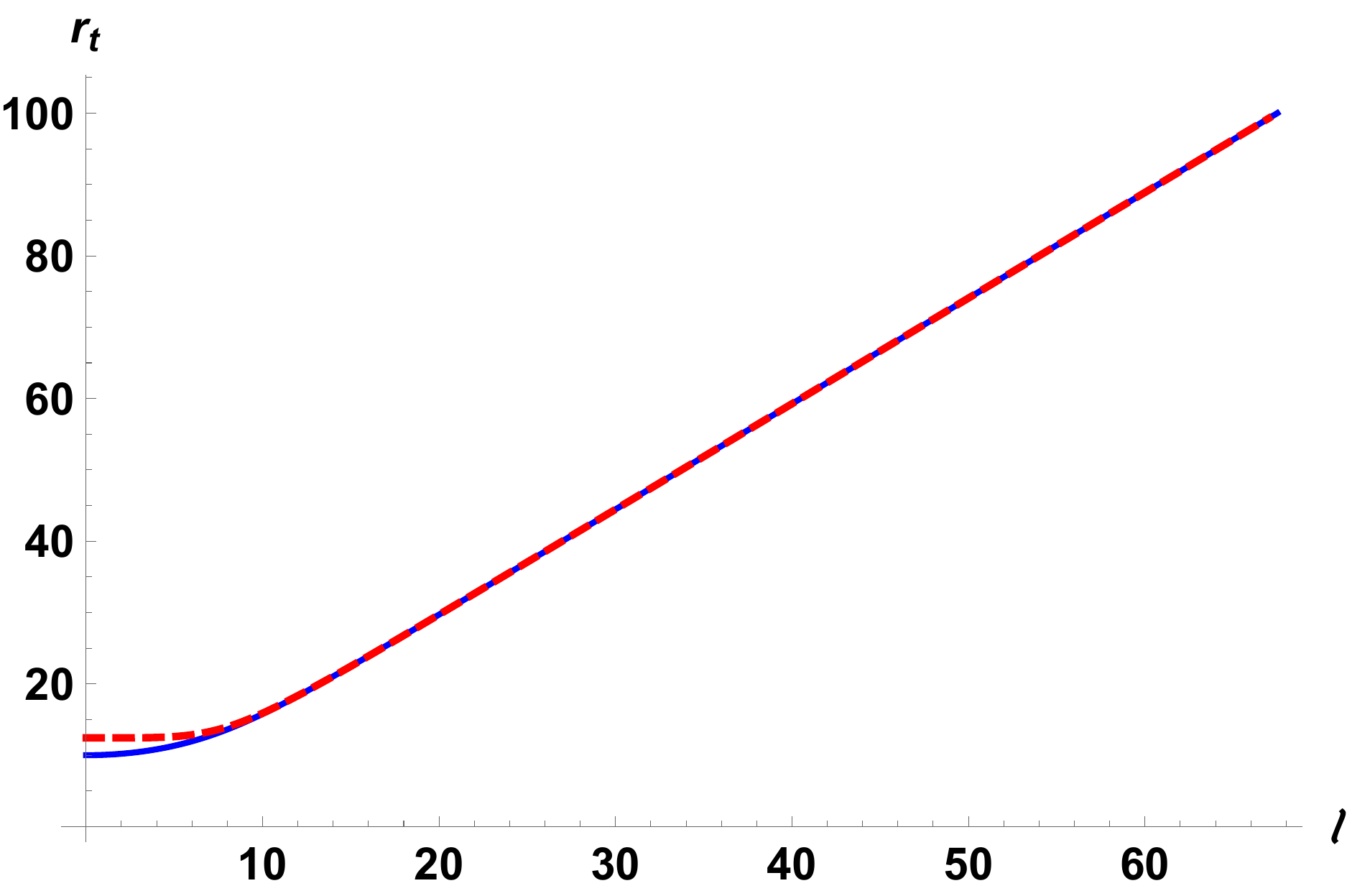}
		\vspace{-5mm}
	\end{center}
	\caption{The turning point $r_t$ as a function of $\ell$ for: {\it Left}) $r_c= 1$ and $d_e =3$ {\it Right}) $r_c= 10$ and $d_e =4$. The solid blue curves are based on eq. \eqref{l-rt-neq 1}, and the dashed red curves are based on the perturbative expression given in eq. \eqref{rt-strip-de neq 1- l, rt>>rc} which is valid for very large entangling regions, $\ell \gg r_c$. 
	}
	\label{fig:l-rt-large-ER}
\end{figure}

\begin{figure}
	\begin{center}
		\includegraphics[scale=0.34]{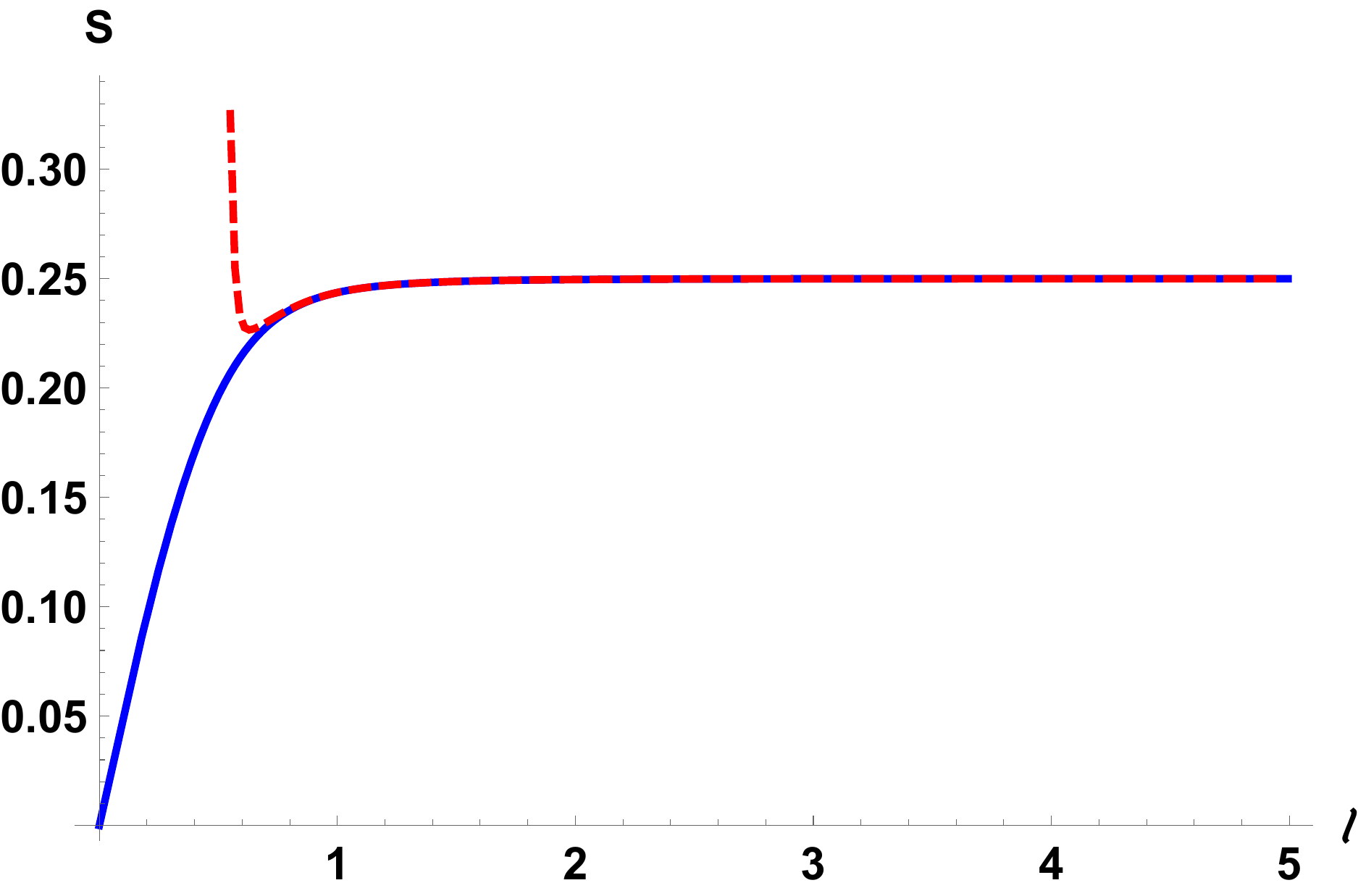}
		\hspace{1cm}
		\includegraphics[scale=0.34]{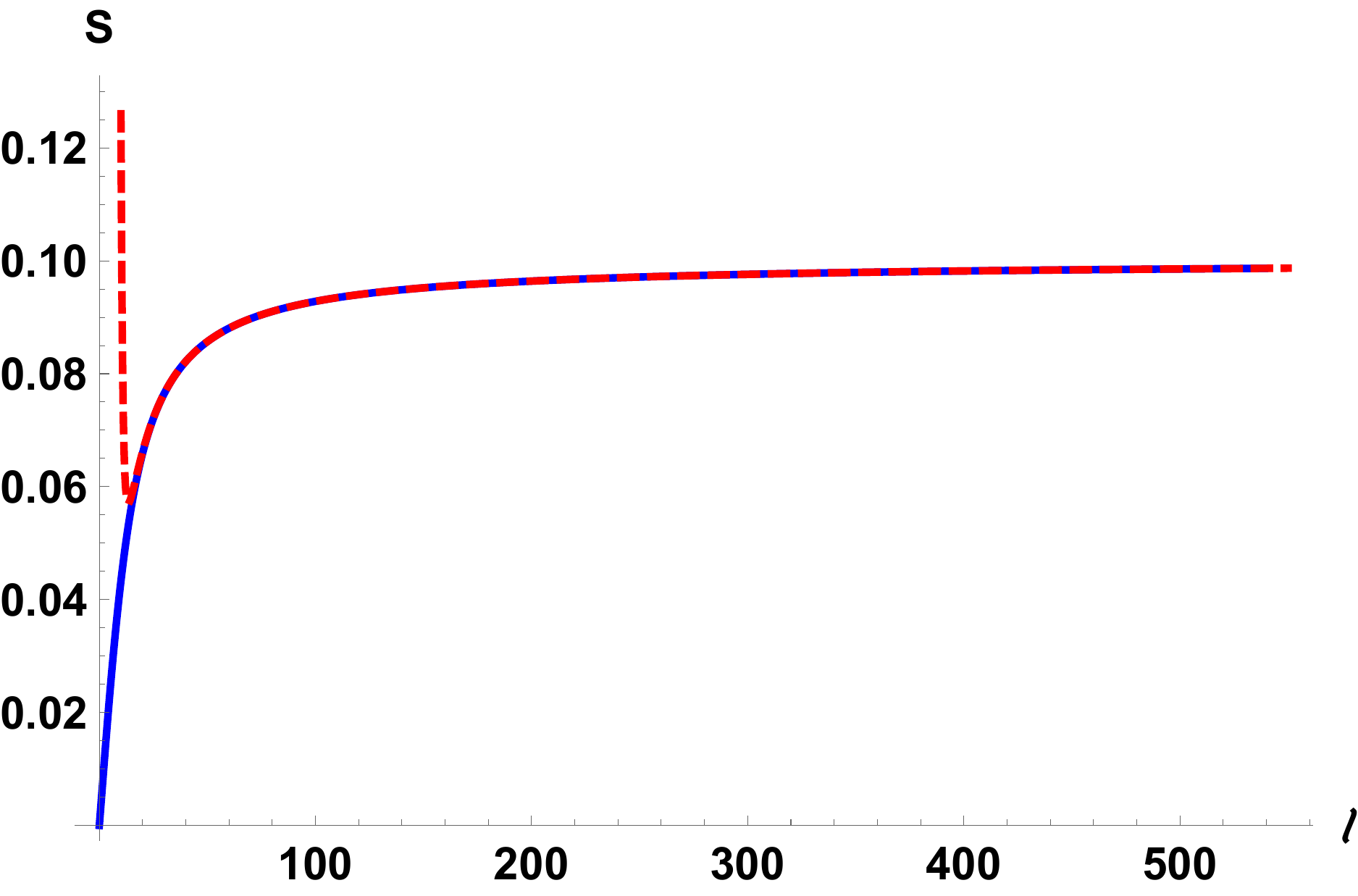}
		\vspace{-5mm}
	\end{center}
	\caption{HEE as a function of $\ell$ for different values of the cutoff: left) $r_c= 1$ and $d_e =5$ right) $r_c= 10$ and $d_e =2$.  The red dashed curves are based on the perturbative expression given in eq. \eqref{EE-strip-de neq 1-l>>rc} which is valid for very large entangling regions, and the solid blue curves are numerical results based on eq. \eqref{EE-strip-de neq 1}. 
	}
	\label{fig:S-large-ER}
\end{figure}

From figure \ref{fig:l-rt}, it is straightforward to see that for very large entangling regions, one has $\ell \approx r_t \gg r_c$. In this limit, one can expand eq. \eqref{l-rt-strip} in powers of $\frac{r_c}{r_t} \ll 1$ as follows
\bea
\frac{\ell}{2 \Upsilon r_t} = 1 - \frac{1}{(d_e +1) \Upsilon} \left( \frac{r_c}{r_t}\right)^{d_e+1} + \frac{1}{2( 3 d_e +1) \Upsilon}  \left( \frac{r_c}{r_t}\right)^{3 d_e+1} + \cdots,
\eea 
By keeping the first two terms on the right hand side of the above equation, one can rewrite it as follows
\bea
\left(\frac{\ell}{2 \Upsilon r_t}\right)^{d_e +1} &=& \Big[
1- \frac{1}{(d_e +1) \Upsilon} \left( \frac{r_c}{r_t}\right)^{d_e +1} + \cdots
\Big]^{d_e +1}
\cr && \cr
&=& 1-  \frac{1}{\Upsilon} \left( \frac{r_c}{r_t}\right)^{d_e +1} + \cdots.
\eea 
Then one can find $r_t$ as follows
\bea
r_t &=& \frac{\ell}{2 \Upsilon} \left[ 1 + \Upsilon^{d_e} \left( \frac{2 r_c}{\ell}\right)^{d_e+1} + \cdots \right]^{\frac{1}{d_e+1}}
\cr && \cr
&=&  \frac{\ell}{2 \Upsilon} \left[ 1 + \frac{\Upsilon^{d_e}}{(d_e+1)} \left(\frac{2 r_c}{\ell} \right)^{d_e+1} + \mathcal{O} \left( r_c^{2(d_e+1)} \right)  + \cdots \right].
\label{rt-strip-de neq 1- l, rt>>rc}
\eea 
In figure \ref{fig:l-rt-large-ER}, $r_t$ is drawn as a function of $l$. It shows that $r_t \rightarrow r_c$ when $\ell \rightarrow 0$. Moreover, $r_t \rightarrow \infty$ when $\ell \rightarrow \infty$, as it was expected.
On the other hand, by expanding eq. \eqref{EE-strip-de neq 1} in powers of $\frac{r_c}{r_t}$, one has
\bea
S &=&\frac{R^{d} L^{d-1}}{2 G_N r_F^\theta (d_e -1)} \Big[ \frac{1}{r_c^{d_e-1}} - \frac{\Upsilon}{r_t^{de-1}}  - \frac{(d_e -1)}{2(d_e +1)} \frac{r_c^{d_e+1}}{r_t^{2 d_e}} 
\cr && \cr
&& \;\;\;\;\;\;\;\;\;\;\;\;\;\;\;\;\;\;\;\;\;\;\;\;\;\;\;\; - \frac{3}{8(3 d_e +1)} \frac{r_c^{3 d_e-1}}{r_t^{2(2d_e -1)}} + \mathcal{O} \left( r_c^{5 d_e +1} \right) + \cdots \Big].
\eea 
Next, by plugging eq. \eqref{rt-strip-de neq 1- l, rt>>rc} into the above expression and expanding it in powers of $\frac{r_c}{\ell} \ll 1$, one obtains 
\bea
S= S_0 + \Delta S,
\label{EE-strip-de neq 1-l>>rc}
\eea 
where $S_0$ is the HEE for the zero cutoff case given in eq. \eqref{EE-strip-de neq 1-zero cutoff}, and
\bea
\Delta S
&=& \frac{R^{d} L^{d-1}}{4 G_N r_F^\theta (d_e+1)}  \left( \frac{2 \Upsilon}{\ell} \right)^{2d_e} \Big[
 r_c^{d_e+1} - \frac{3 (d_e+1)}{4(3d_e +1) }  \left(\frac{2 \Upsilon}{\ell}\right)^{2 d_e} r_c^{3d_e+1} 
\cr && \cr
&&
\;\;\;\;\;\;\;\;\;\;\;\;\;\;\;\;\;\;\;\;\;\;\;\;\;\;\;\;\;\;\;\;\;\;\;\;\;\;\;\;\;\;\;
- \frac{d_e (3 d_e+1)}{3(d_e+1)^2 \Upsilon} \left(\frac{2 \Upsilon}{\ell}\right)^{2 (d_e +1)} r_c^{3 (d_e+1) } 
\cr && \cr
&&
\;\;\;\;\;\;\;\;\;\;\;\;\;\;\;\;\;\;\;\;\;\;\;\;\;\;\;\;\;\;\;\;\;\;\;\;\;\;\;\;\;\;\;
+ \frac{3 d_e}{(3 d_e +1) \Upsilon} \left(\frac{2 \Upsilon}{\ell}\right)^{3 d_e+1} r_c^{2 (2 d_e+1)}
+ \cdots
\Big].
\label{Delta-EE-strip-de neq 1-l>>rc}
\eea 
In figure \ref{fig:S-large-ER}, the above expression for the HEE is drawn as a function of $\ell$ for different values of $r_c$, and compared with the numerical result.
\subsubsection{Very Small Entangling Regions}
\label{Sec: Very Small Entangling Regions}
\begin{figure}
	\begin{center}
		\includegraphics[scale=0.34]{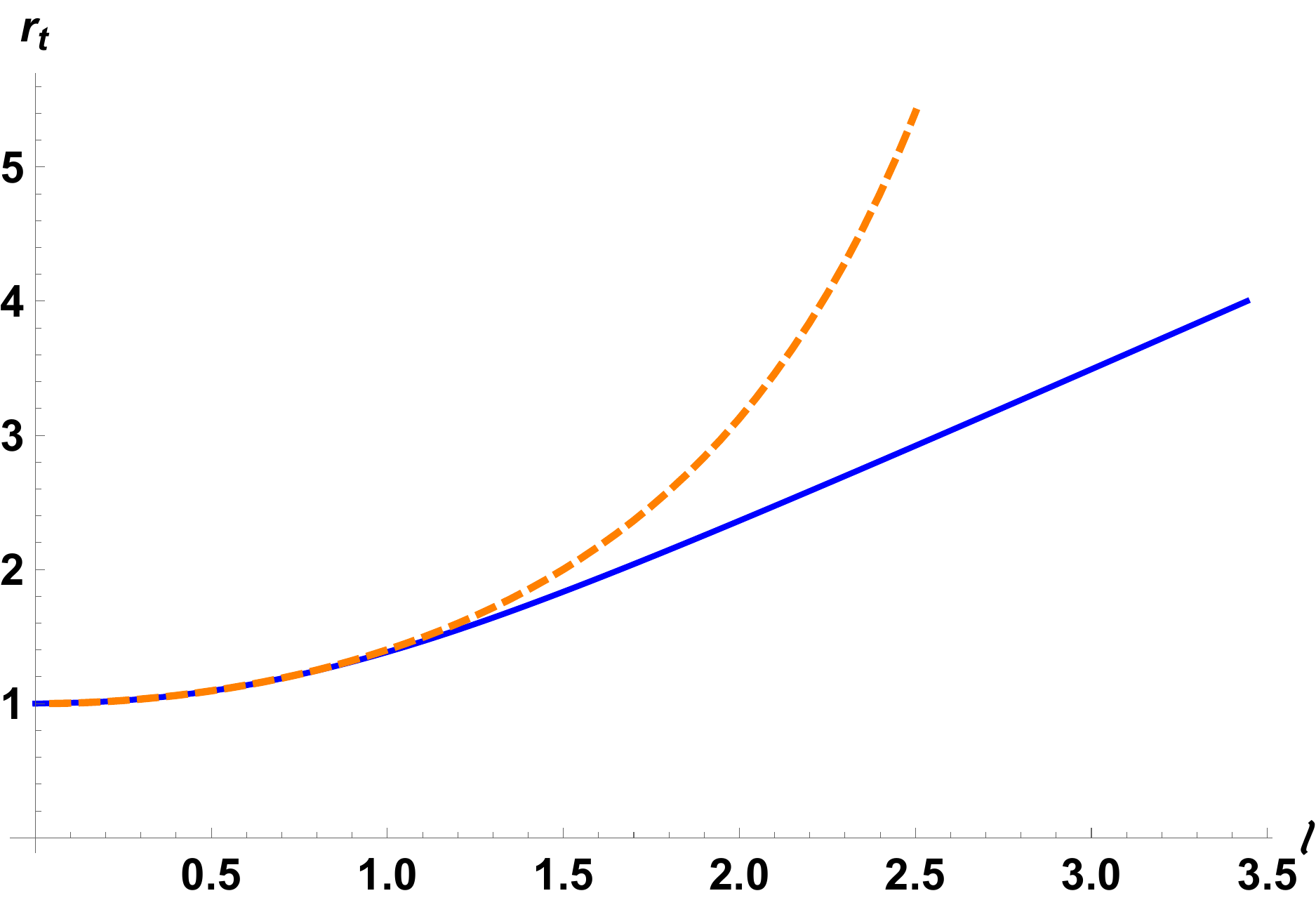}
		\hspace{1cm}
		\includegraphics[scale=0.34]{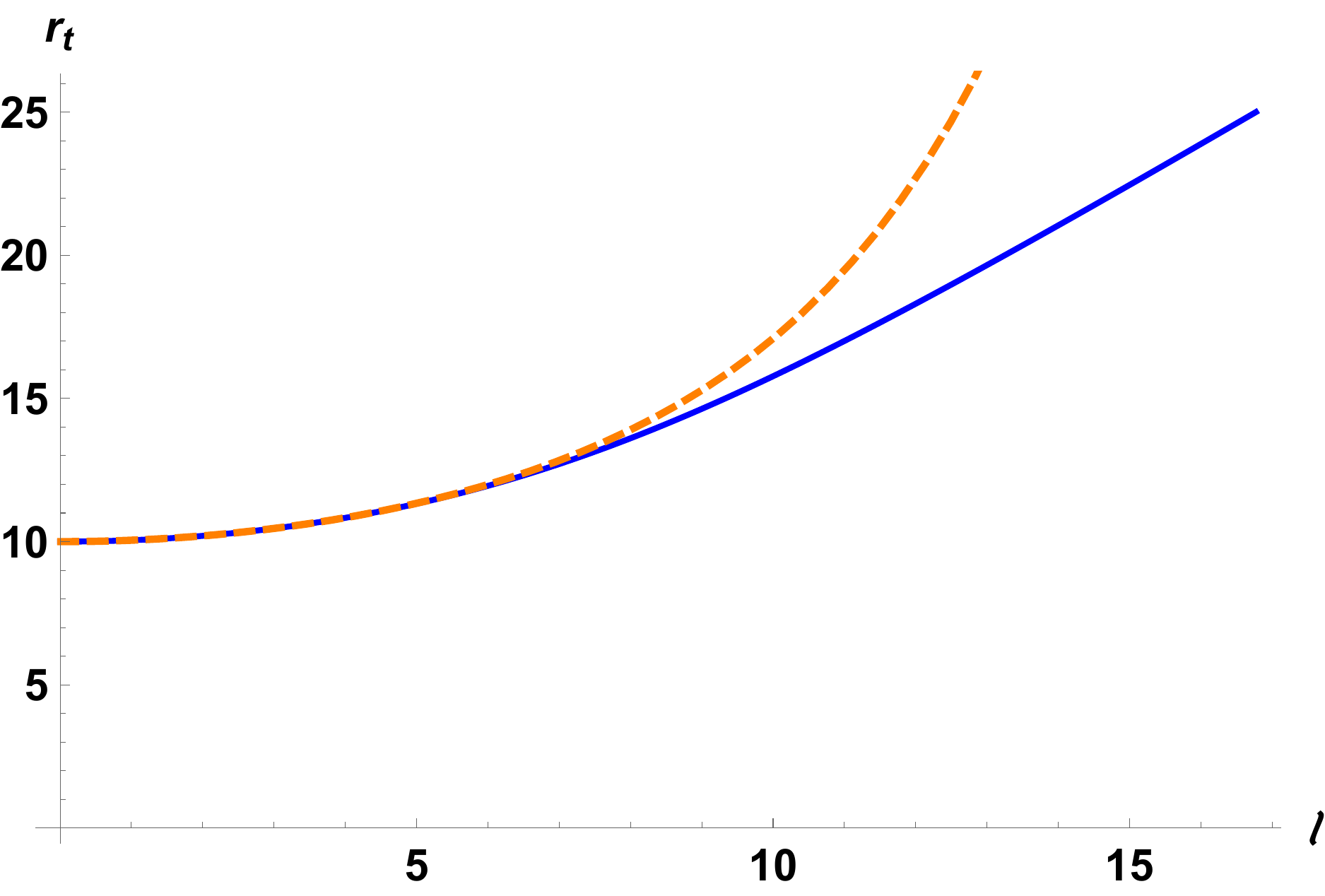}
		\vspace{-5mm}
	\end{center}
	\caption{The turning point $r_t$ as a function of $\ell$ for different values of the cutoff: {\it Left}) $r_c= 1$ and $d_e =3$ {\it Right}) $r_c= 10$ and $d_e =4$. The solid blue curves are based on eq. \eqref{l-rt-neq 1} and the dashed orange curves are based on the perturbative expression in eq. \eqref{rt-strip-de neq 1- l, rt approx rc} which is valid for very small entangling regions, $ \ell \ll r_c$.
	}
	\label{fig:l-rt-small-ER}
\end{figure}

\begin{figure}
	\begin{center}
		\includegraphics[scale=0.34]{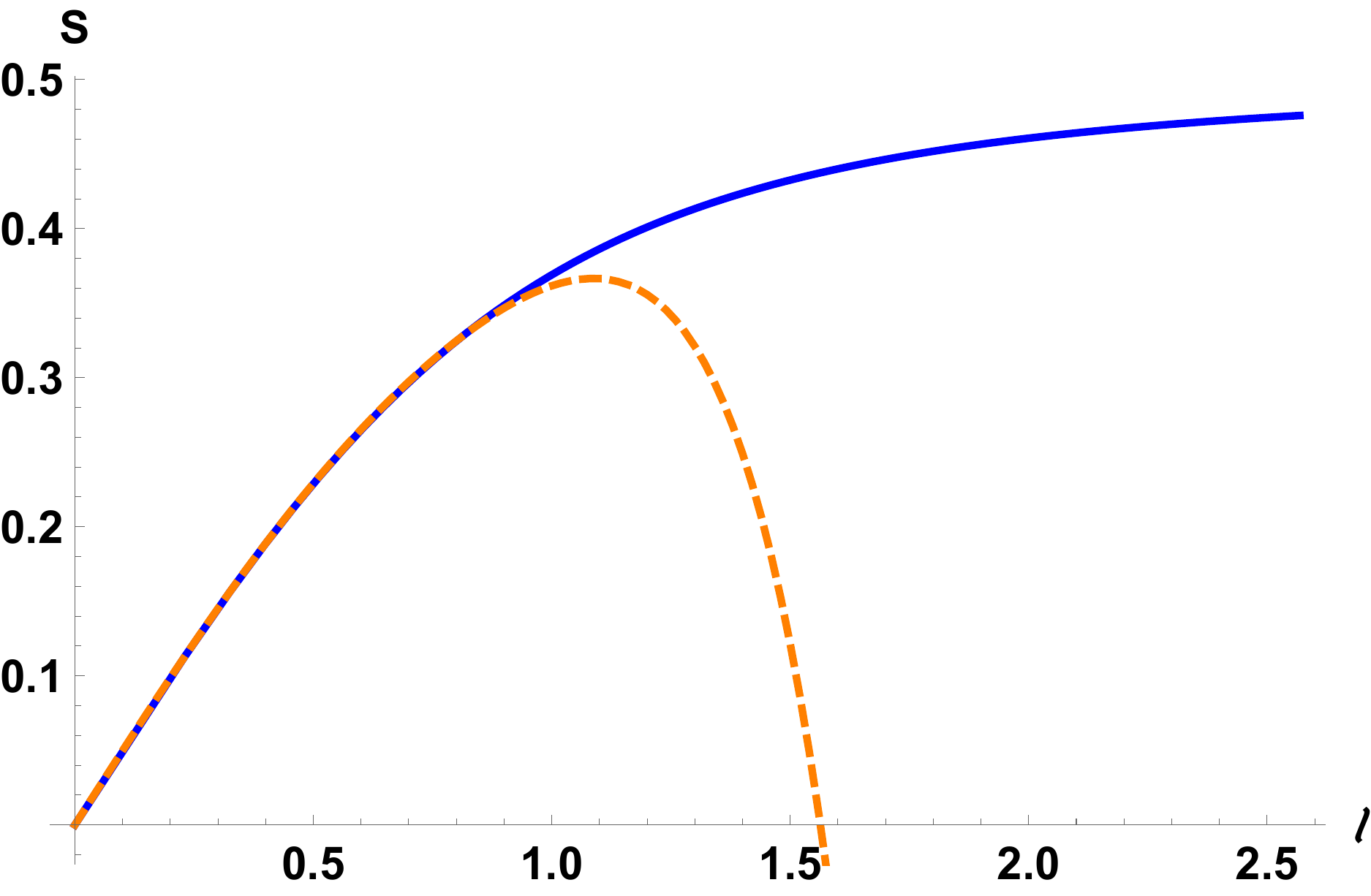}
		\hspace{1cm}
		\includegraphics[scale=0.34]{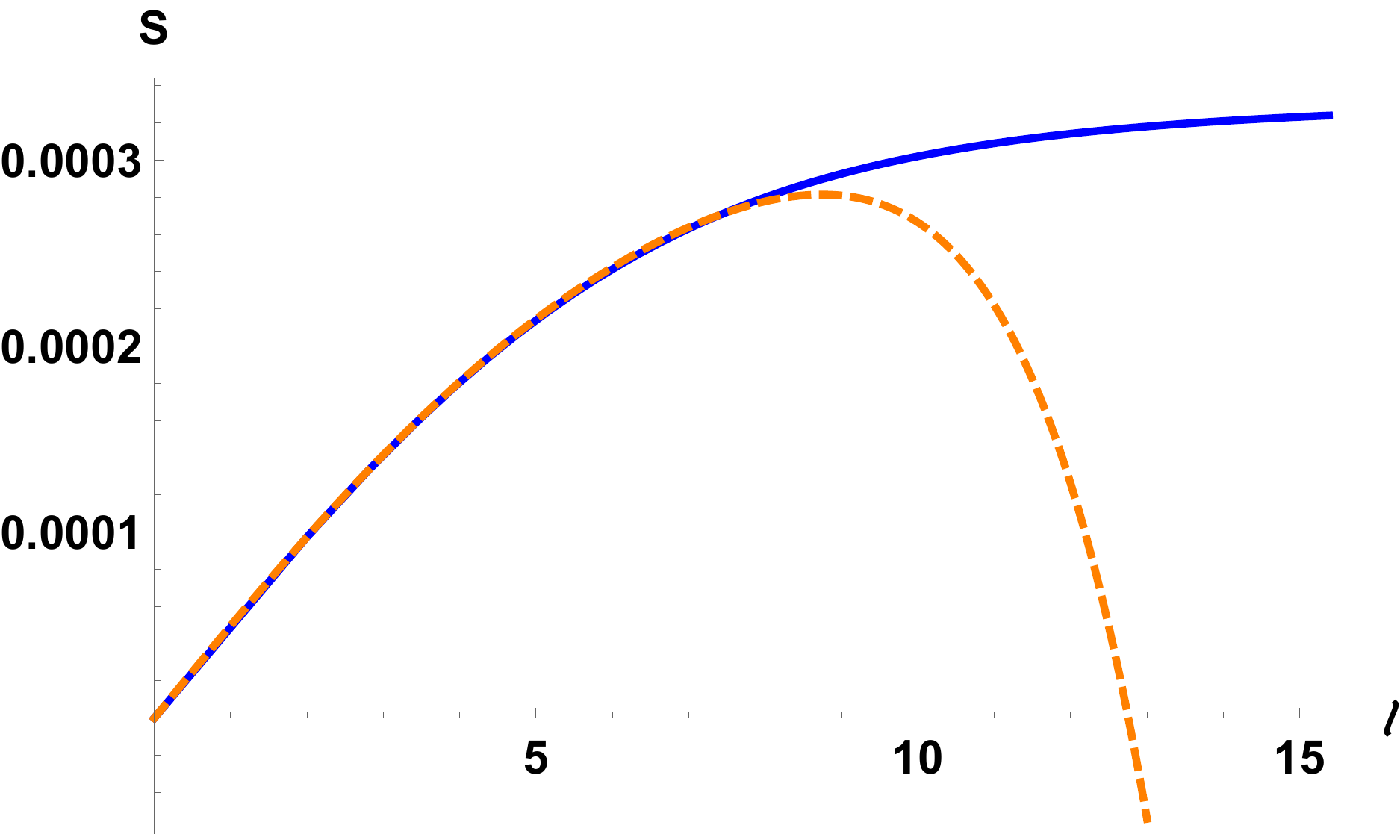}
		\vspace{-5mm}
	\end{center}
	\caption{HEE as a function of $\ell$ for different values of the cutoff: {\it Left}) $r_c= 1$ and $d_e =3$ {\it Right}) $r_c= 10$ and $d_e =4$.  The orange dashed curves are based on the perturbative solution in eq. \eqref{EE-strip-de neq 1-l<<rc} which is valid for very small entangling regions, i.e. $l \ll r_c$, and the solid blue curves are numerical results based on eq. \eqref{EE-strip-de neq 1}. 
	}
	\label{fig:S-small-ER}
\end{figure}

From figure \ref{fig:l-rt}, it is straightforward to see that for very small entangling regions, one has $r_t \approx r_c \gg \ell$. In this limit, one can expand eq. \eqref{l-rt-neq 1} around $r_t = r_c$ as follows
\bea
\frac{\ell}{r_c} = \sqrt{\frac{8 (r_t-r_c)}{d_e r_c}} \left[ 1 + \frac{(5 - 2 d_e)}{12} \frac{(r_t - r_c)}{r_c} + \cdots \right].
\eea 
Next, one can solve the above equation and find $r_t$. Then one can expand it in powers of $\frac{l}{r_c} \ll 1$, and obtain
\bea
r_t &=& r_c \bigg[ 1 + \frac{d_e}{8} \left(\frac{\ell}{r_c}\right)^2 - \frac{d_e^2 (5-2 d_e)}{384} \left(\frac{\ell}{r_c}\right)^4 
\cr && \cr
&& \;\;\;\;\;\;\;\;\;\; 
+ \frac{7 d_e^3 (5 - 2 d_e)^2}{73728} \left( \frac{\ell}{r_c} \right)^6 
 - \frac{5 d_e^4 (5 - 2 d_e)^3}{1179648} \left( \frac{\ell}{r_c} \right)^8
+ \cdots \bigg].
\label{rt-strip-de neq 1- l, rt approx rc} 
\eea 
In figure \ref{fig:l-rt-small-ER}, the above solution is drawn as a function of $\ell$.
On the other hand, by expanding eq. \eqref{EE-strip-de neq 1} around $r_t = r_c$, one has
\bea
S &=&\frac{R^{d} L^{d-1}}{\sqrt{2} G_N r_F^\theta } \frac{1}{ r_c^{d_e - 1}}
\Big[  \left(\frac{r_t -r_c}{r_c} \right)^\frac{1}{2} 
+ \frac{(5- 6 d_e)}{12} \left( \frac{r_t - r_c}{r_c} \right)^{\frac{3}{2}}
\cr && \cr
&& \;\;\;\;\;\;\;\;\;\;\;\;\;\;\;\;\;\;\;\;\;\;\;\;\;\;
+\frac{(1 +2 d_e)(50 d_e - 47)}{480} \left( \frac{r_t - r_c}{r_c}\right)^{\frac{5}{2}}
\cr && \cr
&& \;\;\;\;\;\;\;\;\;\;\;\;\;\;\;\;\;\;\;\;\;\;\;\;\;\;
+\frac{(1+2 d_e)(4 d_e (61+ 105 d_e) -639)}{13440} \left( \frac{r_t - r_c}{r_c}\right)^{\frac{7}{2}}
+\cdots
\Big].
\eea 
Next, by plugging eq. \eqref{rt-strip-de neq 1- l, rt approx rc} into the above expression, and expanding it again in powers of $\frac{\ell}{r_c} \ll 1$, one obtains 
\bea
S &=&\frac{R^{d} L^{d-1}}{4 G_N r_F^\theta } \frac{1}{r_c^{d_e -1}} \Big[
\left( \frac{\ell}{r_c} \right) - \frac{d_e^2}{24} \left( \frac{\ell}{r_c}\right)^3 + \frac{ d_e^2 \left(4 d_e (39 + 5 d_e) - 47 \right)}{30720} \left( \frac{\ell}{r_c} \right)^5 
\cr && \cr
&&
\;\;\;\;\;\;\;\;\;\;\;\;\;\;\;\;\;\;\;\;\;\;\;\;\;\; +\frac{d_e^3 (5071 - 4 d_e (2561 + 113 d_e))}{10321920} \left( \frac{\ell}{r_c} \right)^7
+  \cdots
\Big].
\label{EE-strip-de neq 1-l<<rc}
\eea
In figure \ref{fig:S-small-ER}, the above expression for the HEE is drawn as a function of $\ell$, and compared with the numerical result.
\subsubsection{Very Small Cutoff}
\label{Sec: Very Small Cutoff}
For finite $\ell$ and very small $r_c$, one again has $\ell \approx r_t \gg r_c$, and hence the HEE is the same as in section \ref{Sec: Very Large Entangling Regions}. From eq. \eqref{EE-strip-de neq 1-l>>rc} and \eqref{Delta-EE-strip-de neq 1-l>>rc}, it is obvious that, when $r_c= \epsilon \rightarrow 0$, one has $\Delta S=0$ and hence the HEE reduces to eq. \eqref{EE-strip-de neq 1-zero cutoff}.
\\Before concluding this section, we should emphasize that EE is a quantity defined in the dual QFT and should be expressed in terms of the parameters of the QFT. In particular, one should be able to express the coefficient $c= \frac{R^{d}}{4 G_N r_F^\theta }$ which appears in the HEE. Unfortunately, for arbitrary values of the exponents $z$ and $\theta$, we cannot interpret the coefficient $c$ in the dual QFT. However, in ref. \cite{Alishahiha:2015goa} the two-point function of stress tensor in the dual QFT was calculated for the case $z=1$ as follows
\bea
\langle T_{ab}(x) T_{cd}(y)\rangle=\frac{C_T}{(x-y)^{2(d+1)}}G_{abcd}(x,y),
\eea
where $G^{\;\;b\;\;d}_{a\,\;c}(x,y)=J_a^i(x-y)J^b_j(x-y) P^{\;j\;\;d}_{i\,\;c}$ such that
\be
J^i_j(x)
=
\delta_j^i-2\frac{x_j x^i}{|x|^2},
\;\;\;\;\;\;
P^{\;j}_{i\,\;ab}
=
\frac{1}{2}
\left(
\delta_{ia}\delta^j_{b}+\delta_{ib}\delta^j_{a}
\right)-
\frac{1}{d+1}\delta_i^j\delta_{ab}.
\ee
Moreover, the constant $C_T$ is defined as follows \cite{Alishahiha:2015goa}
\bea
C_T=\frac{R^d}{8\pi Gr_F^{\theta}}\;\frac{(d+2)}{d}\;\frac{\Gamma(d_e+2)}
{\pi^{\frac{d+1}{2}}\Gamma\left( \frac{1+2d_e-d}{2}\right)}.
\label{CT-z=1}
\eea
Having said this, one might rewrite the coefficient $c$ in terms of $C_T$ as follows
\bea
c = c_0 \; C_T
\label{c-CT}
\eea 
where 
\bea
c_0 = \frac{2d \; \Gamma\left( \frac{1+2d_e-d}{2}\right)}{(d+2) \Gamma(d_e+2)},
\label{c0}
\eea
is a dimensionless constant depending on $d$ and $d_e$. 
\\As mentioned before, for $z=1$ and $\theta = 0$, the Lorenz and scaling symmetries are restored in the dual QFT and it becomes a $CFT_{d+1}$. Notice that this $CFT_{d+1}$ lives on $\mathbb{R}^{d+1}$ and is in its vacuum state. 
Therefore, all of our results can be applied for this $CFT_{d+1}$, if one sets $\theta=0$. In particular, 
for $d \geq 2$ and very small cutoff, by setting $\theta=0$ in eq. \eqref{Delta-EE-strip-de neq 1-l>>rc}, one can write $\Delta S$ as follows
\bea
\Delta S &=& \frac{c_0 C_T}{(d+1)}  \left( \frac{2 \Upsilon}{\ell} \right)^{2d} \Big[
r_c^{d+1} - \frac{3 (d+1)}{4(3d +1) }  \left(\frac{2 \Upsilon}{\ell}\right)^{2 d} r_c^{3d+1} 
\cr && \cr
&&
\;\;\;\;\;\;\;\;\;\;\;\;\;\;\;\;\;\;\;\;\;\;\;\;\;\;\;\;\;\;\;\;\;\;\;\;
- \frac{d (3 d+1)}{3(d+1)^2 \Upsilon} \left(\frac{2 \Upsilon}{\ell}\right)^{2 (d +1)} r_c^{3 (d+1) } 
\cr && \cr
&&
\;\;\;\;\;\;\;\;\;\;\;\;\;\;\;\;\;\;\;\;\;\;\;\;\;\;\;\;\;\;\;\;\;\;\;\;
+ \frac{3 d}{(3 d+1) \Upsilon} \left(\frac{2 \Upsilon}{\ell}\right)^{3 d+1} r_c^{2 (2 d+1)}
+ \cdots
\Big],
\label{Delta-EE-strip-de neq 1-l>>rc-CFT}
\eea 
where for $c_0$ and $C_T$ one should set $\theta=0$ in eqs. \eqref{CT-z=1} and \eqref{c0}.
\section{Holographic Mutual Information}
\label{Sec: Holographic Mutual Information}
In this section, we calculate holographic mutual information (HMI) for two parallel disjoint strips and study its dependence on the cutoff $r_c$. 
\begin{figure}
	\begin{center}
		\includegraphics[scale=0.83]{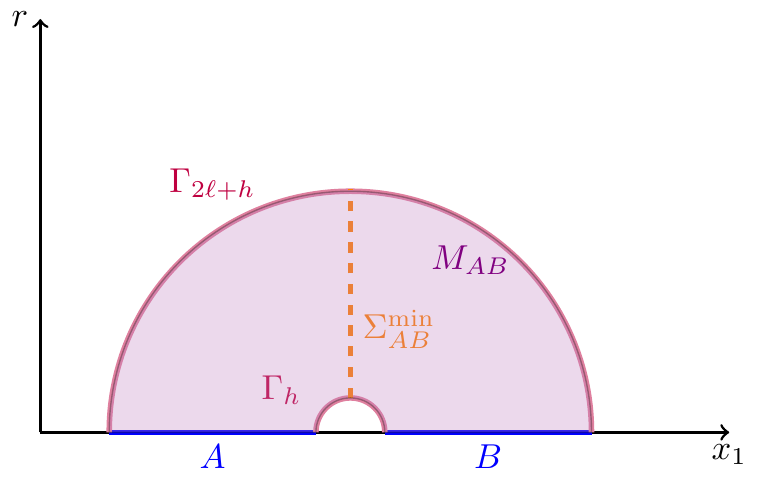}
		\hspace{0.2cm}
		\includegraphics[scale=0.83]{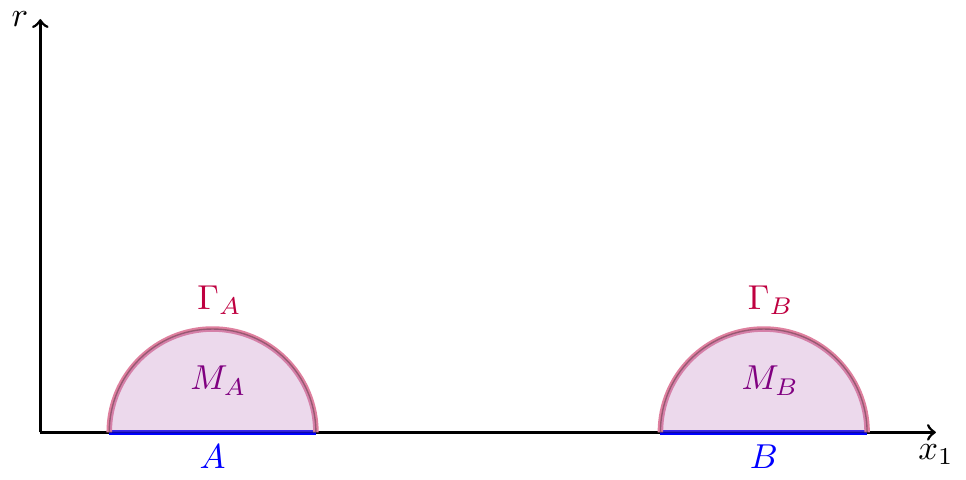}
	\end{center}
	\caption{Illustrations of the connected and disconnected RT surfaces which contribute to the HMI as well as EW for a bipartite system $A \cup B$. {\it Left}) Connected configuration: the two entangling regions $A$ and $B$ are close enough to each other such that the RT surface corresponding to $S_{AB}$ is given by $\Gamma_h \cup \Gamma_{ 2 \ell +h}$ which are shown by the solid purple curves. In this case, the EW is also connected and indicated by the shaded violet region. The minimal surface $\Sigma^{\rm min}_{AB}$ which measures the minimal cross section of EW is indicated by an orange dashed curve. {\it Right}) Disconnected configuration: the two entangling regions are far enough from each other, such that the RT surface corresponding to $S_{AB}$ is disconnected and given by $\Gamma_A \cup \Gamma_B$ shown by purple solid curves. Moreover, the EW is disconnected and given by $M_A \cup M_B$ which are indicated by the shaded violet regions. In this case, the minimal surface $\Sigma^{\rm min}_{AB}$ does not exist, and hence $E_W = 0$. Notice that, these diagrams are schematic, and only for the $d_e=1$ case, the RT surfaces are semicircles.
	}
	\label{fig:HMI-EWCS}
\end{figure}
In ref. \cite{Headrick:2010zt} it was proposed that, there are two types of RT surfaces which contribute in $S_{AB}$: {\it Connected}) in which each RT surface is started from the boundary of one entangling region, say $A$, and ended on the boundary of the other entangling region $B$ (see the left panel of figure \ref{fig:HMI-EWCS}). {\it Disconnected}) in which each RT surface is started from the boundary of one of the entangling regions and is ended on the other boundary of the same entangling region (see the right panel of figure \ref{fig:HMI-EWCS}). It was proposed in ref. \cite{Headrick:2010zt} that
\bea
S_{A B} = {\rm Min} \left( S_{\rm con.} , S_{\rm dis.} \right) ,
\label{SAB}
\eea 
where $S_{\rm con.}$ and $S_{\rm dis.}$ are the HEE for the connected and disconnected configurations, respectively. Therefore, depending on the size of the entangling regions and the distance between them, always one of the connected or disconnected configurations dominates and contributes in the HMI. Consequently, there is a first-order phase transition in the HMI \cite{Headrick:2010zt,Tonni:2010pv}. For convenience, we choose the width of both strips to be equal to $\ell$ and show the distance between them by $h$. One simply obtains
\begin{align}\label{SAUB}
I(A,B)=
\begin{cases}
0 & ~~ \ell \ll  h\\
2 S(\ell) - S(h) - S(2 \ell+h) & ~~ \ell \gg h
\end{cases}
\end{align}
In the following, we consider the $d_e=1$ and $d_e \neq 1$ cases, separately. It should be pointed out that the HMI for HV geometries at zero cutoff was studied in refs. \cite{Huijse:2011ef,Fischler:2012uv,Tanhayi:2015cax}
\subsection{$d_e=1$}
\label{Sec: HMI-de=1}

\begin{figure}
	\begin{center}
		\includegraphics[scale=0.34]{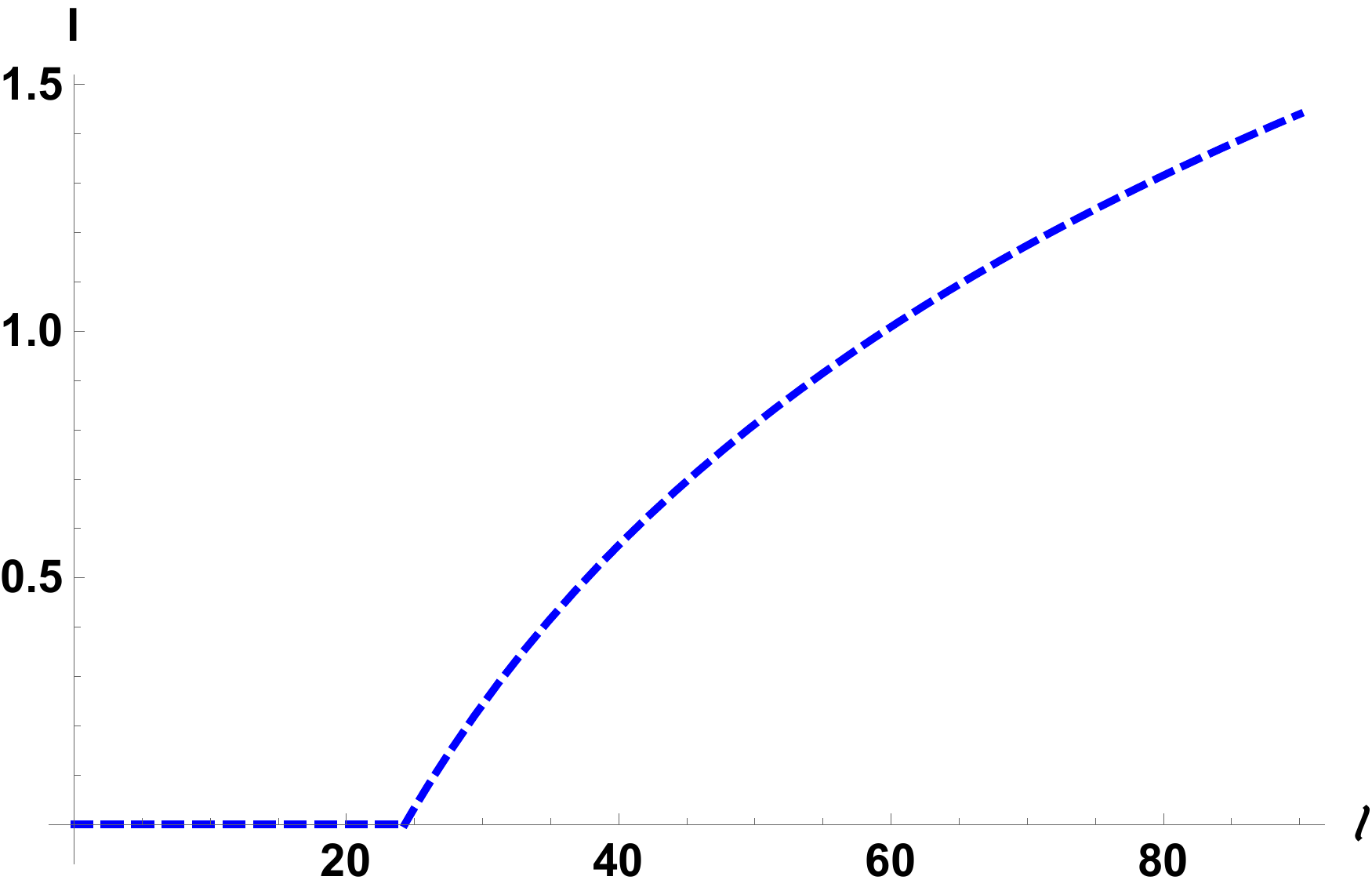}
		\hspace{1cm}
		\includegraphics[scale=0.34]{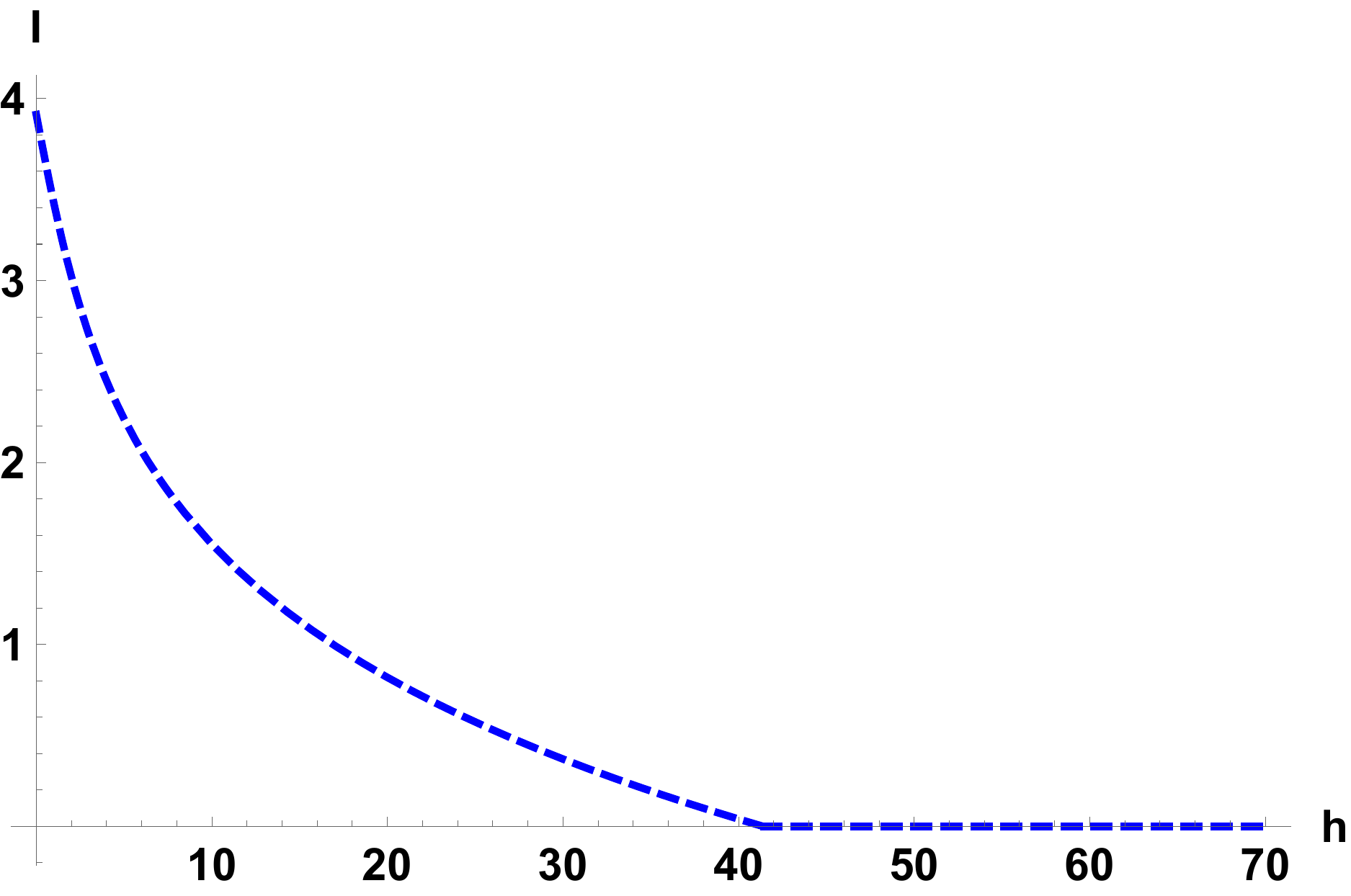}
		\vspace{-5mm}
	\end{center}
	\caption{HMI for $d_e=1$: {\it Left}) as a function of $\ell$ for $r_c=1$ and $h=10$. {\it Right}) as a function of $h$ for $r_c=1$ and $\ell=100$.  HMI shows a first-order phase transition at the point $h_{\rm crit.}$ which depends on the cutoff $r_c$. Here we set $R=1$ and renormalized $I$ as $\tilde{I}= \frac{I}{a}$, where $a= \frac{R^d L^{d-1}}{2 G_N r_F^\theta}$.
	}
	\label{fig:HMI-l-h-de=1}
\end{figure}

From eq. \eqref{EE-strip-de=1-zero cutoff}, for the zero cutoff case one has \cite{Fischler:2012uv}
\begin{align}
I_0(A,B)=
\begin{cases}
0 & ~~ \ell \ll  h\\
\frac{R^d L^{d-1}}{2 G_N r_F^{\theta}} \log\left( \frac{\ell^2}{h(2 \ell +h)}\right)  & ~~ \ell \gg h
\end{cases}
\label{HMI-de=1-zero cutoff}
\end{align}
which is independent of the UV cutoff, and for $\theta=0$ and $d=1$, it reduces to that of a $CFT_2$ in its vacuum state \cite{Headrick:2010zt,Fischler:2012uv}. Moreover, there is a first-order phase transition which happens at the critical value $h_{\rm crit.}$ given by
\bea
h_{\rm crit.}^{(0)} = \ell ( \sqrt{2} -1 ).
\label{hcrit-de=1-zero cutoff}
\eea 
Furthermore, as it was shown in refs. \cite{Swingle:2010jz,Mozaffar:2015xue}, when the entangling regions have common boundaries, i.e. $h \rightarrow 0$, one has $I_0 \rightarrow \infty$. On the other hand, for the finite cutoff case from eq. \eqref{EE-strip-de=1}, one obtains
\begin{align}
I(A,B)=
\begin{cases}
0 & ~~ \ell \ll  h\\
\frac{R^d L^{d-1}}{2 G_N r_F^{d-1}} \log\left(\frac{ \left(\ell+ \sqrt{\ell^2 + 4 r_c^2} \right)^2 }{\left( h + \sqrt{h^2 + 4 r_c^2} \right) \left( (2 \ell+ h) + \sqrt{( 2 \ell+h )^2 + 4 r_c^2}\right)}\right)  & ~~ \ell \gg h
\end{cases}
\label{HMI-de=1}
\end{align}
In figure \ref{fig:HMI-l-h-de=1}, the HMI is drawn as a function of $\ell$ and $h$. The HMI shows a first-order phase transition which happens at
\bea
h_{\rm crit.} = \ell \left(-1 + \frac{\sqrt{2} (\ell^2 + 2 r_c^2)}{\sqrt{\ell^4+ 6 \ell^2 r_c^2 + 8 r_c^4}}\right).
\label{h-crit-de=1}
\eea 
\begin{figure}
	\begin{center}
		\includegraphics[scale=0.34]{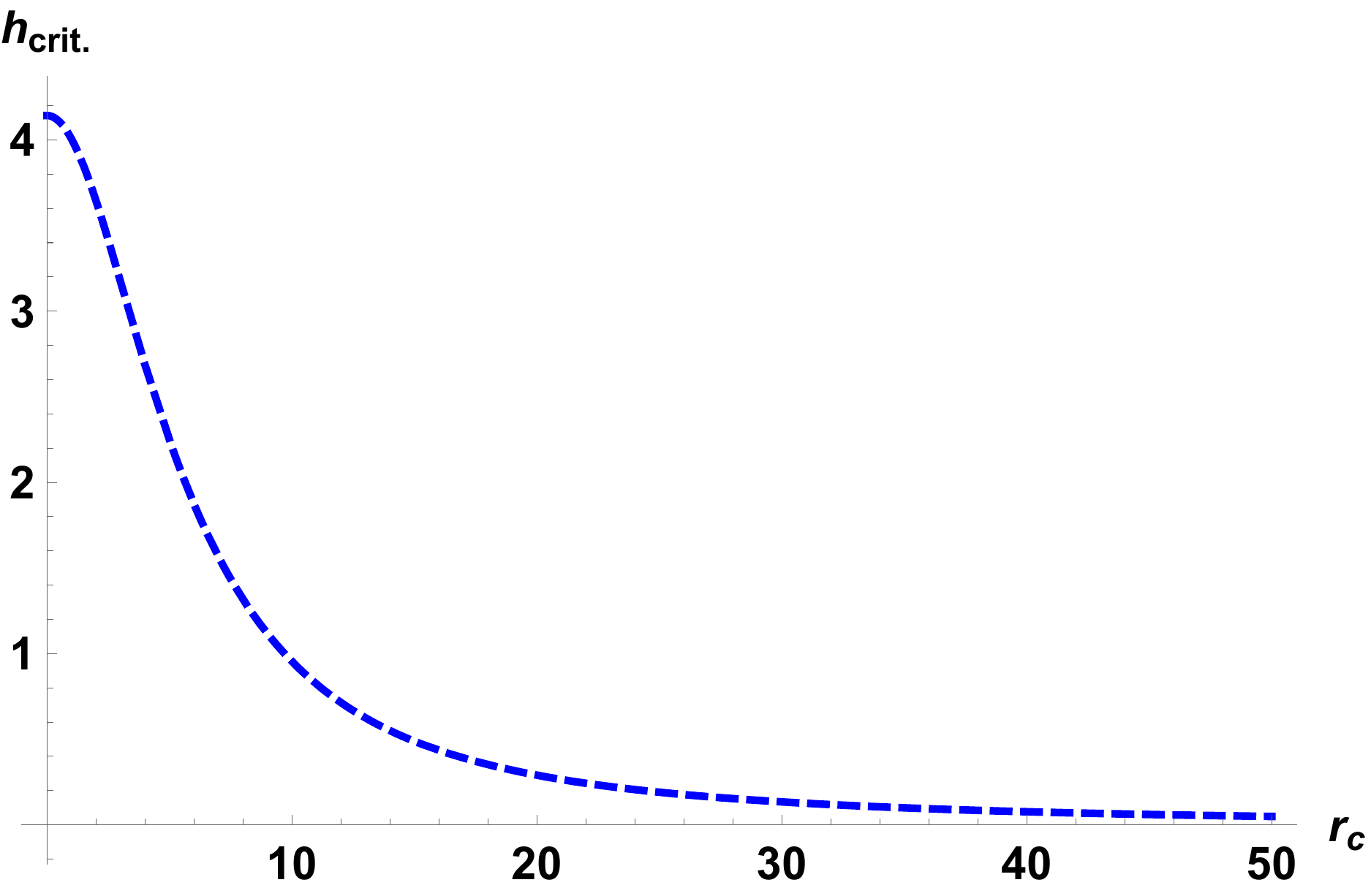}
		\hspace{1cm}
		\includegraphics[scale=0.34]{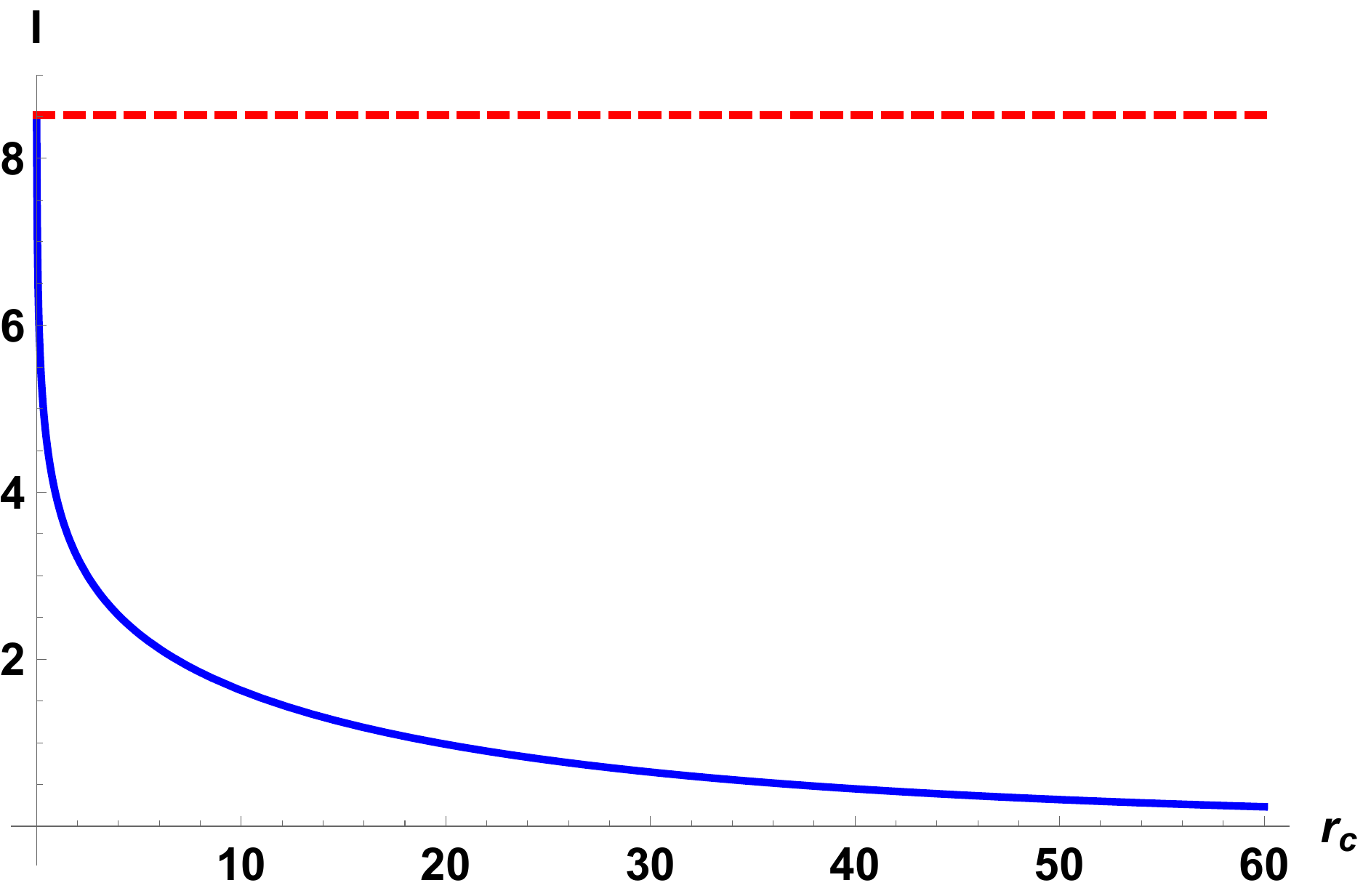}
		\vspace{-5mm}
	\end{center}
	\caption{ {\it Left}) $h_{\rm crit.}$ as a function of $r_c$ for $\ell=10$ and $d_e=1$. It is evident that $h_{\rm crit.}$ is a decreasing function of $r_c$. {\it Right}) HMI as a function of $r_c$ for $l= 10^{2}$, $h=10^{-2}$ and $d_e=1$. The red dashed curve indicates the HMI for the zero cutoff case given in eq. \eqref{HMI-de=1-zero cutoff} and the solid blue curve is the HMI for the finite cutoff case, i.e. eq. \eqref{HMI-de=1}. It is evident that the HMI is a decreasing function of the cutoff $r_c$.
	}
	\label{fig:HMI-rc-hcrit-rc-de=1}
\end{figure}
Therefore, in contrast to the zero cutoff case, $h_{\rm crit.}$ depends on the cutoff $r_c$. In particular, the left side of figure \ref{fig:HMI-rc-hcrit-rc-de=1}, shows that $h_{\rm crit}$ is a decreasing function of $r_c$. Another interesting point is that according to eq. \eqref{HMI-de=1}, the HMI depends on the cutoff $r_c$, which is in contrast to the zero cutoff case, i.e. eq. \eqref{HMI-de=1-zero cutoff}. Furthermore, it is observed from the right panel of figure \ref{fig:HMI-rc-hcrit-rc-de=1} that the HMI is a decreasing function of $r_c$, and goes to zero in the limit $r_c \rightarrow \infty$. Moreover, when $h \rightarrow 0$ from eq. \eqref{HMI-de=1}, one has
\begin{align}
I(A,B)= \frac{R^d L^{d-1}}{2 G_N r_F^{d-1}} \log\left(\frac{ \left(\ell+ \sqrt{\ell^2 + 4 r_c^2} \right)^2 }{ r_c \left( \ell + \sqrt{\ell^2 + r_c^2}\right)}\right).
\label{HMI-de=1-h=0}
\end{align}
Consequently, in contrast to the zero cutoff case, the HMI remains finite when the distance between the two strips goes to zero. In ref. \cite{Asrat:2020uib}, it was observed that for BTZ black holes at finite radial cutoff, the HMI shows the same behavior.
\subsection{$d_e \neq 1$}
\label{Sec: HMI-de neq 1}
\begin{figure}
	\begin{center}
		\includegraphics[scale=0.29]{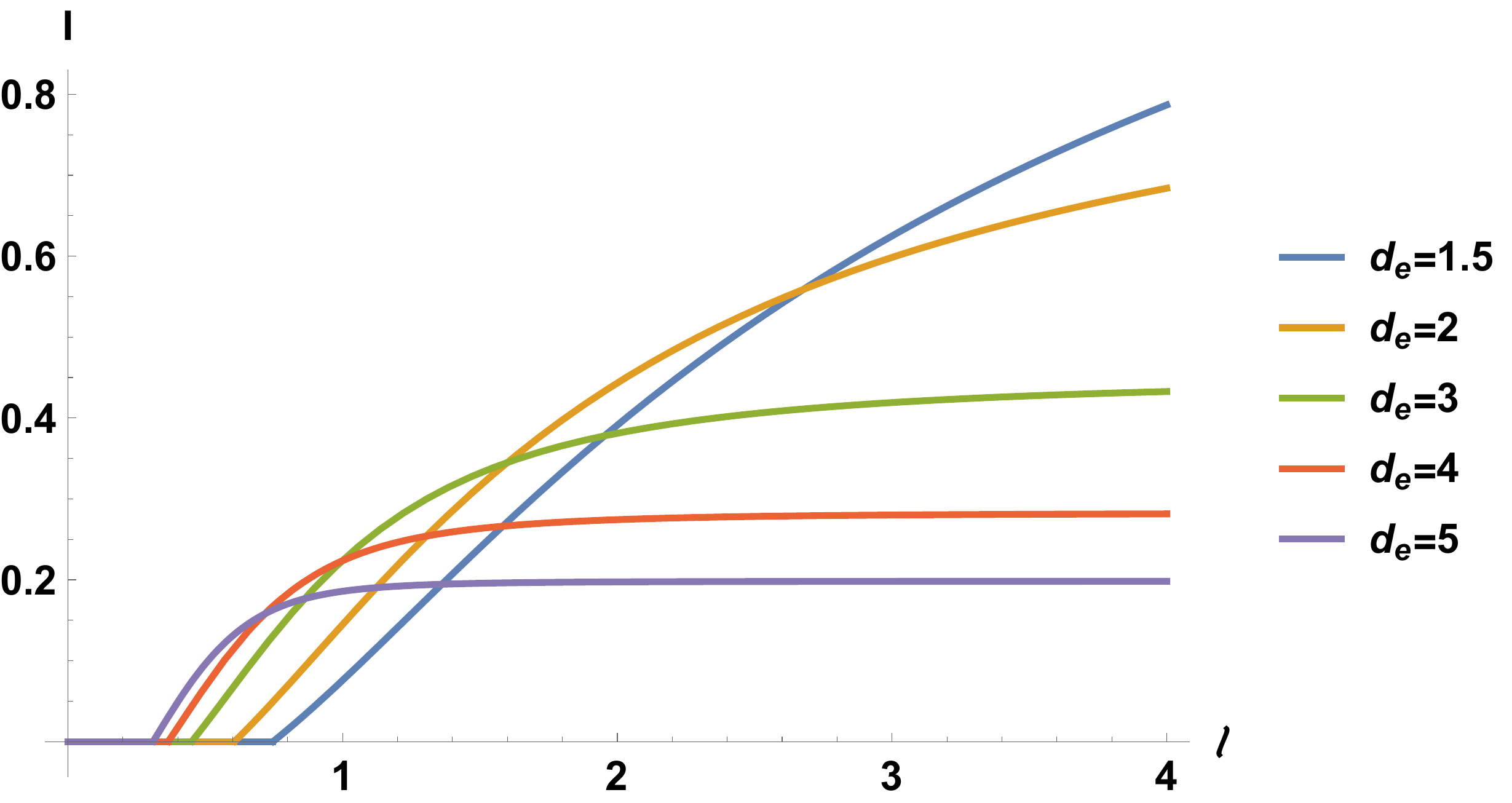}
		\hspace{0.1cm}
		\includegraphics[scale=0.29]{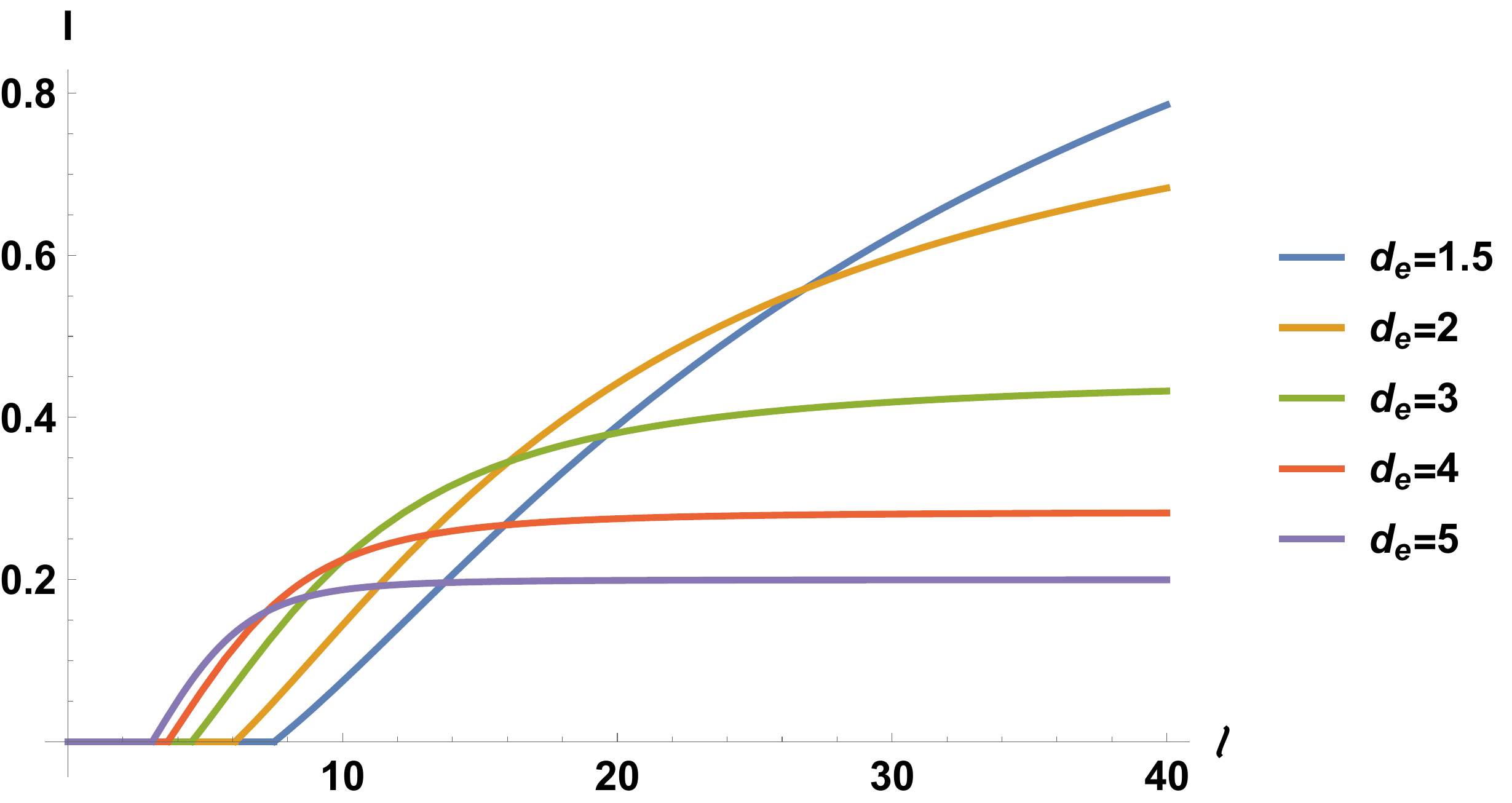}
		\vspace{-5mm}
	\end{center}
	\caption{HMI for $d_e \neq 1$ as a function of $\ell$ for: {\it Left}) $h=0.1$ and $r_c= 1$ {\it Right}) $h=1$ and $r_c = 10$. 
		Here and in the following figures, we renormalized $I$ as $\tilde{I}= \frac{r_c^{d_e -1}I}{a}$, where $a= \frac{R^d L^{d-1}}{2 G_N r_F^\theta}$.
	}
	\label{fig:HMI-l}
\end{figure}

\begin{figure}
	\begin{center}
		\includegraphics[scale=0.33]{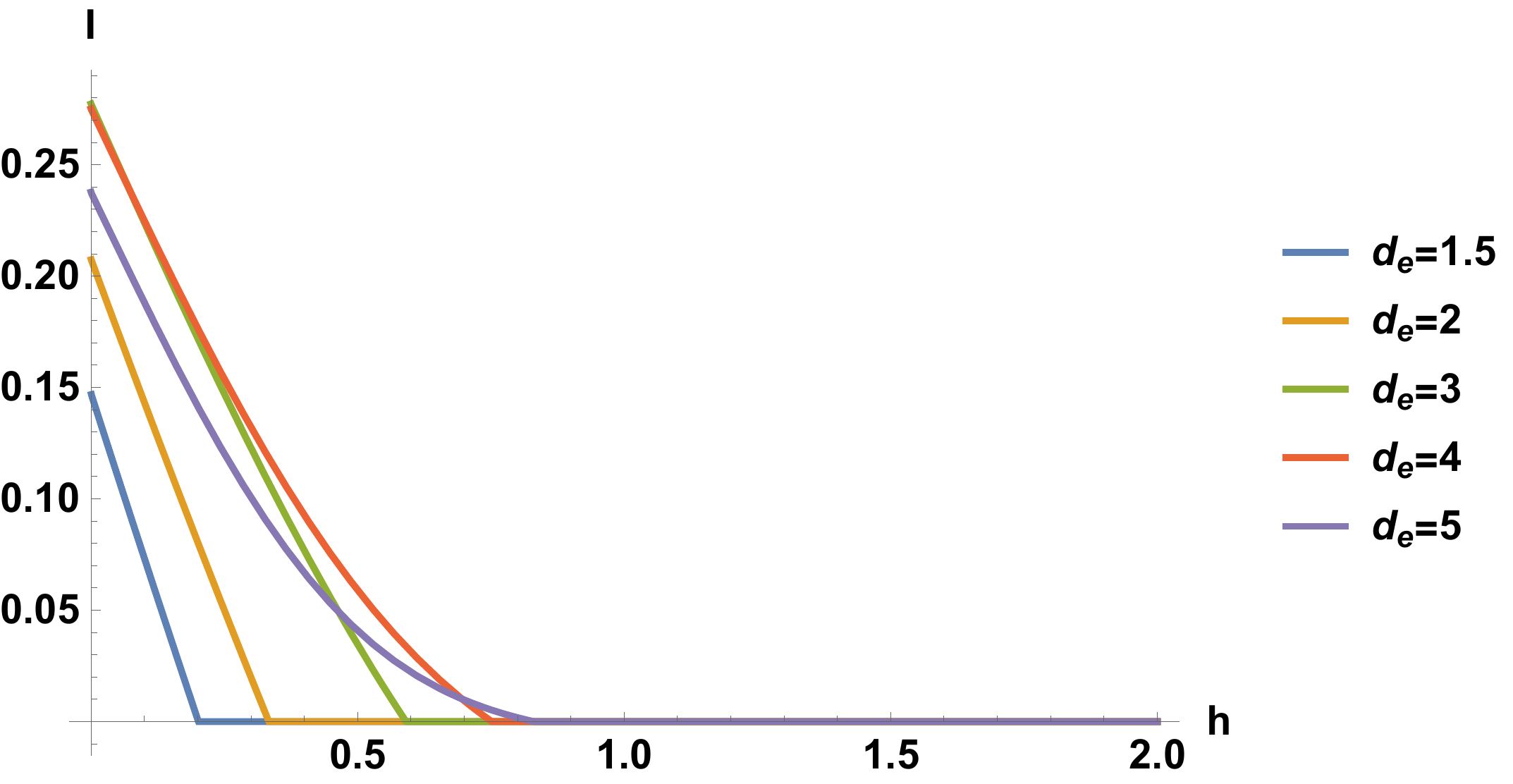}
		\hspace{0.1cm}
		\includegraphics[scale=0.33]{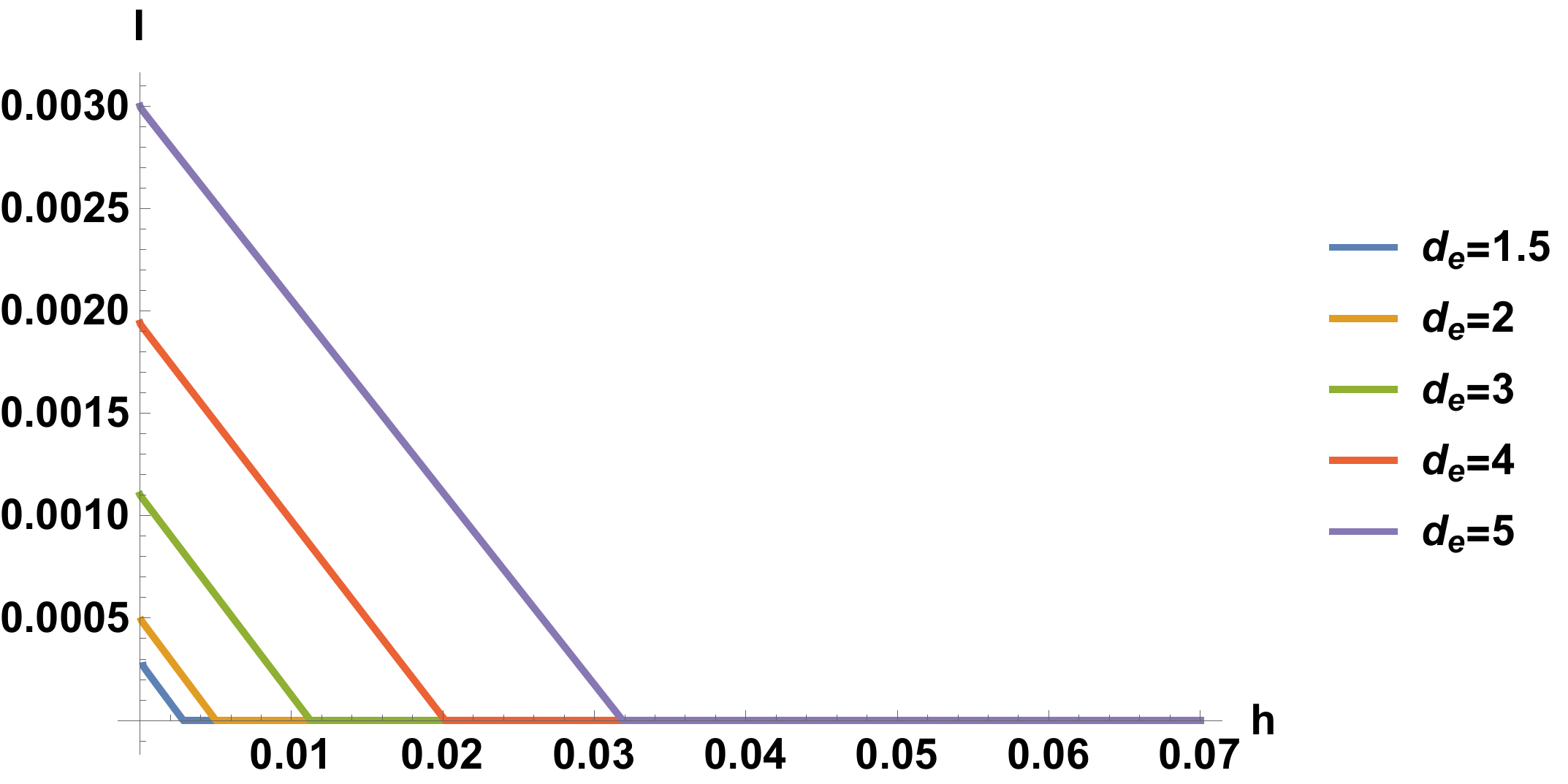}
		\vspace{-5mm}
	\end{center}
	\caption{HMI as a function of $h$ for: {\it Left}) $\ell=1$ and $r_c= 1$ {\it Right}) $\ell=1$ and $r_c = 10$. There is a critical length $h_{\rm crit.}$ at which there is a first-order phase transition in the HMI. Comparison between the left and right panels, shows that for each value of $d_e$, $h_{\rm crit.}$ decreases by increasing $r_c$. Moreover, by increasing $d_e$, $h_{\rm crit}$ becomes larger. Furthermore, at $h=0$, the HMI is finite which is in contrast to the zero cutoff case. 
	}
	\label{fig:HMI-h}
\end{figure}

\begin{figure}
	\begin{center}
		\includegraphics[scale=0.34]{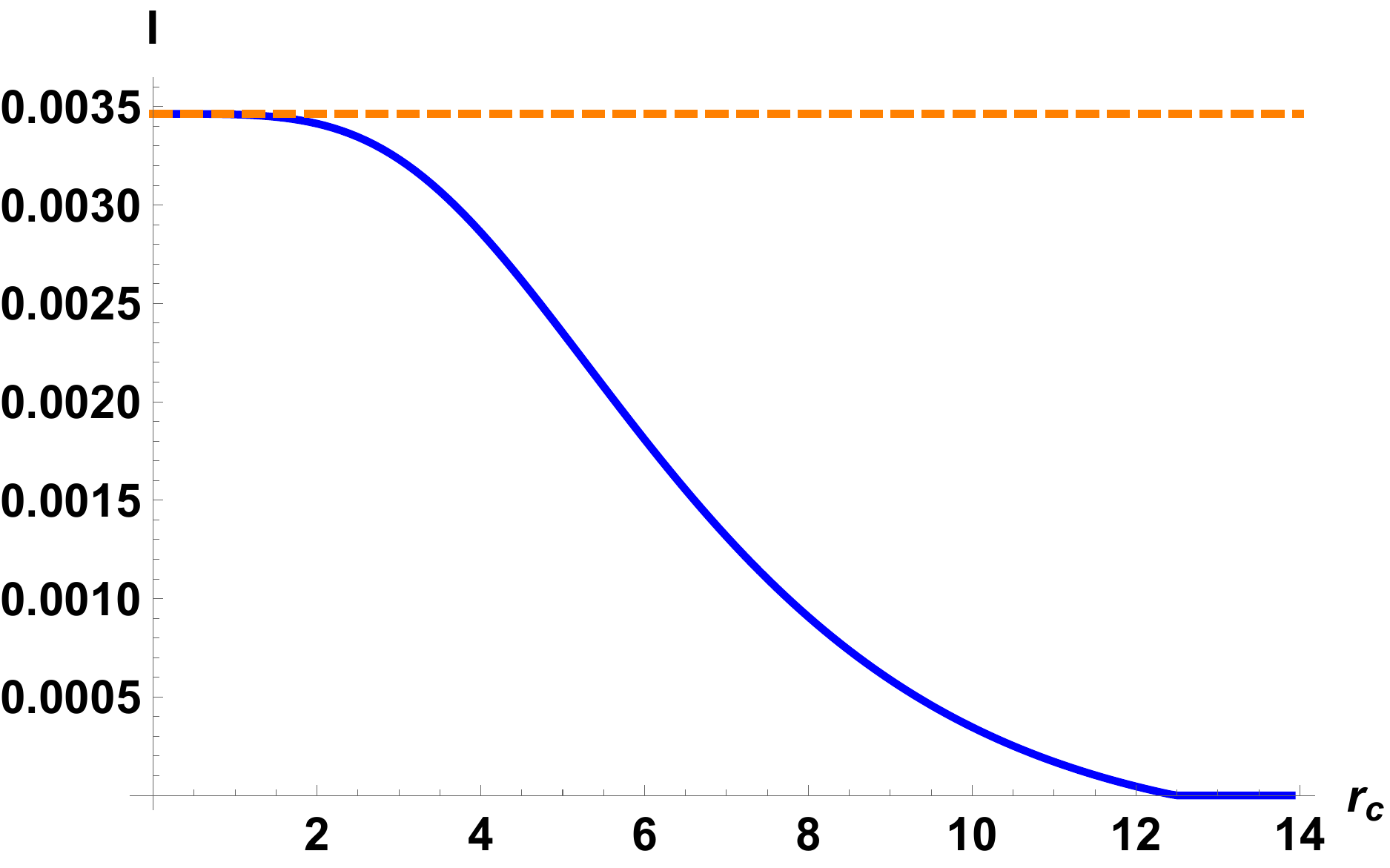}
		\hspace{1.2cm}
		\includegraphics[scale=0.34]{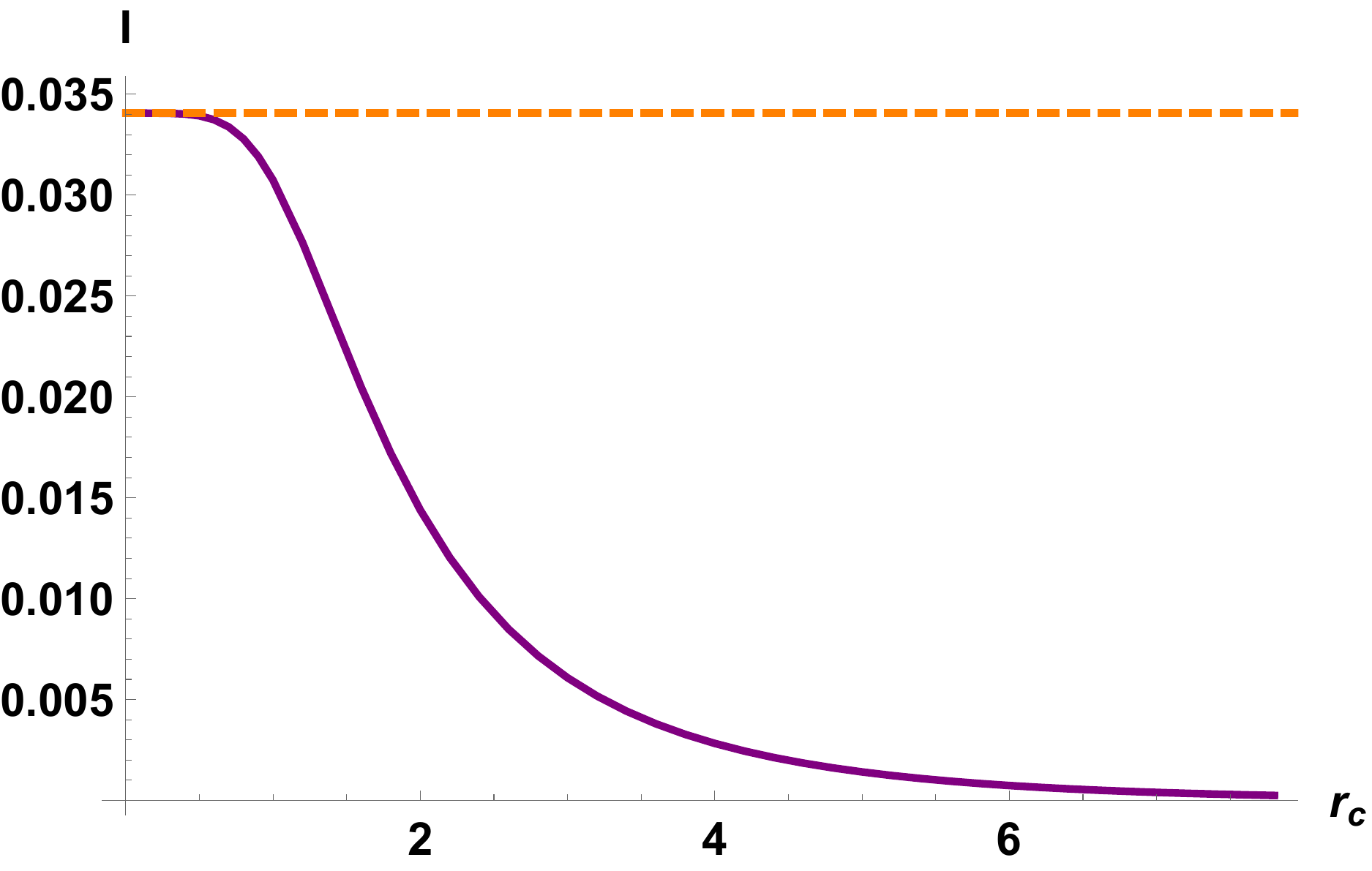}
		\vspace{-5mm}
	\end{center}
	\caption{HMI as a function of $r_c$ for: {\it Left}) $ \ell = 10$, $h=5$ and $d_e=3$. {\it Right}) $ \ell =5$, $h=1$ and $d_e=4$. The orange dashed curves indicate the HMI for the zero cutoff case given in eq. \eqref{HMI-de neq 1-zero cutoff}. It is obvious that the HMI depends on the cutoff, and is a decreasing function of $r_c$, such that it goes to zero when $r_c \rightarrow \infty$. 
	}
	\label{fig:HMI-rc}
\end{figure}

In figures \ref{fig:HMI-l}, \ref{fig:HMI-h} and \ref{fig:HMI-rc} the HMI is calculated numerically from eq. \eqref{EE-strip-de neq 1}. In figure \ref{fig:HMI-l}, the HMI is drawn as a function of $\ell$, and one can see that by increasing $d_e$ the phase transition happens at smaller values of $\ell$. In figure \ref{fig:HMI-h}, the HMI is drawn as a function of $h$. It is evident that when the distance between the two strips goes to zero, i.e. $h \rightarrow 0$, the HMI remains finite which is in contrast to the zero cutoff case (see also eq. \eqref{HMI-de neq 1-zero cutoff}). 
Furthermore, comparison between the left and right panels of figure \ref{fig:HMI-h}, shows that for a specific value of $d_e$, $h_{\rm crit.}$ decreases by increasing $r_c$. 
\footnote{This behavior is more evident in figure \ref{fig:h-crit-rc-de neq 1}.}
On the other hand, in figure \ref{fig:HMI-rc}, the HMI is drawn as a function of $r_c$. It is observed that the HMI depends on the cutoff and is a decreasing function of $r_c$, and it goes to zero in the limit $r_c \rightarrow \infty$.
In the following, we find analytic results for very large and small entangling regions.
\subsubsection{Very Large Entangling Regions}
\label{Sec: HMI-very large ER}
\begin{figure}
	\begin{center}
		\includegraphics[scale=0.34]{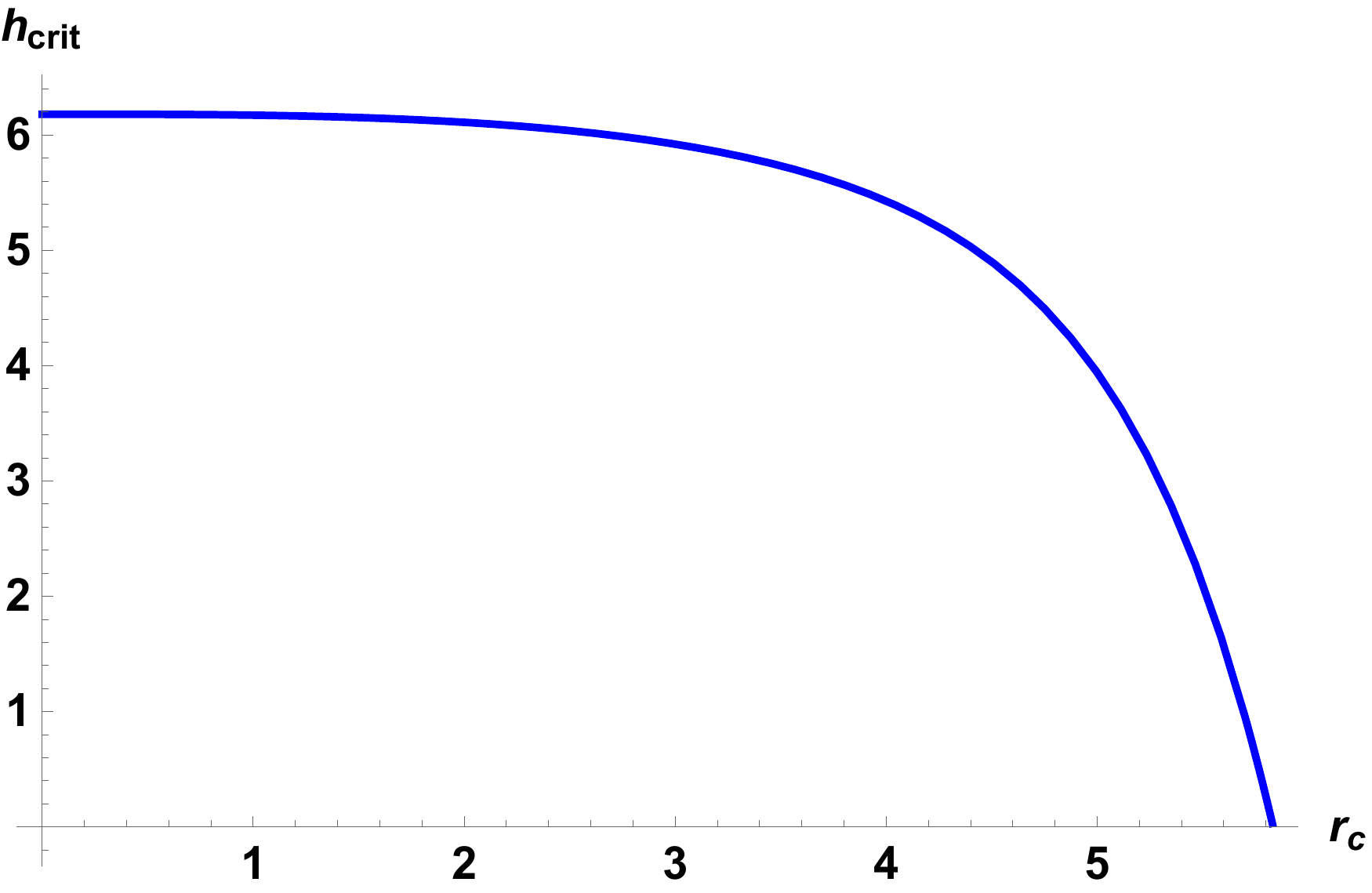}
		\hspace{1cm}
		\includegraphics[scale=0.34]{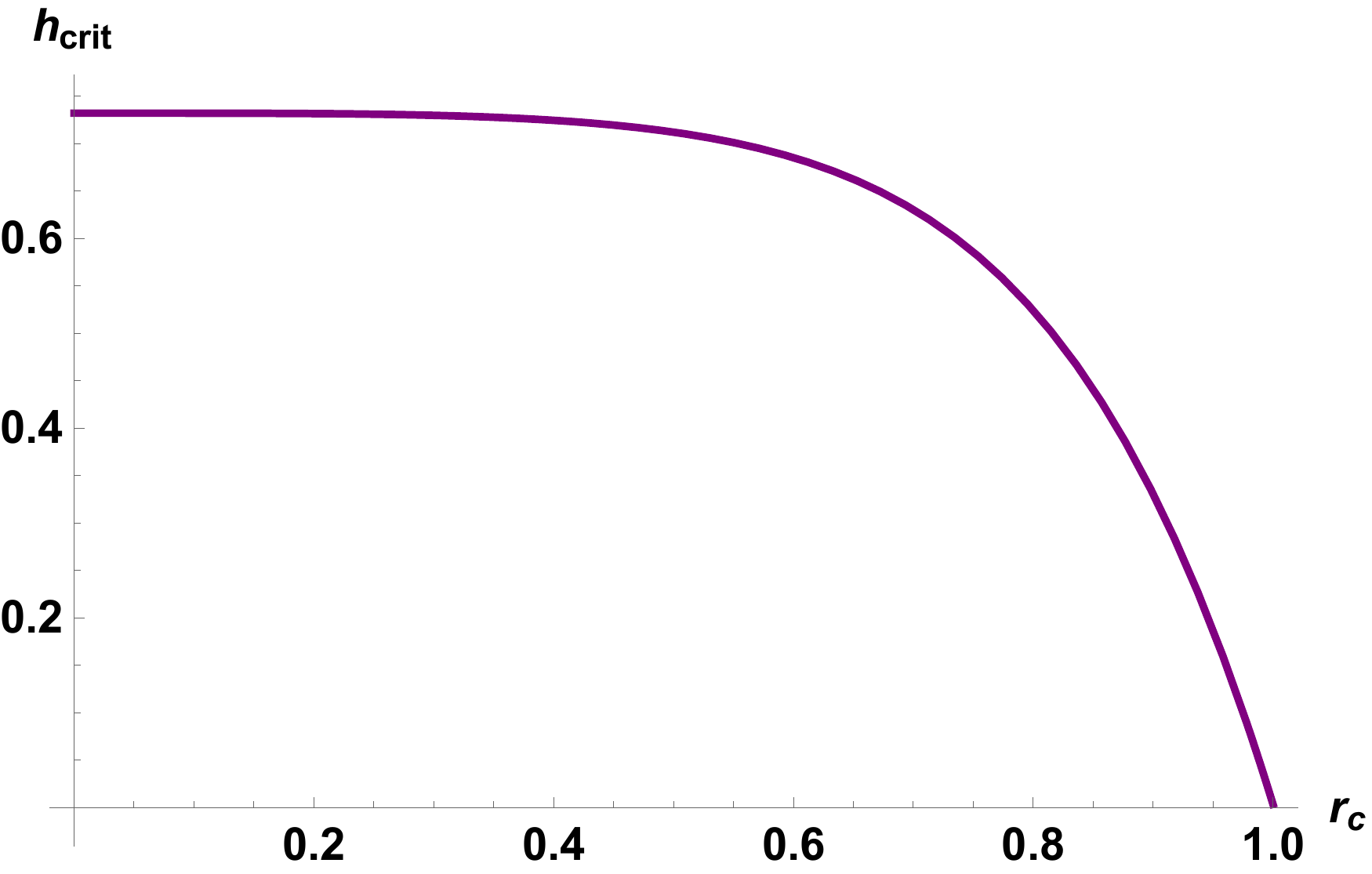}
		\vspace{-5mm}
	\end{center}
	\caption{$h_{\rm crit.}$ as a function of $r_c$ for: {\it Left}) $ \ell = 10$ and $d_e=2$. {\it Right}) $ \ell =1$ and $d_e=3$. Notice that these diagrams are based on eqs. \eqref{h-crit-de=2} and \eqref{h-crit-de=3} which are valid for $\ell \gg r_c$.
	}
	\label{fig:h-crit-rc-de neq 1}
\end{figure}

First we consider very large entangling regions, i.e. $\ell \gg r_c$. From eq. \eqref{EE-strip-de neq 1-l>>rc}, one has 
\begin{align}
I(A,B)=
\begin{cases}
0 & ~~ \ell \ll  h\\
I_0 + \Delta I  & ~~ \ell \gg h
\end{cases}
\label{HMI-large-ER}
\end{align}
where $I_0$ is the HMI for the zero cutoff case and given by \cite{Fischler:2012uv}
\bea
I_0 = - \frac{R^d L^{d-1}}{2 G_N r_F^\theta (d_e-1)} \Upsilon^{d_e} \; \mathcal{I}(d_e-1),
\label{HMI-de neq 1-zero cutoff}
\eea 
in which $\mathcal{I}(n)$ is defined as follows
\bea
\mathcal{I}(n) = 2 \left( \frac{2}{\ell}\right)^{n} -  \left( \frac{2}{h}\right)^{n} -  \left( \frac{2}{2 \ell + h}\right)^{n}.
\eea 
Notice that $I_0$ is independent of the cutoff $r_c$. Moreover, it goes to infinity when $h \rightarrow 0$. 
On the other hand, one has
\bea
\Delta I  &=& \frac{R^d L^{d-1}}{4 G_N r_F^\theta (d_e+1)} \Upsilon^{2d_e} \; r_c^{d_e+1} \Bigg[ \mathcal{I}(2 d_e) - \frac{3(d_e+1)}{4(3de+1)} \; \mathcal{I}(4 d_e)  \; ( \Upsilon r_c)^{2 d_e}
\cr && \cr
&& \;\;\;\;\;\;\;\;\;\;\;\;\;\;\;\;\;\;\;\;\;\;\;\;\;\;\;\;\;\;\;\;\;\;\;\;\;\;\;\;\;\;\;\;
 - \frac{d_e(3 d_e+1)}{3(d_e+1)^2 \Upsilon} \; \mathcal{I}(4d_e+2)  \; ( \Upsilon r_c)^{2(d_e+1)}
\cr && \cr
&& \;\;\;\;\;\;\;\;\;\;\;\;\;\;\;\;\;\;\;\;\;\;\;\;\;\;\;\;\;\;\;\;\;\;\;\;\;\;\;\;\;\;\;\;
+ \frac{3d_e}{(3 d_e+1) \Upsilon} \; \mathcal{I}(5 d_e+1) \; ( \Upsilon r_c)^{3d_e+1} + \cdots
\Bigg]\!.
\label{DeltaI-large-ER}
\eea 
Notice that the above expression is only valid for $\ell \gg r_c$ and $\ell \gg h$. From the above expression, one can see that the HMI depends on the cutoff in contrast to the zero cutoff case. In the left panel of figure \ref{fig:HMI-large-ER}, the perturbative expression \eqref{HMI-large-ER} for the HMI is drawn as a function of $r_c$, and compared with the numerical result based on eq. \eqref{EE-strip-de neq 1}. It is observed that the HMI is a decreasing function of $r_c$. Moreover, in the middle and right panels of figure \ref{fig:HMI-large-ER}, the perturbative expression \eqref{HMI-large-ER} for the HMI is drawn as a function of $\ell$ and $h$, and compared with the numerical result based on eq. \eqref{EE-strip-de neq 1}. On the other hand, by applying eq. \eqref{HMI-large-ER} one can find the critical length $h_{\rm crit}$ as follows:
\begin{figure}
	\begin{center}
		\includegraphics[scale=0.27]{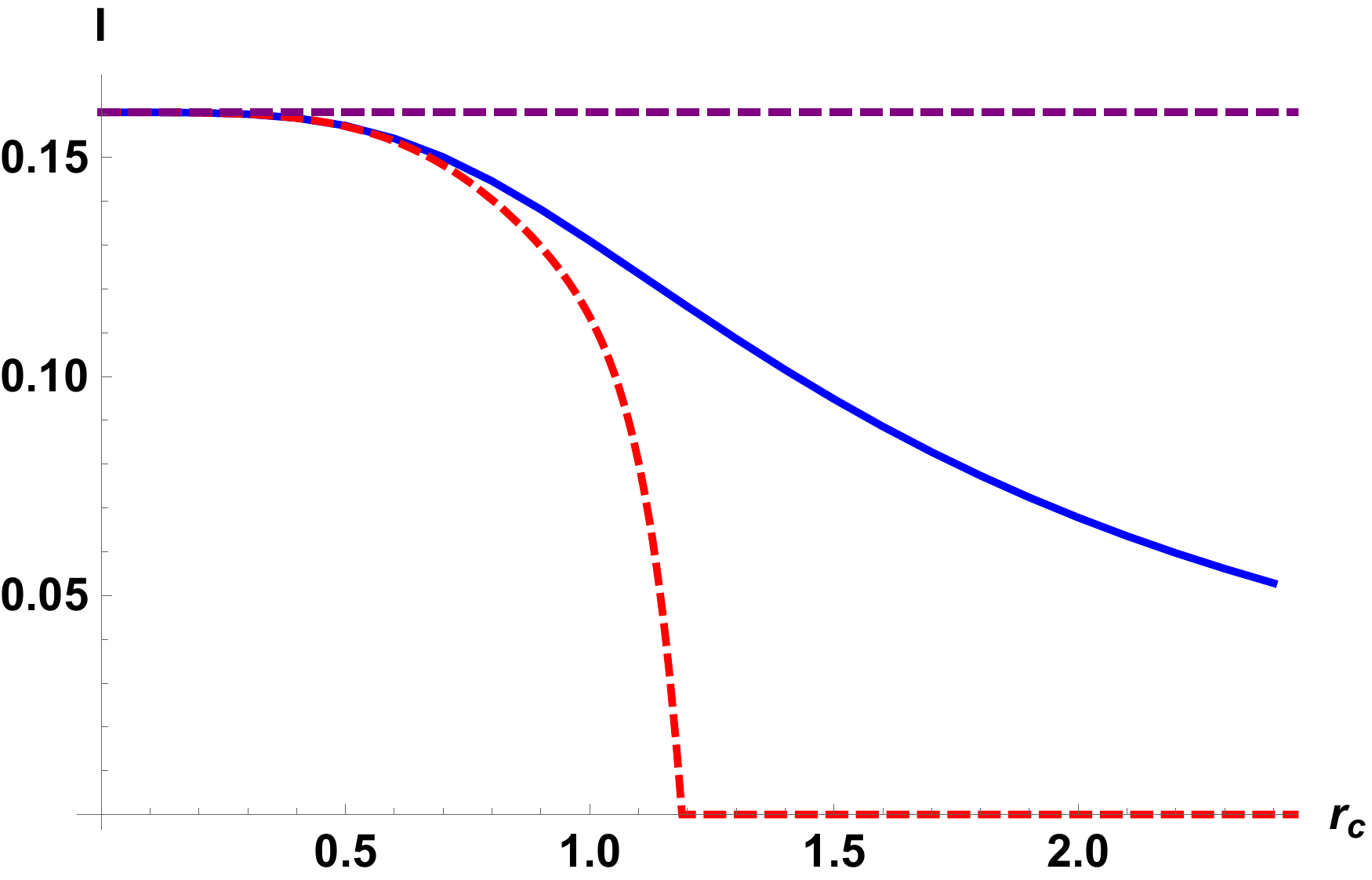}
		\includegraphics[scale=0.27]{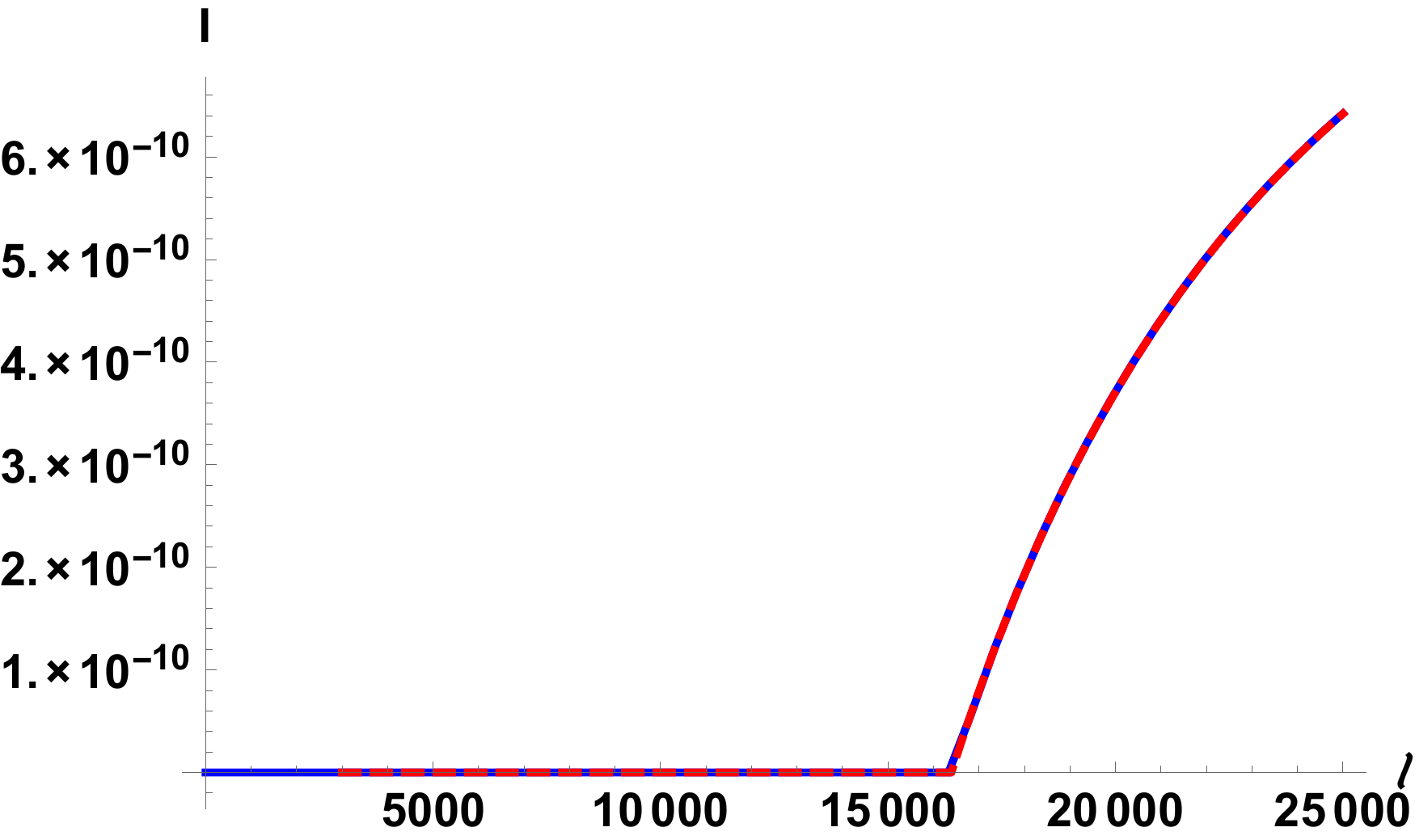}
		\includegraphics[scale=0.27]{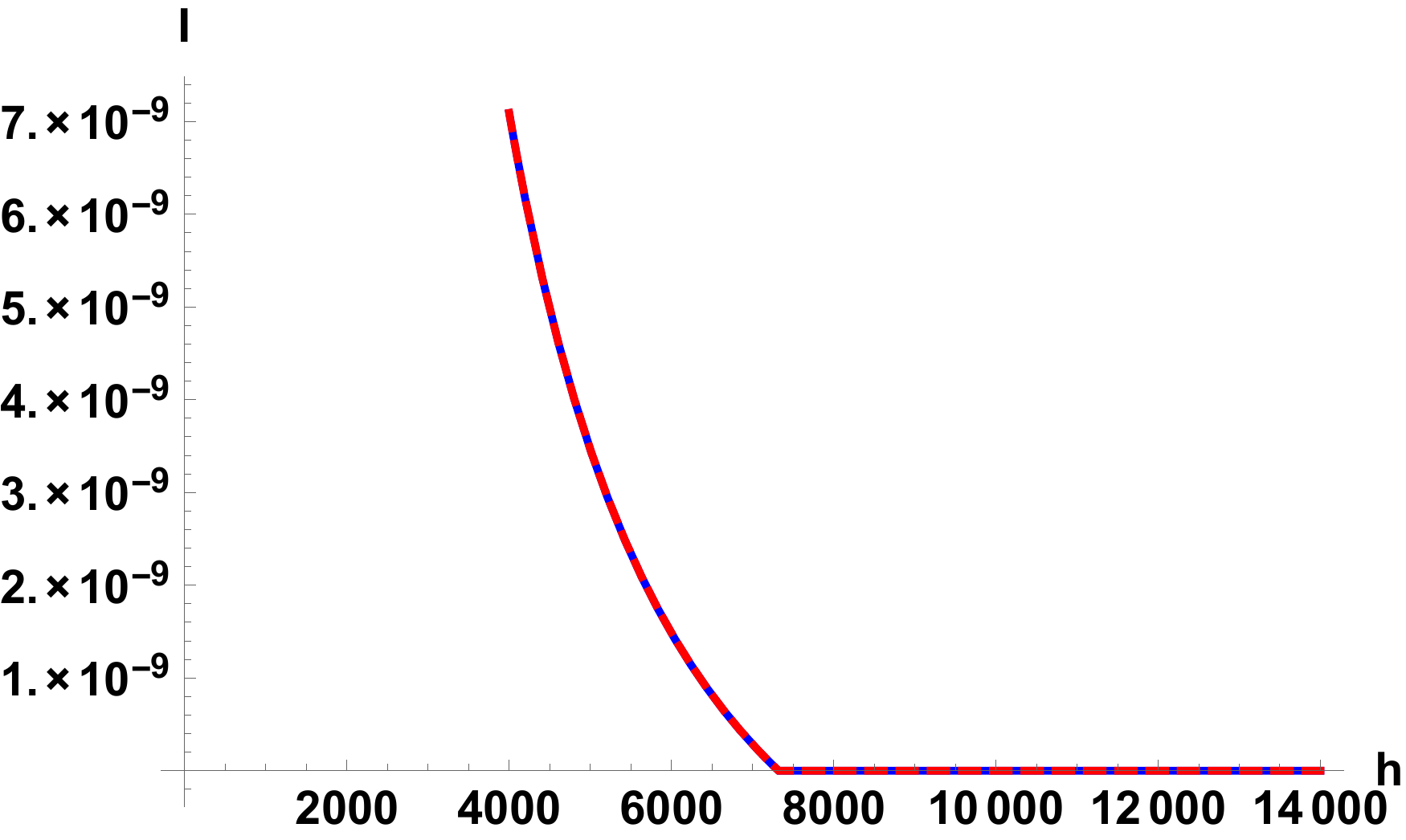}
		\vspace{-5mm}
	\end{center}
	\caption{HMI for very large entangling regions, i.e. $\ell \gg r_c$: {\it Left}) as a function of $r_c$ for $l= 10^{4}$, $h=1$ and $d_e =3$. The purple dashed curve indicates the HMI for the zero cutoff case given in eq. \eqref{HMI-de neq 1-zero cutoff}. The solid blue curves are numerical results based on eq. \eqref{EE-strip-de neq 1}, and the red dashed curves are based on the perturbative expression given in eq. \eqref{EE-strip-de neq 1-l>>rc}, which is valid for very large entangling regions. 
		{\it Middle}) as a function of $\ell$ for $h= 1.2 \times 10^{4}$, $r_c=1$ and $d_e =3$. {\it Right}) as a function of $h$ for $\ell=10^{4}$, $r_c=1$ and $d_e=3$. 
	}
	\label{fig:HMI-large-ER}
\end{figure}
\begin{itemize}
	\item For $d_e =2$, one has
\bea
h_{\rm crit.} &=& h_{\rm crit.}^{(0)} + \ell \bigg[
- 0.844 \left( \frac{r_c}{\ell} \right)^3 - 5.38 \left( \frac{r_c}{\ell}\right)^6
\cr && \cr
&& \;\;\;\;\;\;\;\;\;\;\;\;\;\;\;\;\;\; 
+ 5.16 \left( \frac{r_c}{\ell} \right)^7 + 5.74 \left( \frac{r_c}{\ell} \right)^9
-88.0 \left( \frac{r_c}{\ell} \right)^{10} + \cdots
\bigg],
\label{h-crit-de=2}
\eea 
where 
\bea
h_{\rm crit.}^{(0)} = \frac{\ell}{2} \left( \sqrt{5} - 1\right),
\label{h-crit-de=2-zero-cutoff}
\eea 
is the value of $h_{\rm crit.}$ for the zero cutoff case.
\item For $d_e=3$, one obtains
\bea
h_{\rm crit.} = h_{\rm crit.}^{(0)} + \ell \bigg[
- 0.278 \left( \frac{r_c}{\ell} \right)^4 - 0.757 \left( \frac{r_c}{\ell}\right)^8
+ 0.306 \left( \frac{r_c}{\ell} \right)^{10} + \cdots
\bigg],
\label{h-crit-de=3}
\eea 
where
\bea
h_{\rm crit.}^{(0)} = \ell \left( \sqrt{3} - 1\right).
\label{h-crit-de=3-zero-cutoff}
\eea 
\end{itemize}
Therefore, $h_{\rm crit.}$ depends on the cutoff $r_c$. In figure \ref{fig:h-crit-rc-de neq 1}, eqs. \eqref{h-crit-de=2} and \eqref{h-crit-de=3} are drawn as a function of $r_c$. It is observed that $h_{\rm crit}$ decreases by increasing the cutoff.
\\Furthermore, it should be pointed out that for finite $\ell$ and very small cutoff, again one has $r_c \ll \ell$. Consequently, eqs. \eqref{HMI-large-ER} to \eqref{h-crit-de=3} are still valid when the cutoff is very small. From eq. \eqref{DeltaI-large-ER}, it is obvious that for $r_c = \epsilon \rightarrow 0$, one has $\Delta I=0$ and the HMI becomes equal to that of the zero cutoff case $I_0$, as can be seen in the left panel of figure \ref{fig:HMI-large-ER}.
\subsubsection{Very Small Entangling Regions}
\label{Sec: HMI-very small ER}
\begin{figure}
	\begin{center}
		\includegraphics[scale=0.26]{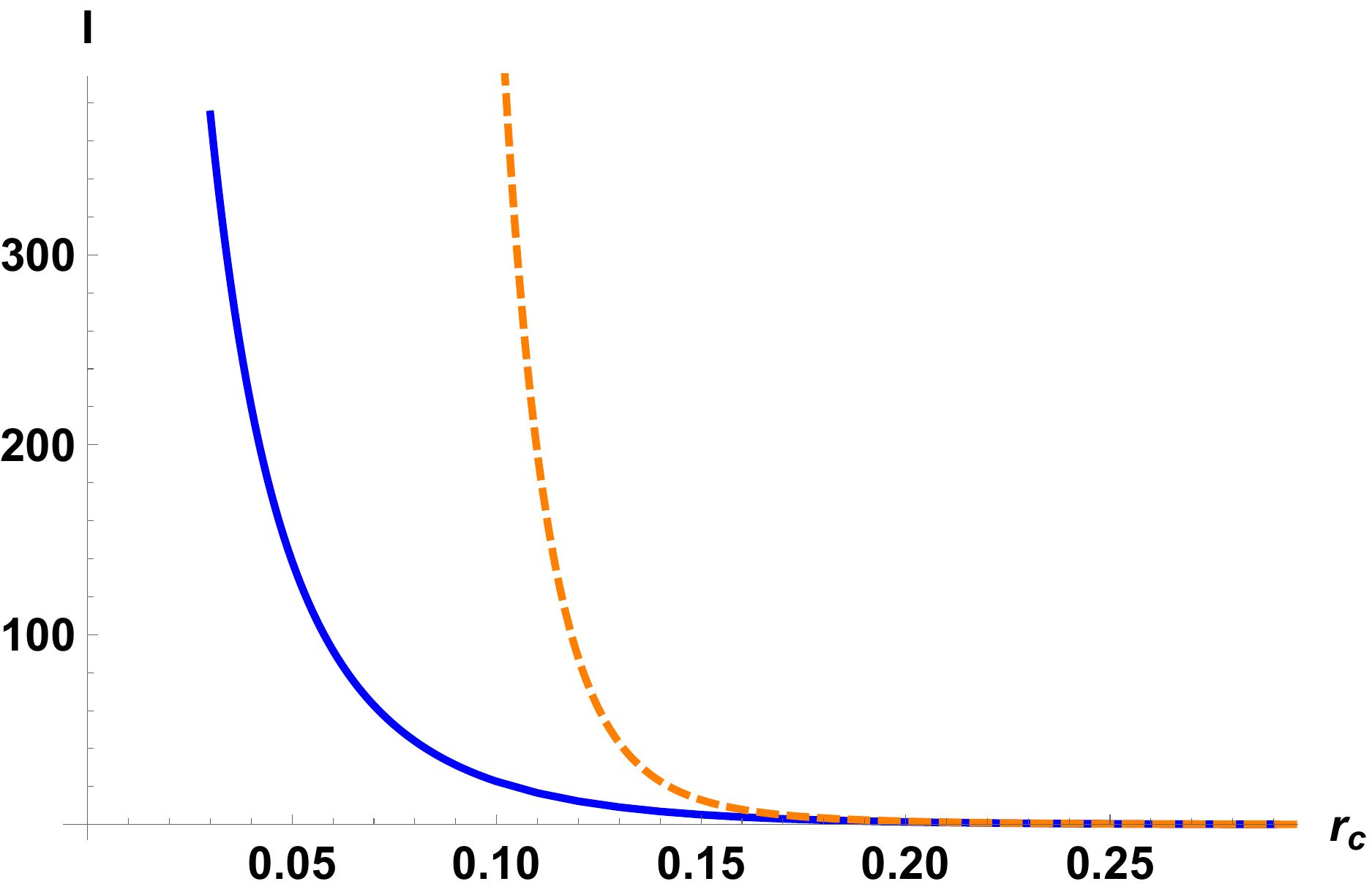}
		\includegraphics[scale=0.26]{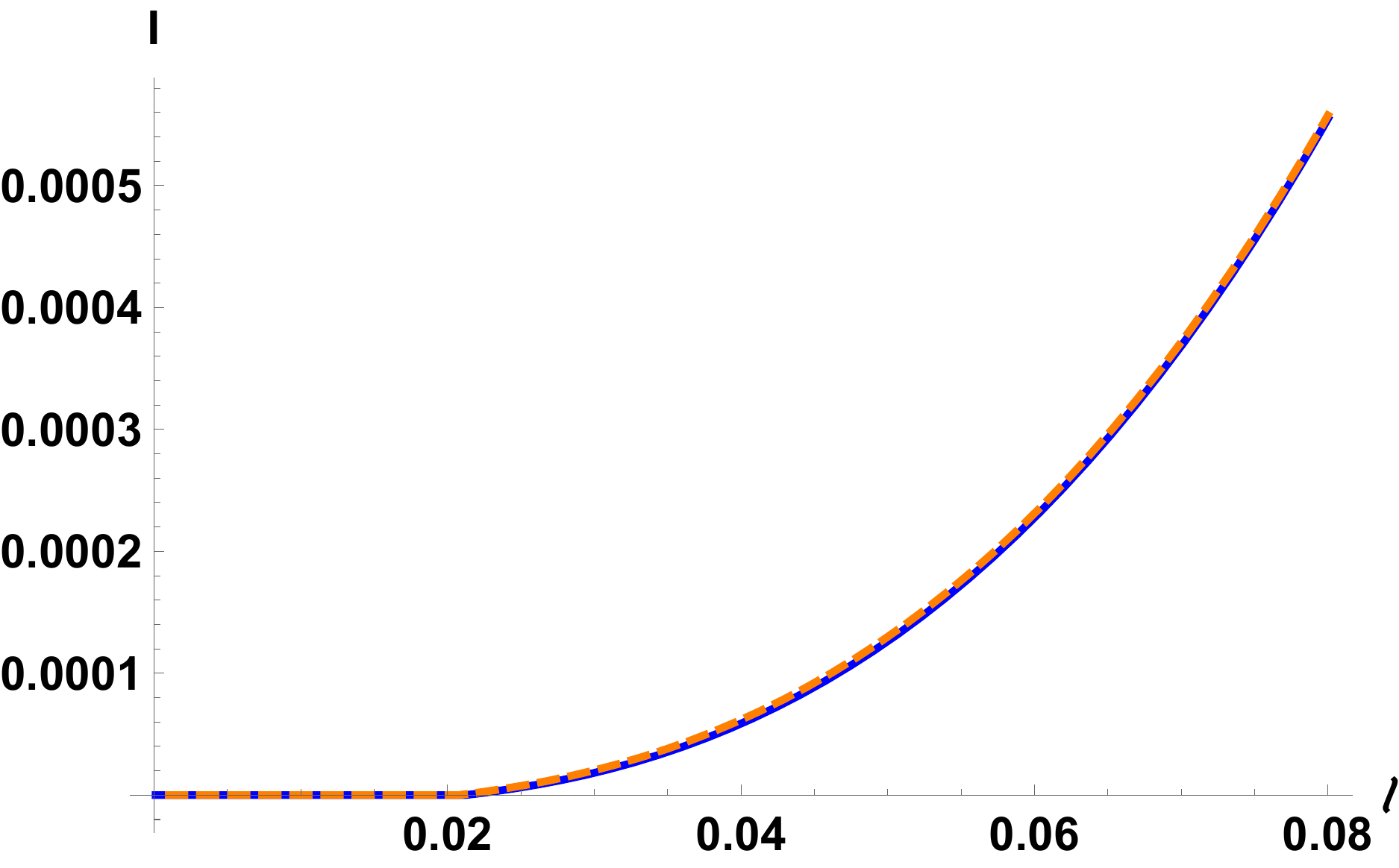}
		\includegraphics[scale=0.26]{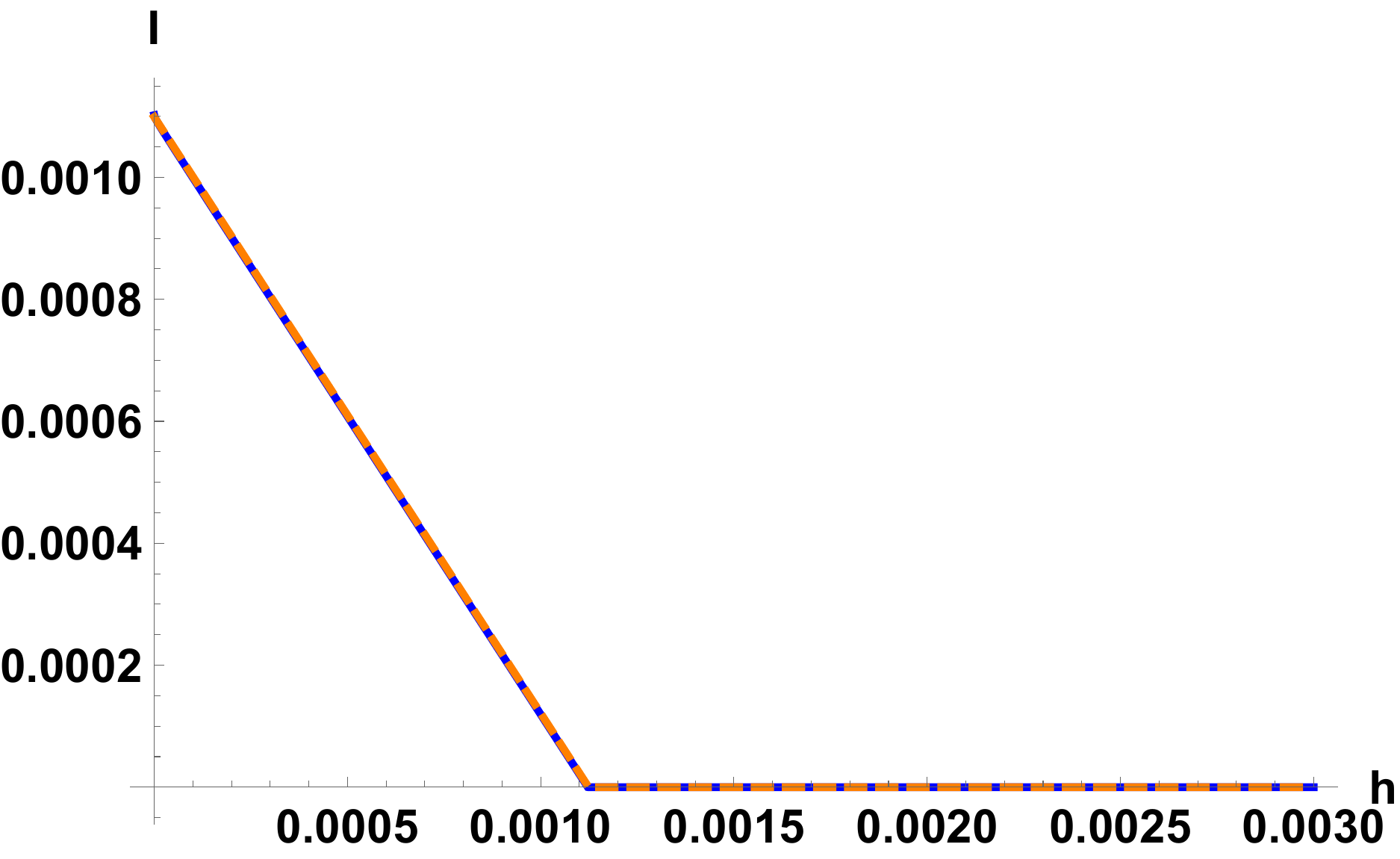}
		\vspace{-5mm}
	\end{center}
	\caption{HMI for very small entangling regions, i.e. $\ell \ll r_c$: {\it Left}) as a function of $r_c$ for $h=10^{-2}$, $\ell=10^{-1}$ and $d_e =3$.  The blue curves are the numerical results based on eq. \eqref{EE-strip-de neq 1}, and the orange dashed curves are based on eq. \eqref{HMI-small-ER-2} which are valid for very small entangling regions. {\it Middle}) as a function of $\ell$ for $h=10^{-5}$, $r_c=1$ and $d_e=3$. {\it Right}) as a function of $h$ for $\ell= 0.1$, $r_c=1$ and $d_e=3$. 
	}
	\label{fig:HMI-rc-l-small-ER}
\end{figure}

For very small entangling regions, i.e. $\ell \ll r_c$, from eq. \eqref{EE-strip-de neq 1-l<<rc}, one can simply write
\begin{align}
I(A,B)=
\begin{cases}
0 & ~~ \ell \ll  h\\
I_s  & ~~ \ell \gg h
\end{cases}
\label{HMI-small-ER-1}
\end{align}
where
\bea
I_s &=& \frac{R^d L^{d-1}}{4 G_N r_F^\theta} \frac{d_e^2}{r_c^{d_e -1}} \Bigg[
-\frac{1}{24} \frac{\mathcal{K}(3)}{r_c^3} + \frac{(4 d_e (39+ 5 d_e) - 47)}{30720} \frac{\mathcal{K}(5)}{r_c^5}
\cr && \cr
&&\;\;\;\;\;\;\;\;\;\;\;\;\;\;\;\;\;\;\;\;\;\;\;\;\;
+ \frac{d_e (5071 - 4 d_e (2561 + 113 d_e))}{10321920} \frac{\mathcal{K}(7)}{r_c^7}
+ \cdots
\Bigg],
\label{HMI-small-ER-2}
\eea 
and $\mathcal{K}(n)$ is defined as follows
\bea
\mathcal{K}(n) = 2 \ell^n - (2 \ell + h)^n - h^n.
\eea 
In figure \ref{fig:HMI-rc-l-small-ER}, the perturbative expression \eqref{HMI-small-ER-1} for the HMI is drawn as a function of $r_c$, $\ell$ and $h$, and compared with the numerical results based on eq. \eqref{EE-strip-de neq 1}.
\section{Entanglement Wedge Cross Section}
\label{Sec: Entanglement Wedge Cross Section}
In this section, we study entanglement wedge cross section for two strips. Consider a spacelike region $A$ in a holographic CFT and its density matrix $\rho_A$. It was proposed in refs. \cite{Czech:2012bh,Wall:2012uf,Headrick:2014cta} that the holographic dual of $\rho_A$ is a codimension zero region $M_A$ in the bulk spacetime which is called Entanglement Wedge (EW). More precisely, EW is the domain of dependence of a spacelike region in the bulk spacetime which is enclosed by the corresponding RT surface $\Gamma_A$ and the region $A$. In the following, when we talk about EW, we mean a canonical time slice of it which on the boundary coincides with the region $A$. Now consider a bipartite system consisting of two disjoint subsystems $A$ and $B$ in the CFT whose RT surface and EW are shown by $\Gamma_{AB}$ and $M_{AB}$, respectively. Next, one can decompose $\Gamma_{AB}$ as follows \cite{Takayanagi:2017knl}
\bea
\Gamma_{AB} = \Gamma^{(A)}_{AB} \cup \Gamma^{(B)}_{AB},
\label{Gamma-AB}
\eea 
and define two regions $\tilde{\Gamma}_A = A \cup \Gamma^{(A)}_{AB}$ and $\tilde{\Gamma}_B = B \cup \Gamma^{(B)}_{AB}$. Now one can find a RT surface $\Sigma^{\rm min}_{AB}$ in the bulk spacetime whose area gives the HEE of $\tilde{\Gamma}_A$ and $\tilde{\Gamma}_B$ \cite{Takayanagi:2017knl}
\bea
S(\tilde{\Gamma}_A) = S(\tilde{\Gamma}_B) =  \frac{Area(\Sigma^{\rm min}_{AB})}{4 G_N}.
\eea 
In this case, it is evident that $\Sigma^{\rm min}_{AB}$ is homologous to $\tilde{\Gamma}_A$ and $\tilde{\Gamma}_B$.
Then one can define the entanglement wedge cross section (EWCS) as in eq. \eqref{EW-1} \cite{Takayanagi:2017knl,Nguyen:2017yqw}. In other words, EWCS measures the minimal cross section of the corresponding EW. 
\\In the following, we calculate the EWCS for two strips with equal widths $\ell$ which are separated by the distance $h$ (see figure \ref{fig:HMI-EWCS}). When the two strips are close enough to each other, the EW is connected (see the left panel of figure \ref{fig:HMI-EWCS}). In this case, it is straightforward to see that
\bea
E_{W} = \frac{R^d L^{d-1}}{4 G_N r_F^\theta} \int_{r_t(h)}^{r_t(2 \ell +h)} \frac{dr}{r^{d_e}},
\label{EW-2}
\eea 
where $r_t ( h)$ and $r_t(2 \ell+ h)$ are the radial coordinates of the turning points for the RT surfaces $\Gamma_h$ and $\Gamma_{ 2 \ell +h}$, respectively. Notice that for the connected configuration, one has $\Gamma_{AB} = \Gamma_h \cup \Gamma_{ 2 \ell +h}$. On the other hand, when the strips are far enough from each other, the EW is disconnected (see the right panel of figure \ref{fig:HMI-EWCS}). In this case, there is no minimal surface $\Sigma^{\rm min}_{AB}$, and hence $E_W=0$.
\subsection{$d_e =1$}
\label{Sec: HEoP-de=1}
\begin{figure}
	\begin{center}
		\includegraphics[scale=0.26]{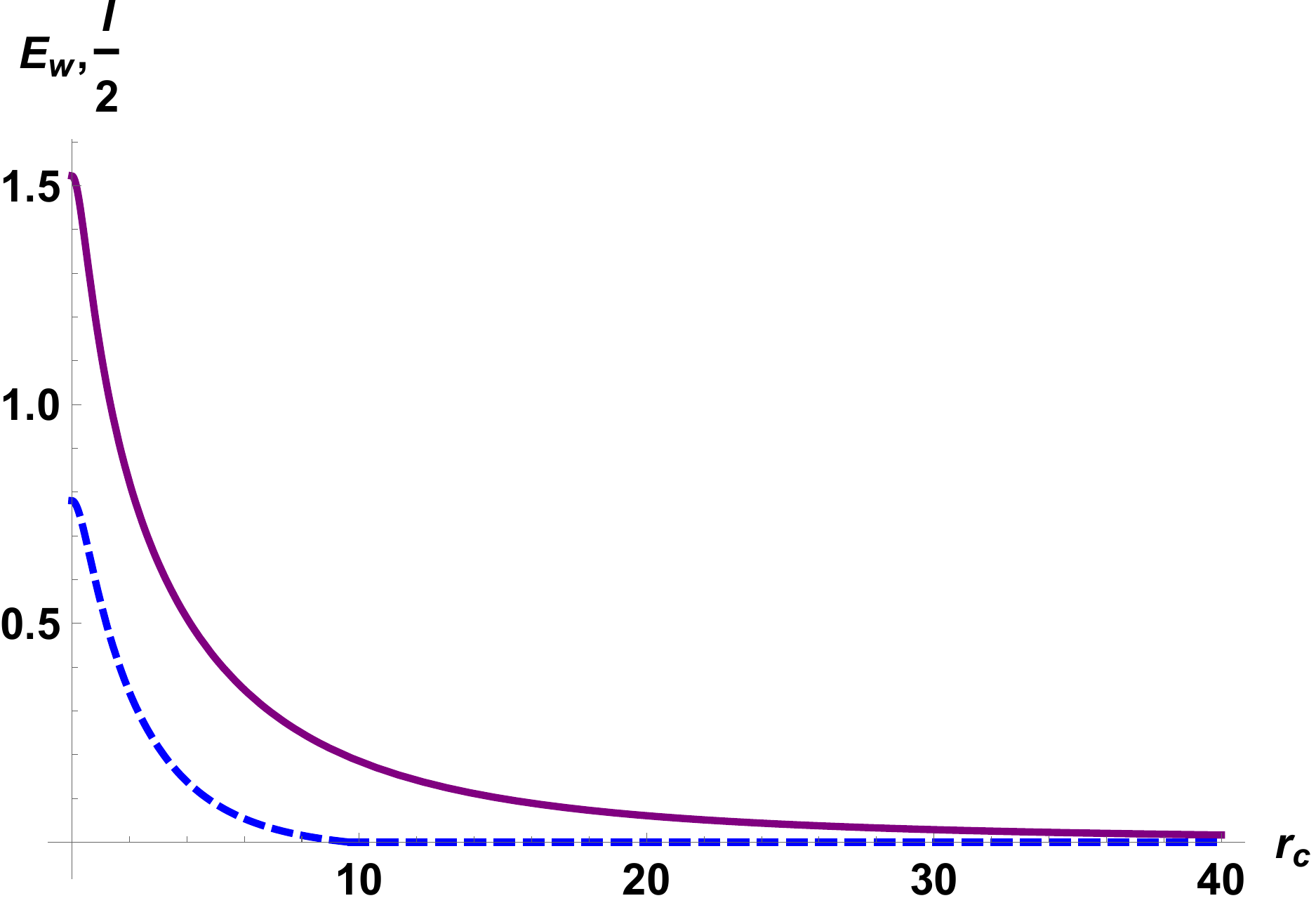}
		\includegraphics[scale=0.26]{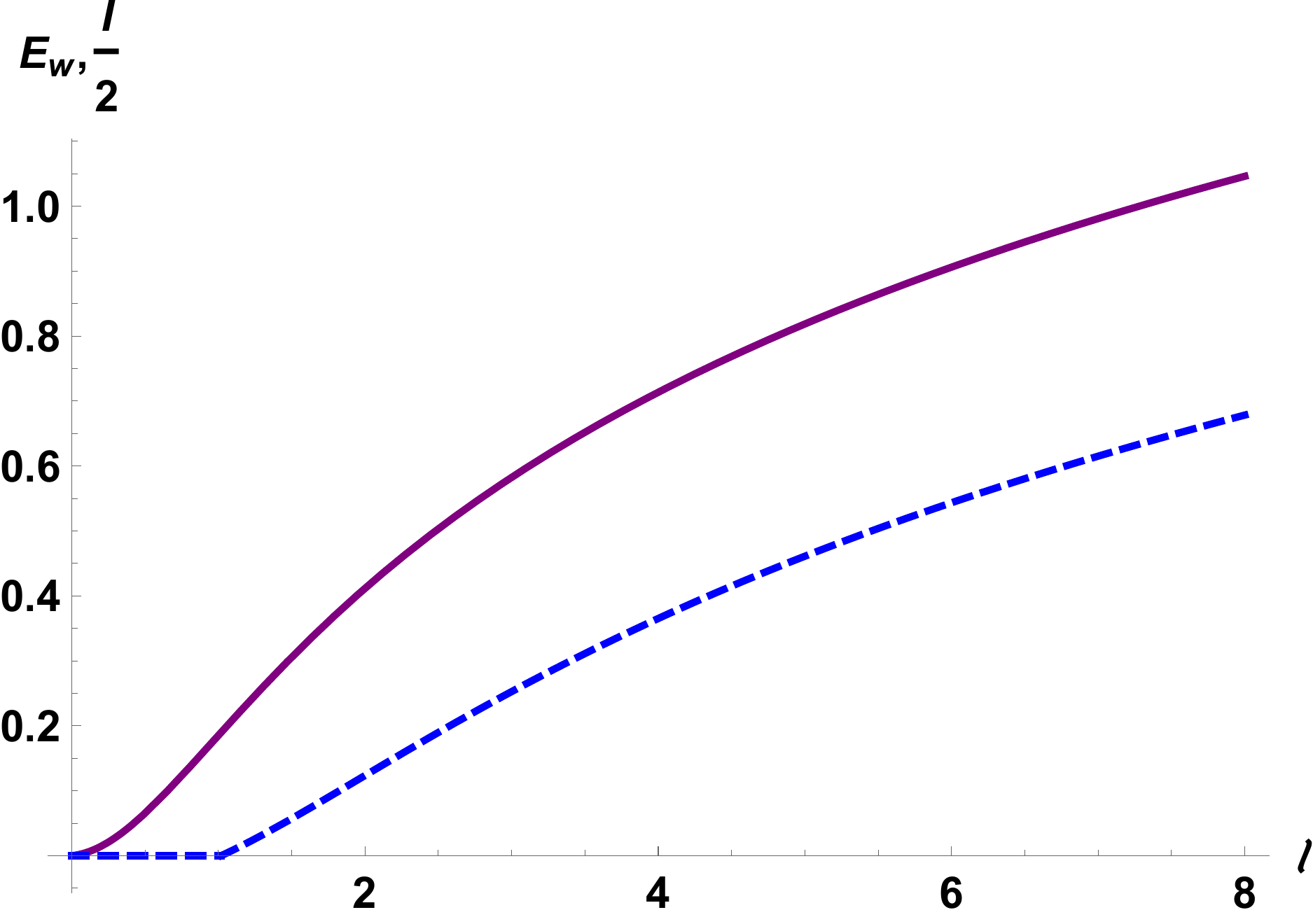}
		\includegraphics[scale=0.26]{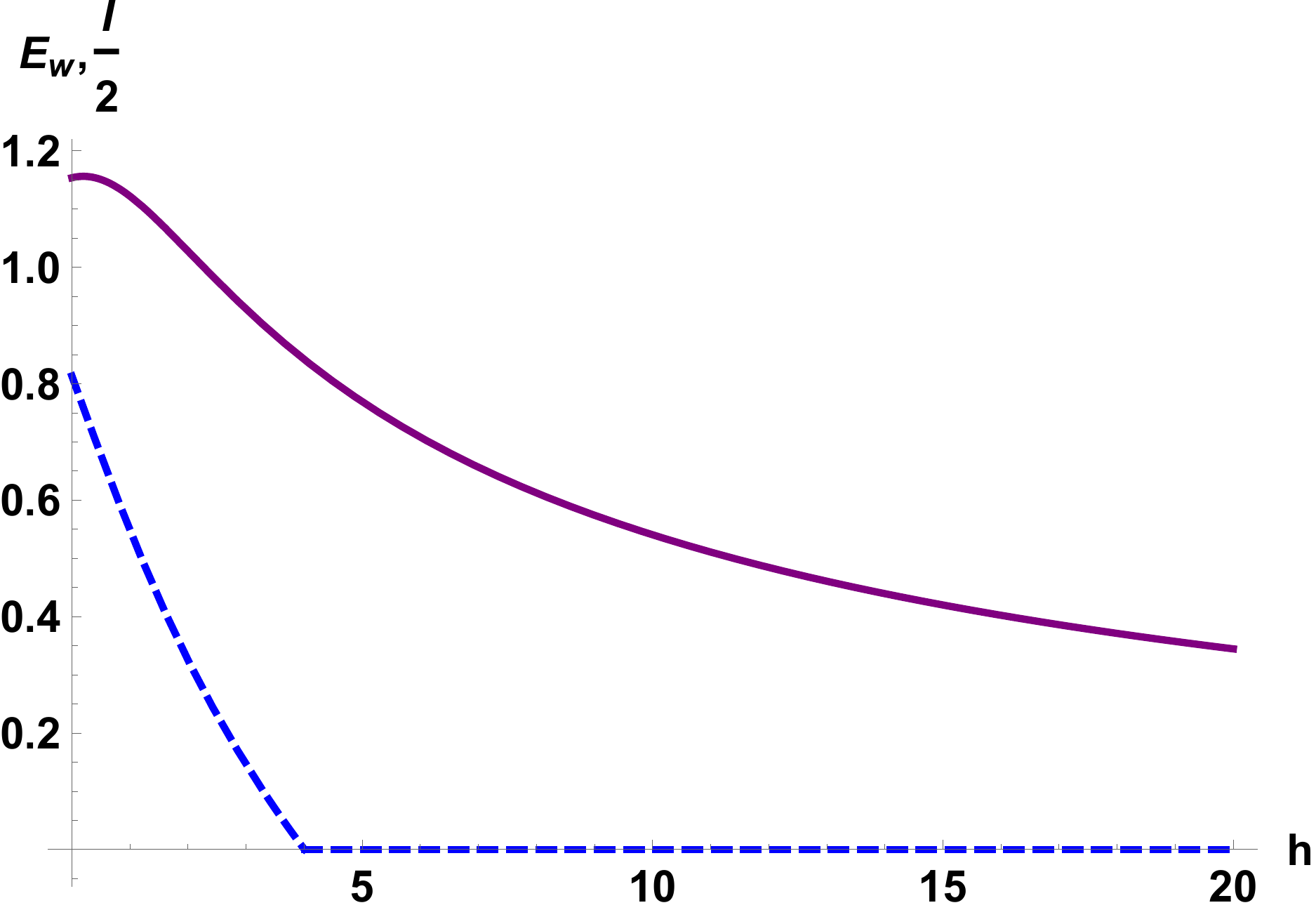}
		\vspace{-5mm}
	\end{center}
	\caption{$E_W$ for $d_e =1$ as a function: {\it Left}) of $r_c$ for $l=10$ and $h=1$. {\it Middle}) of $\ell$ for $h=0.1$ and $r_c=1$. {\it Right}) of $h$ for $\ell=10$ and $r_c=1$. The blue dashed curves are half of the HMI and the purple curves are $E_W$. Here we renormalized $E_W$ as $\tilde{E}_W= \frac{E_W}{a}$, where $a= \frac{R^d L^{d-1}}{2 G_N r_F^\theta}$. 
	}
	\label{fig:EW-de=1}
\end{figure}

For $d_e=1$ and the zero cutoff case, by applying eq. \eqref{rt-zero cutoff} and \eqref{EW-2} one easily obtains
\bea
E^0_{W} = \frac{R^d L^{d-1}}{4 G_N r_F^\theta} \log\left( \frac{2 \ell +h}{h}\right),
\label{EW-de=1-zero cutoff}
\eea 
where for $\theta=0$ and $d=1$ it reduces to that of $AdS_3$ in Poincaré coordinates (see ref. \cite{Takayanagi:2017knl}). On the other hand, for the finite cutoff case, from eqs. \eqref{rt-strip-de=1} and \eqref{EW-2} one can simply find $E_W$ as follows
\bea
E_W = \frac{R^d L^{d-1}}{8 G_N r_F^\theta} \log\left( \frac{(2 \ell +h)^2 + 4 r_c^2}{h^2 + 4 r_c^2}\right).
\label{EW-de=1}
\eea 
From the above expression, one can see that $E_W$ is independent of the dynamical exponent $z$ and depends on the hyperscaling violation exponent $\theta$ only through the coefficient $1/r_F^\theta$. Moreover, the comparison of eqs. \eqref{EW-de=1-zero cutoff} and \eqref{EW-de=1} shows that only in the finite cutoff case, $E_W$ depends on the cutoff $r_c$. In the left panel of figure \ref{fig:EW-de=1}, $E_W$ is plotted as a function of $r_c$. It is observed that $E_W$ satisfies the inequality given in eq. \eqref{EW geq I} for all values of $r_c$. Furthermore, both of the $E_W$ and HMI are decreasing functions of the cutoff, and $E_W$ goes to zero when $r_c \rightarrow \infty$.
\\In the middle and right panels of figure \ref{fig:EW-de=1}, $E_W$ is drawn as a function of $\ell$ and $h$, respectively. It is observed that at the point $h_{\rm crit.}$ where the HMI undergoes a first-order phase transition, $E_W$ is smooth and its concavity changes. In other words, there is a point of inflection in $E_W$ which coincides with $h_{\rm crit}$. It is in contrast to the zero cutoff case, where $E_W$ shows a discontinuous phase transition at $h_{\rm crit.}$. Moreover, similar to the zero cutoff case \cite{Takayanagi:2017knl,Liu:2019qje,BabaeiVelni:2019pkw}, by increasing the distance $h$, $E_W$ goes to zero. On the other hand, from eqs. \eqref{EW-de=1-zero cutoff} and \eqref{EW-de=1}, one can see that in the limit $h \rightarrow 0$, $E^0_W$ diverges. However, $E_W$ remains finite in this limit. Furthermore, in the limit $h \rightarrow \infty$, $E_W$ becomes zero which has the same behavior as in the zero cutoff case. At the end, when the cutoff is very small, i.e. $ r_c \ll \ell, h$, the change in $E_W$, i.e. $\Delta E_W = E_W - E^0_W$,
is given by
\bea
\Delta E_W = \frac{R^d L^{d-1}}{2 G_N r_F^\theta} \Big[
-\left(\frac{1}{h^2} - \frac{1}{(2 \ell+h)^2}\right) r_c^2 +2  \left(\frac{1}{h^4} - \frac{1}{(2 \ell+h)^4}\right) r_c^4 + \cdots
\Big].
\label{DeltaEW-de=1}
\eea 
\subsection{$d_e \neq 1$}
\label{Sec: HEoP-de neq 1}

\begin{figure}
	\begin{center}
		\includegraphics[scale=0.28]{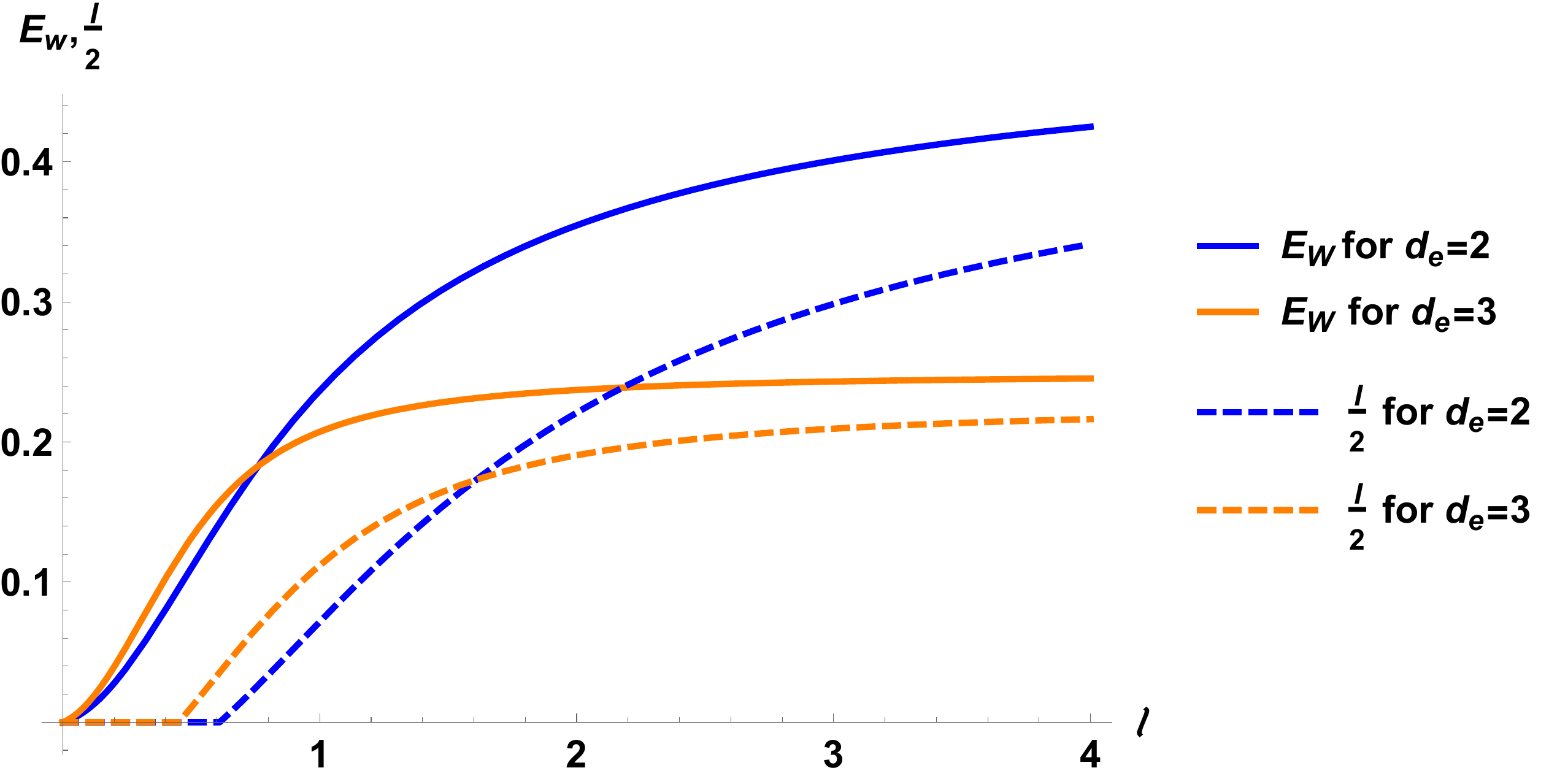}
		\hspace{0.2cm}
		\includegraphics[scale=0.28]{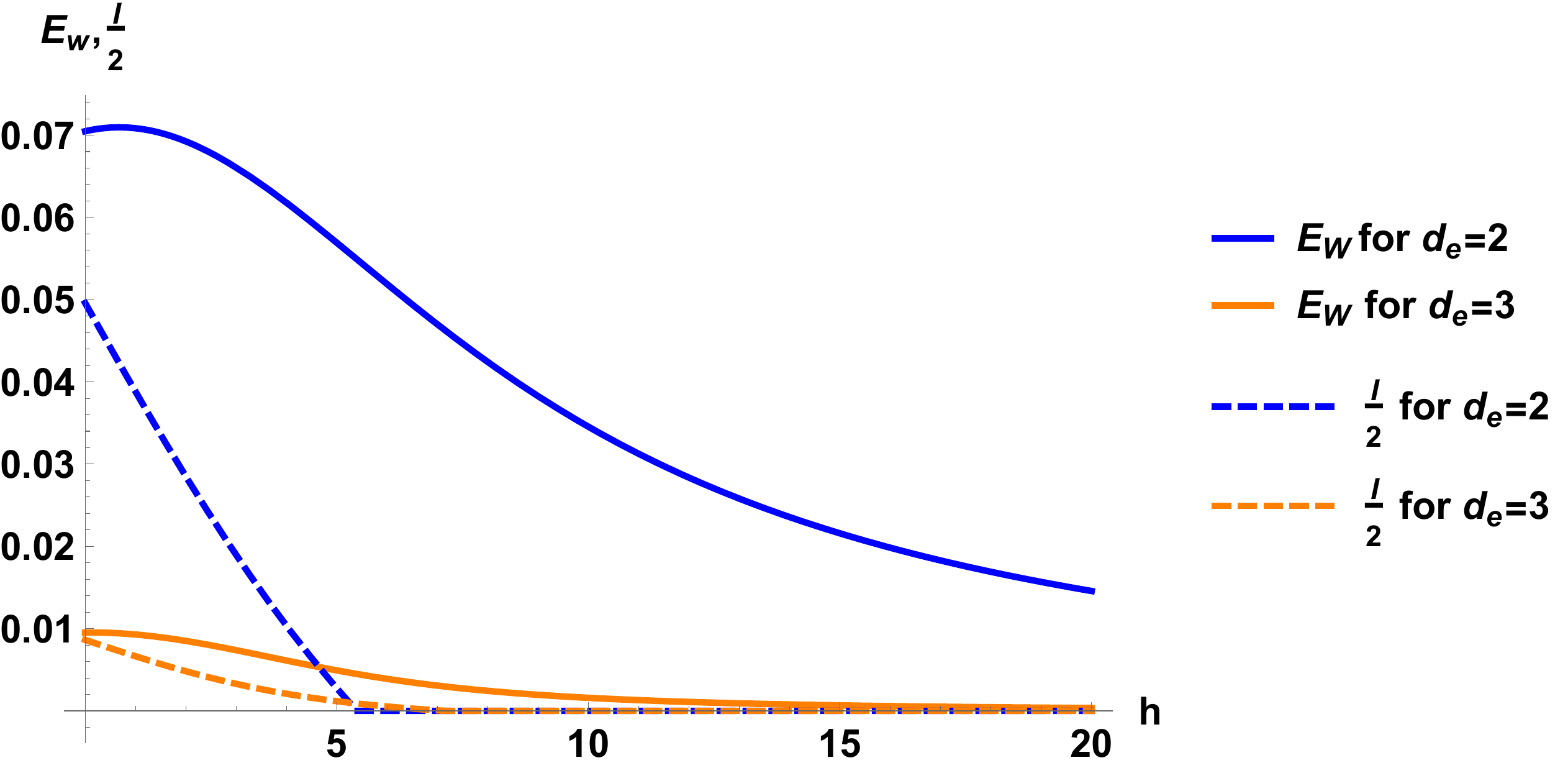}
		\vspace{-5mm}
	\end{center}
	\caption{$E_W$ as a function: {\it Left}) of $\ell$ for $h=0.1$ and $r_c=1$. {\it Right}) of $h$ for $\ell=10$ and $r_c=5$. The dashed curves are half of the HMI and the solid curves are $E_W$. Here we multiplied both $E_W$ and $I$ by $\frac{r_c^{d_e-1}}{a}$, where $a= \frac{R^d L^{d-1}}{2 G_N r_F^\theta}$.
	}
	\label{fig:EW-l-h}
\end{figure}

\begin{figure}
	\begin{center}
	\includegraphics[scale=0.31]{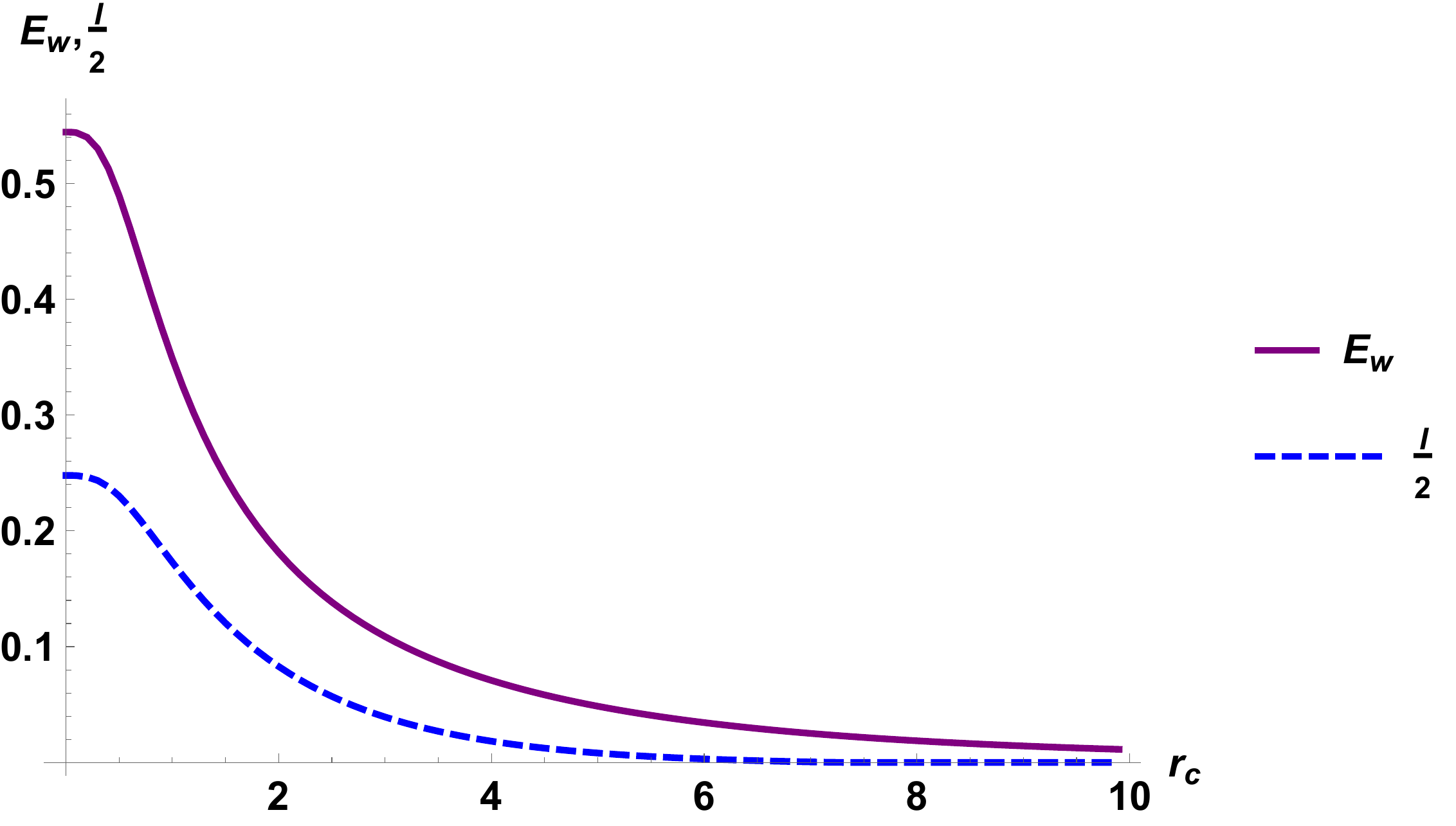}
		\hspace{0.2cm}
	\includegraphics[scale=0.31]{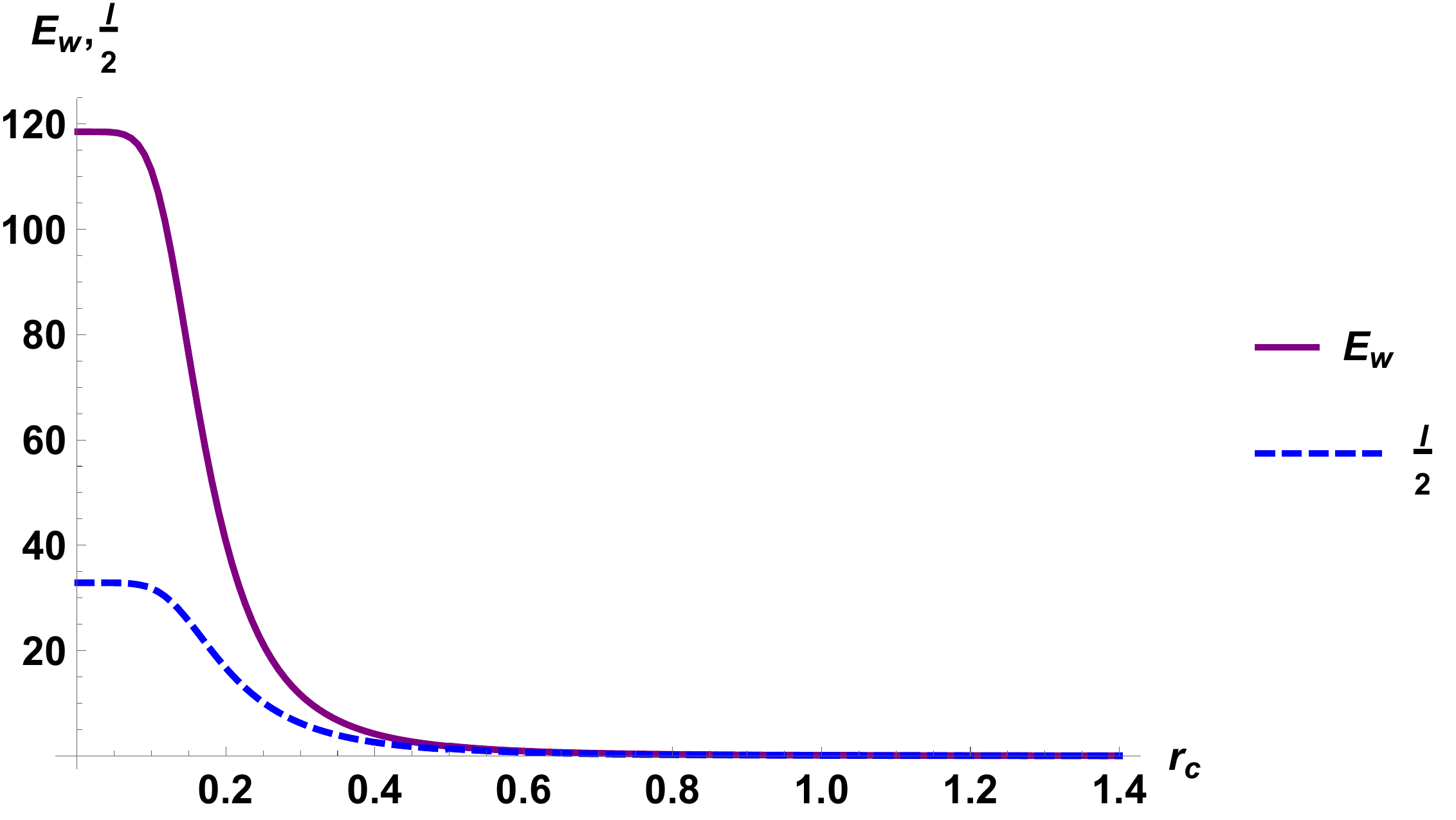}
		\vspace{-5mm}
	\end{center}
	\caption{$E_W$ as a function of $r_c$ for: {\it Left}) $\ell=5$, $h=1$ and $d_e=2$. {\it Right}) $\ell=1$, $h=0.1$ and $d_e=5$. The dashed blue curves are half of the HMI and the solid purple curves are $E_W$. Here we divided both $E_W$ and $I$ by $\frac{R^d L^{d-1}}{2 G_N r_F^\theta}$.
	}
	\label{fig:EW-rc}
\end{figure}

In this case, from eq. \eqref{EW-2}, one can easily write
\bea
E_W = \frac{R^d L^{d-1}}{4 G_N r_F^\theta (d_e -1)} \Bigg[
\frac{1}{r_t(h)^{d_e-1}} - \frac{1}{r_t(2 \ell + h)^{d_e-1}} 
\Bigg].
\label{EW-de-neq-1-1}
\eea 
For the zero cutoff case, by plugging eq. \eqref{rt-zero cutoff} into the above equation, one can simply obtain (see also \cite{BabaeiVelni:2019pkw})
\bea
E^0_W = \frac{R^d L^{d-1}}{4 G_N r_F^\theta (d_e -1)} \left( 2 \Upsilon \right)^{d_e -1} \mathcal{E}(d_e-1),
\label{EW-0-de neq 1}
\eea 
where $\mathcal{E}(n)$ is defined as follows
\bea
\mathcal{E}(n) = \frac{1}{h^n} - \frac{1}{(2 \ell+ h)^n}.
\eea
One can see that similar to the $d_e=1$ case (see eq. \eqref{EW-de=1-zero cutoff}), $E^0_W$ is infinite when $h \rightarrow 0$. Furthermore, in the limit $h \rightarrow \infty$, it becomes zero. Moreover, it is independent of the cutoff.
\\On the other hand, one can numerically calculate eq. \eqref{EW-de-neq-1-1}. In figure \ref{fig:EW-l-h}, $E_W$ is drawn as a function of $\ell$ and $h$. The behaviors of $E_W$ are very similar to those of the $d_e=1$ case in the previous section. In particular, it does not show a discontinuous phase transition when the HMI undergoes a phase transition. Instead, $E_W$  has a point of inflection which coincides with $h_{\rm crit.}$. Moreover, it is finite in the limit $h \rightarrow 0$
\footnote{The same behavior was also observed for BTZ black holes in ref. \cite{Asrat:2020uib}.}
, and goes to zero in the limit $h \rightarrow \infty$. On the other hand, the inequality \eqref{EW geq I} is still satisfied for all values of $\ell$ and $h$.
\\Moreover, in figure \ref{fig:EW-rc}, $E_W$ is drawn as a function of $r_c$. It is observed that it is a decreasing function of the cutoff. In particular, it is finite when $r_c \rightarrow 0$, and goes to zero when $r_c \rightarrow \infty$. Furthermore, the inequality \eqref{EW geq I} is valid for all values of the cutoff.
On the other hand, one can find analytic expressions for $E_W$ in some limits. In the following, we consider the very small cutoff, i.e. $\ell , h \gg r_c$, and very large cutoff, i.e. $\ell, h \ll r_c$, cases, respectively.
\begin{itemize}
	\item {{\bf Very small cutoff  $(\ell , h \gg r_c)$: }} In this limit, one can apply eq. \eqref{rt-strip-de neq 1- l, rt>>rc}, and find the change $\Delta E_W = E_W - E^0_W$
	as follows
	\bea
	\Delta E_W &\!\!=\!\!& \frac{R^d L^{d-1} (2 \Upsilon)^{d_e -1}}{4 G_N r_F^\theta (d_e +1)} \Bigg[
	\! - \Upsilon^{d_e} (2 r_c)^{d_e +1} \mathcal{E}(2 d_e) 
	+ \frac{d_e}{(d_e +1)} \Upsilon^{2 d_e} (2 r_c)^{2 (d_e +1 )} \mathcal{E}(3 d_e +1)
	\cr && \cr
	&& \;\;\;\;\;\;\;\;\;\;\;\;\;\;\;\;\;\;\;\;\;\;\;\;\;\;\;\;\;
	- \frac{d_e (3 d_e +1)}{3 (d_e +1)^2} \Upsilon^{3 d_e} (2 r_c)^{3(d_e +1)} \mathcal{E}(2 (2 d_e+1))
	\cr && \cr
	&& \;\;\;\;\;\;\;\;\;\;\;\;\;\;\;\;\;\;\;\;\;\;\;\;
	+ \frac{d_e (2 d_e+1)(3d_e+1)}{6 (d_e+1)^3} \Upsilon^{4 d_e} (2 r_c)^{4(d_e+1)} \mathcal{E}(5 d_e +3) + \cdots
	\Bigg]\! .
	\eea 
	Consequently, when $r_c = \epsilon \rightarrow 0$, it becomes zero.
	\item {\bf Very large cutoff $(\ell , h \ll r_c) $: } In this limit, one can use eq. \eqref{rt-strip-de neq 1- l, rt approx rc}, and find $E_W$ as follows
	\bea
	E_W &=& \frac{R^d L^{d-1}}{32 G_N r_F^\theta } \frac{d_e}{r_c^{d_e -1}} \Bigg[
	- \frac{\mathcal{E}(-2)}{r_c^2} + \frac{d_e (d_e +5)}{48} \frac{\mathcal{E}(-4)}{r_c^4}
	\cr && \cr
	&& \;\;\;\;\;\;\;\;\;\;\;\;\;\;\;\;\;\;\;\;\;\;\;\;\;
	- \frac{d_e^2 (4 d_e (d_e+1) +175)}{9216} \frac{\mathcal{E}(-6)}{r_c^6}
	\cr && \cr
	&& \;\;\;\;\;\;\;\;\;\;\;\;\;\;\;\;\;\;\;\;\;\;\;\;
	+ \frac{d_e^3 (625 - 4 d_e (39 + d_e (d_e- 12)) )}{147456} \frac{\mathcal{E}(-8)}{r_c^8}
	+ \cdots 
	\Bigg].
	\eea 
Therefore, when $r_c \rightarrow \infty$, one has $E_W \rightarrow 0$.
\end{itemize}
\section{Discussion}
\label{Sec: Discussion}
In refs. \cite{McGough:2016lol,Taylor:2018xcy,Hartman:2018tkw}, it was proposed that the holographic dual of a $T \overline{T}$ deformed $CFT_{d+1}$ is a gravity theory in an $AdS_{d+2}$ spacetime with a radial cutoff. Motivated by this proposal, we considered a Hyperscaling Violating (HV) geometry at zero temperature and finite radial cutoff, which one might expect to be dual to a $T \overline{T}$ deformed QFT in which the Lorentz and scaling symmetries are broken. It should also be emphasized that in ref. \cite{Alishahiha:2019lng}, a HV geometry at finite temperature and radial cutoff was studied and it was proposed that the dual QFT might be considered as a $T \overline{T}$ like deformation of a HV QFT where the deformation operator is given by eq. \eqref{TT-operator-HV}. 
\\We calculated holographically some measures of quantum entanglement in these geometries including holographic 
entanglement entropy (HEE), mutual information (HMI) and entanglement wedge cross section (EWCS) which is proposed to be the holographic dual of entanglement of purification \cite{Takayanagi:2017knl}. These calculations might be a small step toward understanding some properties of the deformed HV QFTs. In particular, we considered entangling regions in the shape of strips, and calculated the HEE numerically. It is observed that the turning point depends on the cutoff, in contrast to the zero cutoff case. Moreover, the HEE is a decreasing function of $r_c$ (see figures \ref{fig:S-rc-de-1} and \ref{fig:S-rc}). Therefore, one might conclude that the quantum correlations among the degrees of freedom decrease by increasing the cutoff. On the other hand, we found analytic expressions for very small and large entangling regions as well as very small cutoff. 
\\Furthermore, we studied the HMI between two disjoint parallel strips, and its dependence on the cutoff. The HMI shows interesting behaviors in comparison to the zero cutoff case:
\begin{itemize}
	\item  It is a decreasing function of the cutoff $r_c$, and goes to zero in the limit $r_c \rightarrow \infty$ (see figure \ref{fig:HMI-rc-hcrit-rc-de=1} and \ref{fig:HMI-rc}). It is in contrast to the zero cutoff case where the HMI is independent of the cutoff (see eqs. \eqref{HMI-de=1-zero cutoff} and \eqref{HMI-de neq 1-zero cutoff}).
	\item It still shows a first-order phase transition, and the critical length $h_{\rm crit.}$ becomes larger by increasing $d_e$ (see figure \ref{fig:HMI-h}). Moreover, $h_{\rm crit.}$ depends on the cutoff and decreases by increasing $r_c$ (see figure \ref{fig:HMI-rc-hcrit-rc-de=1} and \ref{fig:h-crit-rc-de neq 1}). It is in contrast to the zero cutoff case where $h_{\rm crit.}$ is independent of the cutoff (see eqs. \eqref{hcrit-de=1-zero cutoff}, \eqref{h-crit-de=2-zero-cutoff} and \eqref{h-crit-de=3-zero-cutoff}).
	\item When the distance between the two entangling regions becomes zero, i.e. $h \rightarrow 0$, the HMI does not diverge and remains finite (see figures \ref{fig:HMI-l-h-de=1} and \ref{fig:HMI-h}). This behavior is in contrast to the zero cutoff case where the HMI diverges in the limit $h \rightarrow 0$ (see eqs. \eqref{HMI-de=1-zero cutoff} and \eqref{HMI-de neq 1-zero cutoff}).
	\item Since the HEE is independent of the dynamical exponent $z$, the HMI also shows the same behavior.
\end{itemize}
On the other hand, we considered EWCS for two disjoint parallel strips. It was observed that it has the following properties:
\begin{itemize}
	\item It is a decreasing function of the cutoff and goes to zero in the limit $r_c \rightarrow \infty$ (see figure \ref{fig:EW-de=1} and \ref{fig:EW-rc}). It is in contrast to the zero cutoff case where $E_W$ is independent of $r_c$ (see eqs. \eqref{EW-de=1-zero cutoff} and \eqref{EW-0-de neq 1}).
	\item It is a smooth function of both $\ell$ and $h$ (see figures \ref{fig:EW-de=1} and \ref{fig:EW-l-h}), and at the point $h_{\rm crit.}$ where the HMI undergoes a phase transition, it does not show a discontinuous phase transition. It is in contrast to the zero cutoff case. However, at $h_{\rm crit.}$ the concavity of $E_W$ changes. 
	\item By increasing the distance $h$ between the two strips enough, $E_W$ goes to zero, similar to the zero cutoff case (see figures \ref{fig:EW-de=1} and \ref{fig:EW-l-h}). However, in the limit $h \rightarrow 0$, $E_W$ remains finite, which is in contrast to the zero cutoff case where $E_W$ diverges in this limit (see eqs. \eqref{EW-de=1-zero cutoff} and \eqref{EW-0-de neq 1}).
	\item The inequality given in eq. \eqref{EW geq I} is satisfied for all vales of the cutoff $r_c$ as well as $\ell$ and $h$.
	\item It is independent of the dynamical exponent $z$.
\end{itemize}
Furthermore, for $z=1$ and $\theta=0$, the Lorentz and scaling symmetries are restored and the background in eq. \eqref{metric-vacuum} becomes an $AdS_{d+2}$ spacetime in Poincar\'{e} coordinates. Since all of the aforementioned quantities are independent of the dynamical exponent $z$, 
all of our results can be applied for an $AdS_{d+2}$ bulk spacetime in Poincaré coordinates if one sets $\theta=0$.
\\It is also interesting to calculate the EE in HV QFTs and compare it with the results obtained from holography.
\footnote{We would like to thank the referee for her/his helpful comments.}
Recall that in a $T \overline{T}$-deformed CFT, one can calculate the EE by the replica method in which one makes a replicated manifold $\mathcal{M}^n$ from the boundary manifold $\mathcal{M}$ on which the CFT lives \cite{Donnelly:2018bef,Chen:2018eqk,Banerjee:2019ewu,Murdia:2019fax,Jeong:2019ylz,Grieninger:2019zts}. In two dimensions, by applying the factorization property, one can calculate $\langle T \overline{T} \rangle_{\mathcal{M}^n}$ and $\langle T \overline{T} \rangle_{\mathcal{M}}$ in terms of the one-point functions of the components of the stress tensor, and hence find the EE \cite{Chen:2018eqk,Jeong:2019ylz}. In higher dimensions, since the conformal symmetry is less restrictive in comparison to two dimensions, the calculation is very difficult. However, in some situations where the boundary manifold is maximally symmetric such as for $S^{d+1}$, one can calculate EE by applying the factorization property together with the trace flow equation \cite{Donnelly:2018bef,Banerjee:2019ewu,Murdia:2019fax,Grieninger:2019zts}. On the other hand, as mentioned before, in large-N HV QFTs, it is expected that the factorization property still holds. However, since the symmetry group is smaller than the conformal group, one cannot apply the CFT techniques. Therefore, the calculation of EE in a HV QFT living on a flat space seems to be a very difficult task. However, it might still be possible to do this calculation for the case in which the boundary manifold is $S^{d+1}$. It would be very interesting to further study this case. 

\section*{Acknowledgment}

We would like to thank Mohsen Alishahiha very much for his very helpful comments on the draft. We also would like to thank the referee for her/his helpful comments which enriched this work. CP would like to thank Kalyana Rama for very useful discussions. She would also like to express her deep sense of gratitude to Partha Mukhopadhyay for all his very much kind help, which made it feasible for her to carry on this research. The work of FO is supported by Iran Science Elites Federation (ISEF). The work of CP is supported by DAE research fellowship in India.




\end{document}